\documentclass[11pt,a4paper]{article}

\usepackage{tikz}
\usepackage{amsthm,mathrsfs,stmaryrd,bm}
\usepackage[vcentermath]{youngtab}
\usepackage{simplewick}
\usepackage{makecell}	
\usepackage{booktabs}
\usepackage{framed,pifont}
\usepackage{enumitem} 
\usepackage{makecell}

\usepackage{algorithm}
\usepackage{algpseudocode}

\usepackage{graphics} 
\usepackage{graphicx}
\usepackage{epic}
\usepackage{eepic}
\usepackage{slashed}
\usepackage{braket}

\usepackage{feynmp}
\usepackage[retainorgcmds]{IEEEtrantools}
\usepackage{tensor}

\usepackage{verbatim}
\usepackage{latexsym}
\usepackage{amsmath}
\usepackage{wasysym}
\usepackage{amsfonts}
\usepackage{upgreek}
\usepackage{dsfont}
\usepackage{epsf}
\usepackage{epsfig}
\usepackage{amssymb,bm,epsf,epsfig}

\usepackage{jheppub}

\usepackage{color}
\definecolor{rougef}{rgb}{0.56,0,0}
\definecolor{vertf}{rgb}{0,0.5,0}
\definecolor{bleuf}{rgb}{0,0,0.8}
\definecolor{violetf}{rgb}{0.5,0,0.5}

\usepackage{xcolor}
\usepackage{soul}

\newcommand{\es}[2] {\begin{equation} \label{#1} \begin{split} #2 \end{split} \end{equation}}

\usepackage[vcentermath]{youngtab}

 \csname
@addtoreset\endcsname{equation}{section}

\def\3s{{s \choose 3}}
\def\4s{{s \choose 4}}
\def\5s{{s \choose 5}}
\def\6s{{s \choose 6}}

\def\12{\dfrac{1}{2}}

\def\2{\ell_2}

\def\be{\begin{equation}}
\def\ee{\end{equation}}
\def\bea{\begin{eqnarray}}
\def\eea{\end{eqnarray}}
\def\ba{\begin{array}}
\def\ea{\end{array}}
\def\bec{\begin{center}}
\def\ec{\end{center}}

\def\d{\delta}

\def\diff{\mathrm{d}}
\def\ex{\mathrm{e}}
\def\ii{\mathrm{i}}
\def\vol{\mathrm{vol}}
\def\R{\mathbb{R}}
\def\C{\mathbb{C}}
\def\Z{\mathbb{Z}}
\def\ds{\diff s^2}
\def\de{\partial}

\def\Vfive{\mathcal{U}}
\def\Vfour{U}
\def\nc{\mathrm{F}_{\mathrm{cig}}}
\def\ns{\mathrm{F}_{\mathrm{NS5}}}
\def\cy{\mathrm{F}_{\mathrm{CY}}}
\def\CY{{\mathrm{CY}}}
\def\gammaflavor{\tau}
\def\NS5{\mathrm{NS5}}
\def\Dp{\mathrm{D}p}

\def\el{\mathrm{e}}
\def\ma{\mathrm{m}}

\title{On Supergravity and Noncritical Strings}

\author[a]{Andrea Dei,}
\author[b]{Pietro Ferrero,}
\author[c]{Leonardo Rastelli}

\affiliation[a]{Kadanoff Center for Theoretical Physics, Enrico Fermi Institute, and Department of Physics
University of Chicago, 5640 S. Ellis Ave., Chicago IL 60637}
\affiliation[b]{Simons Center for Geometry and Physics, SUNY, Stony Brook, NY 11794, USA}
\affiliation[c]{C. N. Yang Institute for Theoretical Physics, Stony Brook University, Stony Brook, NY 11794, USA}

\emailAdd{adei@uchicago.edu}
\emailAdd{pferrero@sgcp.stonybrook.edu}
\emailAdd{leonardo.rastelli@stonybrook.edu}

\abstract{
Motivated by the long-term goal of finding holographic descriptions for ${\cal N}=1$ and ${\cal N}=2$ super QCD, we revisit the subject of ``noncritical'' superstring theory. Focusing on string models with 6d super Poincar\'e invariance, we provide a careful worldsheet derivation of the leading-order effective theories for the lowest modes. We identify them as {\it seven}-dimensional, {\it maximally supersymmetric} gauged supergravities: the $SO(4)$ theory for  noncritical IIA and the $ISO(4)$ theory for noncritical IIB. The same theories also arise as consistent truncations on $S^3$ of the 10d IIB and IIA  supergravities, respectively, where the chirality flip is as expected from T-duality. These effective supergravities should be interpreted in the following sense. 
The noncritical string can be viewed as a special case of a discrete series of backgrounds labelled by an integer $k$ (which counts  the number of NS5 branes in a certain duality frame); the ``noncritical’' value is $k=2$,
while for $k \to \infty$ one recovers a weakly-curved 10d  target space. The effective supergravities described here give an accurate description of the interactions among the lowest modes for $k \to \infty$, with higher derivative corrections suppressed by powers of $1/k$. 
We discuss BPS solutions of the 7d gauged supergravities and their uplift to 10d solutions.
In particular,  we  find a novel class of solutions with RR flux,
  parametrized
 by a function of three variables that solves an elegant PDE. While we cannot solve the PDE in closed form except in trivial cases, we confirm that our solutions correspond to a 10d IIA Hanany-Witten setup with continuous distributions of both ``color’’ D4 branes and ``flavor’’ D6 branes.  
}

\begin{document}

\begin{fmffile}{diagram}

\maketitle

\setcounter{tocdepth}{2}

\section{Introduction}\label{sec:intro}

In this work we study the interplay between the worldsheet and spacetime descriptions of noncritical superstring theory~\cite{Polyakov:1981rd, Kutasov:1990ua, Kutasov:1991pv, Ooguri:1995wj, Giveon:1999zm, Giveon:1999px, Giveon:1999tq, Hori:2001ax, Murthy:2003es}, focusing on the $d=6$ case.\footnote{As we will elaborate momentarily, we mean the case with $\mathbb{R}^{1, 5}$ Poincar\'e symmetry and sixteen real supercharges.} Our main results are  a careful derivation of the effective supergravity  that captures the lightest modes, a thorough understanding of  half-BPS solutions with NSNS charges, and  a preliminary exploration of RR backgrounds. Noncritical superstrings are an old but still underappreciated subject, and the intricate  web of relationships
between noncritical worldsheet backgrounds,  gauged supergravities and consistent truncations that we explore here would be  worth studying 
in its own right. Our long term motivation is  however more physical: we wish to extend the gauge/string duality to more realistic models.

\subsection*{Motivation: noncritical holography}

The promise implicit in the seminal
work of 't Hooft's~\cite{tHooft:1973alw} is that {\it any} large $N$ gauge theory 
should admit a  dual string theory description in the large $N$ limit, with the string coupling scaling as $g_s \sim 1/N$. 
This vision has been realized in many beautiful examples, which are however all rather close cousins 
of the  paradigmatic case of ${\cal N}=4$ super Yang-Mills theory, as 
they all arise from the decoupling limit of D3 branes\footnote{For definiteness, we focus this discussion to four-dimensional gauge theories.} in critical string theory. While one can partially break supersymmetry ({\it e.g.}~by placing the branes at a local singularity), all models that are under analytic control retain in some way the additional degrees of freedom associated to the six transverse directions. Extending the gauge/string duality to genuinely more minimal theories, such as {\it pure} ${\cal N}=1$ SYM (and, in the fullness of time, to ordinary Yang-Mills)
remains an outstanding challenge.
To be clear, there are several well-known constructions ({\it e.g.}~\cite{Maldacena:2000yy,Apreda:2001qb,Bertolini:2002yr,DiVecchia:2002ks}) that yield {\it e.g.}~${\cal N}=1$ SYM at low energies\footnote{See also \cite{Marotta:2002gc, Nunez:2003cf, Burrington:2004id, Casero:2006pt, Casero:2007jj, Casero:2007pz, Hoyos-Badajoz:2008znk} for the inclusion of fundamental flavor.} (to stick for concreteness to this important example),
but within the supergravity approximation
they inevitably depart from it at energies larger than $\Lambda_{\rm QCD}$. A full-fledged dual of the ``pure'' theory will necessitate a spacetime description that 
incorporates all $\alpha'$-corrections, or equivalently a worldsheet description that 
is intrinsically strongly coupled as a two-dimensional sigma model. Indeed, unlike the canonical example of ${\cal N}=4$ SYM, there  is no exactly marginal deformation that allows to reach the weakly-curved  large volume limit, and the worldsheet CFT is expected to be an isolated fixed point.

The most direct route to the pure ${\cal N}=1$ and ${\cal N}=2$ SYM theories,\footnote{Alternative approaches to this problem are possible for the case with eight supercharges as well: for pure $\mathcal{N}=2$ SYM, \cite{Gauntlett:2001ps, Bigazzi:2001aj, DiVecchia:2002ks} consider D5 branes wrapped on $S^2$, while \cite{Bertolini:2000dk, Petrini:2001fk} use fractional D3 branes. The inclusion of flavor was considered, {\it e.g.}, in \cite{Paredes:2006wb} for the former setup and in \cite{Bertolini:2001qa} for the latter.} 
and to their variations with fundamental flavor, is within noncritical string theory. ``Noncritical'' is a bit of a traditional
misnomer. Noncritical backgrounds are  perfectly critical as cancellation of the worldsheet Weyl anomaly goes, but with a Liouville mode carrying a larger amount $c_{\rm L}$ of the anomaly, such that the
``matter'' central charge $c_{\rm M}$
associated to the ``spacetime'' dimensions ({\it e.g.}~to the standard Poincar\'e invariant background $\mathbb{R}^{d-1, 1}$, with $c_{\rm M} = d$)
can be smaller, $d = 26-c_{\rm L}$ for the bosonic string and $d = 10-c_{\rm L}$ for the superstring.
 In the bosonic case, as well as in the superstring
case with ${\cal N}=1$ worldsheet supersymmetry, 
one is restricted to $c_{\rm M} \leq 1$, as superLiouville theory is otherwise not well-defined.
Fortunately, there is a generalization with ${\cal N}=2$ worldsheet supersymmetry that makes perfect sense for any {\it even} spacetime dimension $d$. The appropriate ${\cal N}= 2$ version of Liouville theory is the Kazama-Susuki coset $SL(2, \mathbb{R})_k/U(1)$,
which is equivalent to the supersymmetric cigar sigma model~\cite{FZZ_unpublished, Ivanov:1983wp, Kazama:1988qp, Hori:2001ax, Hikida:2008pe}. The worldsheet CFTs
\begin{equation} \label{noncritical}
\mathbb{R}^{1, d-1}\times  \frac{SL(2, \mathbb{R})_k}{U(1)}\, ,
\end{equation}
where the level $k$ of the coset is adjusted to cancel the total Weyl anomaly, $k = 4/(8-d)$,
define consistent superstring models for
any even $d \leq 8$. (The case $d=8$ is just the  usual $\mathbb{R}^{1, 9}$ ``critical'' background in disguise.)
 A suitable GSO 
projection makes the spacetime supersymmetric,
with $2^{d/2 +1}$ real supercharges. 

The role of the Liouville mode in the putative string  dual
of large $N$ Yang-Mills has long been emphasized by Polyakov~\cite{Polyakov:1997tj, Polyakov:1998ju, Polyakov:2000fk}. While formulating a suitable worldsheet theory
remains a tall order in the
fully non-supersymmetric case, a blueprint for how this could be accomplished for ${\cal N}=1$ and ${\cal N}=2$ SYM
has been proposed in \cite{Kuperstein:2004yk, Klebanov:2004ya}.\footnote{Various extensions and generalizations have also been considered, see {\it e.g.} \cite{Alishahiha:2004yv,Kuperstein:2004yf,Bigazzi:2005md,Iatrakis:2010jb}.} The idea is  to imitate the logic that leads to the standard AdS/CFT duality for ${\cal N}=4$ SYM.
One should study the decoupling limit of a stack of D3 branes, but 
in the noncritical backgrounds that we have just described, rather than in ten-dimensional flat space. To realize ${\cal N}=1$ SYM,
the starting point is the noncritical IIB background 
with $d=4$.  A D3 brane is defined by a certain exact boundary state for ${\cal N}=2$ superLiouville (semiclassically, it is a brane localized near the tip of the cigar geometry) tensored with the standard Neumann boundary state for $\mathbb{R}^{1,3}$. One checks that the massless spectrum and supersymmetry are those of the ${\cal N}=1$ vector multiplet \cite{Ashok:2005py,Fotopoulos:2005cn}; the absence of ``geometric'' transverse directions to the D3 branes dovetails with the absence of scalar fields in the multiplet.
One can then attempt to imitate the familiar logic~\cite{Maldacena:1997re}. In the limit $\alpha' \to 0$, the massive open string modes decouple, while on the dual closed string side one 
 is instructed to take the  near-horizon limit of the backreacted geometry. The string dual of pure 
${\cal N}=1$ SYM is then identified with the closed string sigma model in the corresponding RR background. Similarly, to realize pure ${\cal N}=2$ SYM one considers a stack of D3 branes in the noncritical background \eqref{noncritical} with $d=6$. The boundary state consists again of the localized brane at the tip of the cigar, tensored with Neumann boundary conditions on $\mathbb{R}^{1, 3} \subset \mathbb{R}^{1, 5}$ and Dirichlet boundary conditions on the transverse $\mathbb{R}^2$ directions. The massless spectrum and supersymmetry are now those of the ${\cal N}=2$ vector multiplet, with the two real scalars corresponding to the two transverse directions. Again, one is led to identify the dual string theory with the closed string sigma model on the near-horizon backreacted background.
In both the ${\cal N}=1$ and ${\cal N}=2$ cases, fundamental flavor can be included by adding D5 branes extending in the cigar directions --
see Tables \ref{tab:cigar_branes_6d} and \ref{tab:cigar_branes_8d} and  for a summary of the brane configurations.

\begin{table}[!ht]
\begin{center}
\begin{tabular}{|c||c|c|c|c|c|c|}
\hline
 & \multicolumn{4}{c|}{$\R^{1,3}$} & \multicolumn{2}{c|}{$SL(2,\mathbb{R})_1/U(1)$}\\
\hline
$N_c$ D3 & $\times$ & $\times$ & $\times$ & $\times$ & \hspace{1cm} & \hspace{1cm} \\
\hline
$N_f$ D5 & $\times$ & $\times$ & $\times$ & $\times$ &  $\times$ & $\times$ \\
\hline 
\end{tabular}
\caption{Distribution of D-branes in (4+2)-dimensional type IIB noncritical string theory corresponding to a 4d $\mathcal{N}=1$ $SU(N_c)$ gauge theory with
$N_f$ fundamental and $N_f$ antifundamental chiral multiplets.}
\label{tab:cigar_branes_6d}
\end{center}
\end{table}
\begin{table}[!ht]
\begin{center}
\begin{tabular}{|c||c|c|c|c|c|c|c|c|}
\hline
 & \multicolumn{4}{c|}{$\R^{1,3}$} & \multicolumn{2}{c|}{$\R^{2}$} & \multicolumn{2}{c|}{$SL(2,\mathbb{R})_2/U(1)$}\\
\hline
$N_c$ D3 & $\times$ & $\times$ & $\times$ & $\times$ & \hspace{0.3cm} &  \hspace{0.3cm} & \hspace{1cm} & \hspace{1cm} \\
\hline
$N_f$ D5 & $\times$ & $\times$ & $\times$ & $\times$ & \hspace{0.3cm} & \hspace{0.3cm}& $\times$ & $\times$ \\
\hline 
\end{tabular}
\caption{Distribution of D-branes in (6+2)-dimensional type IIB noncritical string theory corresponding to a 4d $\mathcal{N}=2$ $SU(N_c)$ gauge theory   with $N_f$ fundamental hypermultiplets.}
\label{tab:cigar_branes_8d}
\end{center}
\end{table}

While this is a compelling physical picture, there are obvious technical 
challenges in turning
this cartoon into a precise proposal. The requisite spacetime backgrounds have string-scale curvature, so that
two-derivative supergravity is not expected to be a valid approximation.
 To circumvent this problem, one can aim for a full-fledged worldsheet description of the backreacted closed string backgrounds. However, a worldsheet approach  faces  
 the usual predicament that RR backgrounds are unwieldy in the standard RNS formulation, and an alternative worldsheet  formalism ({\it e.g.}~using pure spinors~\cite{Berkovits:2000fe, Berkovits:1999im}) might need to be developed. Perhaps the way forward will be a  spacetime approach capable of systematically incorporating $\alpha'$-corrections, such as 
 string field theory,  along the lines of~\cite{Cho:2018nfn, Cho:2023mhw}. 

In this paper, we will achieve a thorough understanding of the effective supergravity that captures the low-lying modes of the noncritical string \eqref{noncritical}, starting  with the $d=6$ case. As we have just emphasized, supergravity alone cannot be the final tool, but it should still  be very useful as a stepping stone. 
In fact, the first motivation of this work was to put on a firmer footing the approach of~\cite{Kuperstein:2004yk, Klebanov:2004ya} to noncritical holography. In particular, starting from a somewhat {\it ad hoc} six-dimensional bosonic effective action (in our notation, this is the $d=4$ case, with the two extra dimensions corresponding to the cigar directions), Klebanov and Maldacena~\cite{Klebanov:2004ya}
found some intriguing solutions in rough qualitative agreement with the physics of ${\cal N}=1$ SQCD in the conformal window. Their strategy was to
make an ansatz in terms of first order equations,
derived from a ``fake superpotential'',\footnote{The term was introduced in \cite{Freedman:2003ax}, see also \cite{Celi:2004st} for comments on the relation between this approach and ``honest'' supergravity in the case of 5d $\mathcal{N}=2$ supersymmetry.}
which in a more complete analysis should correspond to actual supersymmetry variations.
Our goal is to obtain the correct and complete effective supergravity from the underlying microscopic string theory, and to derive and study the corresponding BPS equations.

\subsection*{Supergravity for the $\boldsymbol{d=6}$ noncritical superstring}

In this work, we will focus on the $d=6$ noncritical superstring, with an eye towards holography for ${\cal N}=2$ SQCD.  The $d=4$ case,  relevant for ${\cal N}=1$ SYM, is morally similar but presents several additional complications and is the subject of ongoing work~\cite{wip}. 
Following~\cite{Giveon:1999zm, Giveon:1999px, Giveon:1999tq}, we will find it both technically useful and conceptually insightful 
to view the ``noncritical'' background as
a special case of an infinite discrete family,
defined by the  worldsheet CFTs 
\begin{equation}
\R^{1,5} \times \left( \frac{SL(2,\R)_k}{U(1)} \times \frac{SU(2)_k}{U(1)} \right) \Bigl/ \Z_{k} \, ,  \qquad k \geq 2 \,.
    \label{10d-ws-CFT-intro}
\end{equation}
Apart from the six flat directions, the background comprises both the supersymmetric cigar
$SL(2,\R)_k/ U(1)$ CFT and 
the supersymmetric parafermion CFT $SU(2)_k/U(1)$,
which semiclassically has the geometric interpretation
of a sigma model into the disk.
The total matter central charge has the correct critical value  for any $k$. 
For $k=2$, the parafermion CFT trivializes and one recovers the noncritical background (\ref{noncritical}) with $d=6$.
By contrast, for $k$ large both the cigar
and the disk sigma models are weakly coupled and one recovers the flat 10d critical string.
This family of backgrounds arises naturally by considering the near-horizon limit of  $k$ parallel NS5 branes, separated in a symmetric configuration 
around a (contractible) circle \cite{Giveon:1999px, Giveon:1999tq}. More precisely,
the CFT (\ref{10d-ws-CFT-intro}) is T-dual to such NS5 brane configuration \cite{Sfetsos:1995ac,Sfetsos:1998xd}, see Figure \ref{fig:frames}.  Although we are most interested in the noncritical value $k=2$, it is both easy and useful to  keep $k$ generic. For large $k$, there is a clear sense in which two-derivative supergravity provides a good effective field theory description  for the lowest  modes, with higher-derivative corrections suppressed by inverse powers of $k$.\footnote{Looking ahead at the application to noncritical holography, we can envision a family of RR backgrounds where $k$ is a parameter, which become weakly-curved for large $k$; equivalently, the corresponding 2d worldsheet theories are weakly-coupled for large $k$. For each $k$, there is a dual 4d ${\cal N}=2$ field theory, which reduces to ${\cal N}=2$ SYM for $k=2$.} 

One of our main results is a careful derivation
of the effective supergravity. As we are only interested in the continuum of delta-function normalizable states,
the worldsheet analysis can be performed by replacing the cigar background $SL(2)/U(1)$ with its asymptotic region as $\rho \to \infty$,
where it simplifies to the free theory of a linear dilaton
$\mathbb{R}_\rho$ times a compact boson $S^1$.
For both the IIB theory (the one relevant for the holographic dualities of 4d gauge theories discussed above) and  the IIA theory, the effective theory that describes the lowest modes turns out to be a {\it seven}-dimensional, {\it maximally supersymmetric} gauged supergravity. 

Both features (7d and maximal supersymmetry) may appear surprising at first sight. 
For starters, one might have  anticipated an
{\it eight}-dimensional supergravity description, as the noncritical background comprises the six Minkowski directions and the two cigar directions $\mathbb{R_\rho} \times S^1$. What goes wrong with dimensionally reducing
the backgrounds (\ref{10d-ws-CFT-intro}) on the disk $SU(2)/U(1)$ to find an eight-dimensional effective theory?
The problem is that while the full-fledged worldsheet theory
(\ref{10d-ws-CFT-intro}) enjoys 16 spacetime supercharges, the corresponding solution of 10d type II supergravity breaks supersymmetry completely. 
 Supersymmetry is instead manifest in the T-dual picture,\footnote{The fact that T-duality can break manifest supersymmetry at the supergravity level is not unfamiliar~\cite{Kiritsis:1993pb, Hassan:1995je, Sfetsos:1995ac, Alvarez:1995np, Hassan:1999bv}: it happens whenever the Lie derivative of the Killing spinors with respect to  the T-duality direction is non-zero.} corresponding to a configuration of NS5 branes,
which asymptotically  reduce to the CHS~\cite{Callan:1991at, Callan:1991dj} background $\mathbb{R}^{1, 5} \times \mathbb{R}_\rho \times S^3$. Dimensionally reducing on $S^3$ yields a valid, but {\it seven}-dimensional, supergravity description.
The corresponding supergravities have been identified before, as certain gaugings of the maximally supersymmetric 
theory. To wit, the reduction of 10d IIA on $S^3$ yields a 7d sugra  with $ISO(4)$ gauging, while the reduction of 10d IIB yields the 7d sugra with $SO(4)$ gauging. Because of the T-duality involved, we identify the $ISO(4)$ and $SO(4)$ gauged supergravities as the appropriate descriptions of noncritical IIB and IIA, respectively.
In fact, in both cases there is a full-fledged nonlinear consistent truncation ansatz that allows to uplift any 7d solution to a 10d solution. A strong hint that such a consistent truncation must be possible comes from the worldsheet analysis, where the vertex operators for the lowest modes are seen to form a consistent subsector.

The other perhaps surprising feature is the counting of supercharges. The point is that 
while the $SO(4)$ and $ISO(4)$ 7d gauged supergravities are maximally supersymmetric, they do not admit maximally supersymmetric Minkowski or AdS  solutions. The simplest background is instead the half-BPS linear dilaton solution $\mathbb{R}^{5, 1} \times \mathbb{R}_\rho$. This agrees with the 
the counting of supercharges from the worldsheet description of the noncritical string.

 We
perform a detailed analysis of the worldsheet spectrum.
On the worldsheet, only 6d super Poincar\'{e} invariance (with 16 supercharges) is manifest, and some care is needed in organizing the spectrum of BRST classes from a 7d perspective.
Several technical subtleties enter the comparison between the worldsheet and the supergravity descriptions, especially in the RR sector, but 
when the dust settles we
find a perfect match. In particular, there is a sense in which the lowest states that we focus on are ``massless'', namely gauge invariance, even if, as we have emphasized, there is no parametric separation of scales to the higher levels.

We also review the construction of boundary states \cite{DiVecchia:1997vef, Billo:1998vr, Fotopoulos:2005cn, Ashok:2005py, Murthy:2006xt} and identify which of low-lying closed string vertex operators have  non-zero one-point functions.
We find that it is the same set for both the ``color'' D3 branes and the ``flavor'' D5 branes, but with different asymptotic rates of fall off in the radial cigar direction $\rho \to \infty$.

\subsection*{Supergravity solutions}

The consistent truncation ansatz gives a powerful recipe to find 10d solutions by uplifting solutions of the 7d gauged supergravities. For purely NSNS backgrounds,
these solutions have been obtained before.
They correspond to the backreacted geometries of various (smeared) configurations of NS5 branes. 
We revisit these solutions from our perspective of noncritical string theory. 
We also discuss an 8d effective action (of the type introduced in~\cite{Kuperstein:2004yk, Klebanov:2004ya}, but limited to NSNS fields) for the noncritical string, showing how to relate the 8d ``fake superpotential'' approach to the genuine 7d BPS equation arising from the gauged supergravities. 

Our real interest is in the putative RR backgrounds corresponding to the configurations of branes that realize ${\cal N}=2$ SYM with flavor. 
We consider a truncation of the 7d $ISO(4)$ supergravity that retains the fields sourced by the color and flavor branes and manifests the $SU(2)_R$ R-symmetry expected on the gauge theory side.
Making an ansatz that preserves super Poincar\'{e} invariance in $\mathbb{R}^{1, 3}$, we
are able to reduce the full set of BPS equations to a single master equation, the elegant nonlinear PDE  (\ref{PDE4d}) for a single function $K (r, \vec y)$ of the three transverse coordinates. For any such solution $K$, we can write the general uplift to a BPS solution of 10d IIA supergravity. 

Unfortunately, we have only been able to find very special, rather trivial explicit solutions of our master equation. We can however observe some general characteristics that hold for any $K$. We confirm that our class of 10d solutions does  indeed correspond to a IIA Hanany-Witten setup, comprising 
NS5 branes, ``color'' D4 branes and ``flavor'' D6 branes.   A first, inevitable feature is that a large number $k$ of NS5 branes are present, while to engineer ${\cal N}=2$ SQCD we need $k=2$. But this is something that we were expecting all along: the two-derivative approximation corresponds to large $k$, and in fact we wish to regard $1/k$ as the parameter that controls the higher derivative corrections. Note that in  our 10d solutions the $SU(2)_R$ symmetry is realized geometrically: the $k \to \infty$ NS5 branes are (continuously) distributed on a segment,  rather than on a circle, as in the exact worldsheet backgrounds (\ref{10d-ws-CFT-intro}) that only preserve the Cartan of $SU(2)_R$. Of course, there is no difference for the $k=2$ case which is our ultimate target. From our perspective it is useful to preserve $SU(2)_R$ for any $k$, but by making this choice we have lost (for $k>2$) the connection with the exact CFT
(\ref{10d-ws-CFT-intro}).
It would be very interesting to find an exact worldsheet description for the specific linear distribution of NS5 branes that appears in our solutions. 

A second feature is that
all our solutions inevitably contain both D4 and D6 branes, as both types of RR flux originate from the 10d uplift of the same 7d RR one-form. A calculation of the fluxes (valid for generic $K$) shows that the ratio of the numbers of D4 and D6 branes is irrational, so that they cannot both be quantized.
We interpret this unpleasant feature in terms of a partial smearing of the D4 branes along the $S^2$ that realizes geometrically $SU(2)_R$. Finally, 
we find that the asymptotic behavior of RR fields
in our solutions is very reminiscent of what we obtain in Section \ref{sec:boundarystates} from a boundary states analysis, despite the caveat that our worldsheet analysis  was performed for the noncritical value $k=2$.

\bigskip

\noindent
The detailed outline of the paper is best apprehended from the table of contents. In Section~\ref{sec:spectrum}, after a short self-contained review of $d=6$ noncritical superstrings, we perform a detailed worldsheet analysis of the low-lying  spectrum, which we organize from a 7d perspective.
In Section \ref{sec:boundarystates} we review the relevant boundary states of the $SL(2)/U(1)$ cigar CFT, determine which of the low-lying physical closed string states have nonzero overlaps with them and investigate the asymptotic  behavior of these disk one-point functions.
In Section \ref{sec:7dsugra} we identify the effective 7d actions for the low-lying states with two maximally supersymmetric gauged supergravities,
namely the $SO(4)$ gauged sugra for  noncritical IIA and $ISO(4)$ gauged sugra for noncritical IIB. We perform a detailed match with the worldsheet analysis and explain how the same result could have been anticipated by KK reduction on $S^3$ in a T-dual picture. In Section \ref{sec:NSNS} we review 
 BPS solutions with only NSNS flux from our noncritical perspective. We also explain the connection of the 7d BPS equations with an 8d ``fake superpotential'' approach.
In Section \ref{sec:RR} we begin an exploration of BPS solutions with RR flux.
We look for solutions of the 7d BPS equations that uplift to a 10d IIA Hanany-Witten setup. We find a general class of them,
parametrized by solutions of a single nonlinear PDE,
and investigate their general properties.
We conclude in Section \ref{sec:outlook} with a brief outlook. We relegate to five appendices some of the more technical material.

\section{Noncritical superstring theory}
\label{sec:spectrum}

\subsection{A brief overview}

The main theme of this paper is $d$-dimensional noncritical superstring theory \cite{Kutasov:1990ua,Kutasov:1991pv}, which we now briefly review. This is a rich topic that has been studied by several authors from a variety of different perspectives, and we will not attempt an exhaustive overview here. Rather, we shall focus on aspects that are instrumental to our main goal: finding a supergravity theory that captures the interactions between string states in a subsector at the lowest mass level, mimicking the relation between ten-dimensional superstring theory and supergravity. In this section we will review some relevant aspects of noncritical superstring theory in general dimension $d \leq 8$, but eventually specialize to $d=6$ which is the main focus of this paper. Most of the calculations presented here have already appeared -- in some version -- in the literature, and we shall give appropriate references when relevant. Besides reviewing relevant material, the purpose of this section is to present (for $d=6$) the calculation of the spectrum at the lowest level from a new, seven-dimensional, perspective. Namely, we interpret the spectrum as arising from the linearized fluctuations of a seven-dimensional theory whose Poincaré symmetry is spontaneously broken to six dimensions by the linear dilaton background. Such interpretation sheds light on the underlying maximal supersymmetry of the spectrum, allows to distinguish the type IIA and IIB GSO projections and ultimately leads to uniquely identifying the corresponding two-derivative supergravity.

\paragraph{Linear dilaton and cigar.}
Let us consider a worldsheet CFT containing $d$ free bosons $X_{\alpha}$, $d$ free fermions $\psi_\alpha$, with $\alpha=0,\ldots,d-1$, as well as the usual ghost system of the ten-dimensional critical superstring. For generic $d$, the total central charge  
\begin{equation}
    \frac{3 d}{2} - 15 < 0
\end{equation}
is not zero and the theory is not critical. The conformal anomaly can be cancelled by adding a Liouville mode \cite{Polyakov:1981rd}. Following \cite{Kutasov:1990ua,Kutasov:1991pv} we also require $\mathcal{N}=2$ worldsheet supersymmetry, which combined with the GSO projection is enough to ensure spacetime supersymmetry~\cite{Banks:1987cy}. One is thus led to consider a worldsheet CFT obtained by tensoring $\mathcal{N}=2$ superLiouville theory, containing bosonic fields $(\rho,\theta)$ and fermions $(\psi^{\rho},\psi^{\theta})$, together with the $(X_{\alpha},\psi_{\alpha})$ free fields. The simplest worldsheet CFT with this field content is the so-called linear-dilaton background, where we think of the Liouville field $\rho$ as a geometric direction and the dilaton field varies linearly with $\rho$,
\begin{align}
    \Phi=-\frac{Q}{2}\rho\,,
\end{align}
in such a way that the $\rho\to +\infty$ region is weakly coupled. We shall discuss how to deal with the strongly coupled region at negative $\rho$ momentarily. The corresponding geometric background is
\begin{align}\label{lineardilatonmetricd+2}
    \diff s^2=\diff s^2(\R^{1,d-1})+\diff\rho^2+\diff\theta^2\,,\qquad \theta\sim \theta+\frac{4\pi}{Q}\,,
\end{align}
where the periodicity of $\theta$ is dictated by the choice of conventions made in what follows. To cancel the Weyl anomaly, we must tune the parameter $Q$ of the linear dilaton in such a way that the total central charge of the system is zero. Given that
\begin{align}
    c_{\text{matter}}=\frac{3d}{2}\,, \qquad
    c_{\text{ghosts}}=-15\,, \qquad
    c_{\text{sLiouville}}=3(1+Q^2)\,,
\end{align}
we find that this is a consistent background for string propagation, that is
\es{}{
 c_{\text{tot}} =c_{\text{matter}}+c_{\text{ghosts}}+c_{\text{sLiouville}}=0\,,
}
when \cite{Kutasov:1990ua}
\begin{align}
    Q=\sqrt{\frac{8-d}{2}}\,.
\end{align}
Critical superstring theory corresponds, in this language, to the case $d=8$, where $Q=0$ and $(\rho,\theta)$ become two additional flat direction. The other cases that are interesting for us are those that preserve $\mathcal{N}=2$ worldsheet supersymmetry, which only happens for {\it even} $d$, leaving us with the three interesting cases $d=2,4,6$, the latter being the focus of this paper. 

\begin{table}
\begin{center}
\begin{tabular}{|c||c|c|c|c|}
\hline
 $d$ & 2 & 4 & 6 & 8\\
 \hline
 $Q$ & $\sqrt{3}$ & $\sqrt{2}$ & 1 & 0 \\
 \hline
 $k$ & $2/3$ & 1 & 2 & $\infty$\\
 \hline
\end{tabular}
\caption{Values of the parameters $Q$ and $k$ for the noncritical superstring theories of interest. The value $d=8$ corresponds to the critical case.}
\label{tab:dQk}
\end{center}
\end{table}

A problem of the description above is that, as already pointed out, it develops a strong coupling singularity as $\rho \to -\infty$. This can be cured by replacing the linear dilaton theory with the so-called ``cigar CFT'' \cite{Witten:1991yr, Dijkgraaf:1991ba}, {\it i.e.}~the Kazama-Susuki coset $SL(2,\R)_k/U(1)$ \cite{Kazama:1988qp}, which is dual (via mirror symmetry) to $\mathcal N =2$ Liouville theory \cite{FZZ_unpublished,Hori:2001ax,Hikida:2008pe}. The level $k$ of the coset\footnote{Here $k$ is the ``supersymmetric level'', meaning that we consider the $\mathcal{N}$=2 supersymmetric cigar CFT. This corresponds to the bosonic $SL(2,\R)_{k+2}$ WZW model plus three fermions, where a supersymmetric $U(1)$ current is gauged.} is related to the parameter $Q$ of the linear dilaton model by
\begin{align}\label{c_SL2U1}
    c\left[\tfrac{SL(2,\R)_k)}{U(1)}\right]=3\left(1+\tfrac{2}{k}\right)=3(1+Q^2)=c_{\text{sLiouville}}\,,\qquad \iff \qquad
    k=\frac{2}{Q^2}\,.
\end{align}
\noindent The values of $Q$ and $k$ that are relevant for the interesting supersymmetric models are summarized in Table \ref{tab:dQk}. The corresponding $d+2$-dimensional target space geometry is
\begin{align}\label{cigartargetspace}
\begin{split}
    \diff s^2&=\diff s^2(\R^{1,d-1})+\diff\rho^2+\tanh^2\tfrac{Q\rho}{2}\diff\theta^2\,,\qquad \theta\sim \theta+\frac{4\pi}{Q}\,,\\
    \Phi&=-\log \cosh \tfrac{Q\rho}{2}\,.
\end{split}
\end{align}
The geometry is that of a cigar that caps off smoothly at $\rho=0$ (where curvature and string coupling are finite), which fixes the periodicity of $\theta$. On the other hand, as $\rho \to +\infty$ the geometry is well approximated by a cylinder and the background reduces to the linear-dilaton one, introduced in \eqref{lineardilatonmetricd+2}.

\paragraph{Embedding in critical string theory and dualities.}

While the models introduced above represent consistent backgrounds for the propagation of noncritical strings, they are inherently $d+2$-dimensional, so they are not immediately embedded into critical string theory (except for the trivial case $d=8$). However, for the purposes of this paper it will be particularly useful to have such explicit embedding, and we now review how this can be achieved. The key idea can be described as follows, following \cite{Aharony:1998ub,Giveon:1999zm,Giveon:1999px}. For a fixed value of $d$, imagine to keep $k$ (or $Q$) generic, and tensor the linear dilaton or cigar model with a new CFT $\mathcal{M}$, which admits a parameter that we can use to set the total central charge to zero. The new worldsheet CFT is then of the form
\begin{align}\label{cigargeometry10d}
    \R^{1,d-1}\times \frac{SL(2)_k}{U(1)}\times \mathcal{M}\,. 
\end{align}
Requiring that the background \eqref{cigargeometry10d} is ten-dimensional we find
\begin{align}\label{dimM}
    \text{dim}(\mathcal{M})=2(n-1)=8-d\,,\qquad 2n\equiv 10-d\,. 
\end{align}
We are assuming that $\mathcal{M}$ admits a semi-classical (geometrical) limit and by $\text{dim}(\mathcal{M})$ we mean the dimension of its target space geometry. The condition that the total central charge vanishes in \eqref{cigargeometry10d},
\es{ctot_embed}{
c_{\text{tot}} =c_{\text{matter}}+c_{\text{ghosts}}+c\left[\tfrac{SL(2,\R)_k)}{U(1)}\right]+c_{\mathcal{M}}=0\,,
}
fixes the central charge $c_{\mathcal{M}}$, 
\begin{align}\label{c_M}
    c_{\mathcal{M}}=\frac{3}{2}(8-d)-3Q^2=\frac{3}{2}(8-d)-\frac{6}{k}\,.
\end{align}
When $k$ approaches one of the values listed in Table \ref{tab:dQk} the central charge of $\mathcal{M}$ vanishes, and the worldsheet theory is described by \eqref{cigartargetspace}. This is often interpreted as ``losing'' the $8-d$ spacetime dimensions described in \eqref{dimM}: hence the name ``noncritical'', intended as a string theory defined in $d+2<10$ dimensions. Despite the name, this is still a ``critical'' theory in the sense that the total central charge (including the ghosts) vanishes, as described by \eqref{ctot_embed}. On the other hand, when considering the ten-dimensional string theory \eqref{cigargeometry10d}, $k$ is a tunable parameter and we can consider the limit $k\to \infty$. In this limit, the CFT $\mathcal{M}$ is weakly coupled and indeed its central charge \eqref{c_M} in this regime is the appropriate one for a non-linear sigma model with a target space of dimension as in \eqref{dimM}.

The proposal of \cite{Giveon:1999zm} is to take $\mathcal{M}$ to be a $\mathcal{N}=2$ Landau-Ginzburg (LG) SCFT with $n+1$ chiral superfields $z_a$, $a=1,\ldots,n+1$, with superpotential $W=F(z_a)$, where $F$ is a quasi-homogenous polynomial with weight one under 
\begin{align}\label{z_rescaling}
    z_a\to \lambda^{w_a}\,z_a\,,
\end{align}
that is
\begin{align}
    F( \lambda^{w_a}\,z_a)=\lambda\,F(z_a)\,,\quad \lambda\in \C\,.
\end{align}
The central charge of this theory is
\begin{align}
    c_{\mathcal{M}}=3(n-1-2\bar{w})\,,\qquad \bar{w}\equiv -1+\sum_{a=1}^{n+1}w_a\,,
\end{align}
so eq. \eqref{c_M} fixes
\begin{align}\label{barw}
    Q^2=2\bar{w}\,.
\end{align}
String dualities provide useful alternative descriptions of this setup \cite{Giveon:1999zm}, which can be reached after performing certain T-dualities. Let us then refer to the one described so far as the duality frame $\nc$ and explore the other two frames that will be relevant for this paper. A convenient way to present the other frames is to start with critical string theory at a Calabi-Yau (CY) singularity
\begin{align}
    \R^{1,d-1}\times \CY_n\,,
\end{align}
where $\CY_n$ is a non-compact singular CY $n$-fold, described as the hypersurface singularity $F(z_1,\ldots,z_{n+1})=0$ in $\C^{n+1}$. Sending the string coupling $g_s\to 0$ with fixed string length $\ell_s$ then leads to a theory without gravity describing the interaction between the modes localized near the singularity of $\CY_n$ (located at $z_a=0$) \cite{Giveon:1999zm}. We refer to this frame as the CY frame $\cy$. The proposal of \cite{Giveon:1999zm} is that this should be dual to the string background
\begin{align}\label{FNS5}
    \R^{1,d-1}\times \R_{\rho}\times \mathcal{N}\,,
\end{align}
where $\mathcal{N}=\CY_n/\R_+$, the action of $\R_+$ being given by \eqref{z_rescaling} with $\lambda\in \R_+$. Here $\rho$ is a linear dilaton direction, which makes it closer to the setup we started with. We refer to this frame as $\ns$, since as we shall discuss in the case of interest it can be realized with a certain configuration of NS5 branes.

\begin{figure}[!ht]
\centering
    \includegraphics[width=\textwidth]{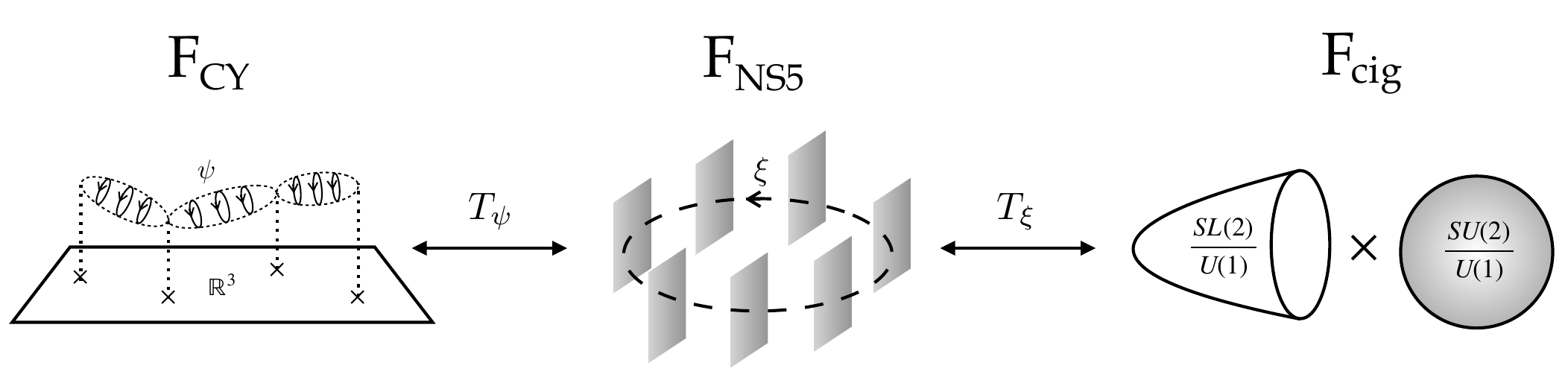}
\caption{T-duality frames in the case $d=6$. In the frame $\cy$ we have type II string theory on a Taub-NUT space with $k$ monopoles. T-duality along the circle parametrized by $\psi$ in the picture gives a set of parallel NS5 branes localized at points in a circle in $\R^4$ \cite{Ooguri:1995wj} in the frame $\ns$. A further T-duality along the direction of such circle, parametrized by $\xi$ in the picture, gives the worldsheet CFT \eqref{10d-ws-CFT} in the frame $\nc$ \cite{Sfetsos:1998xd}.}
\label{fig:frames}
\end{figure}

The final duality, which brings us back to the frame $\nc$ we started with, is realized by observing that choosing $\lambda\in \R_+$ in the quotient above still leaves a leftover $U(1)$ symmetry given by \eqref{z_rescaling} with $|\lambda|=1$. The analysis of \cite{Giveon:1999zm} focuses on cases where 
\begin{align}\label{N_vs_N/U(1)}
    \mathcal{N}\simeq U(1)\times \frac{\mathcal{N}}{U(1)}\,,
\end{align}
where the product above needs not be a direct product, and the $U(1)$ used in this decomposition is the leftover $U(1)$ action just discussed. We then arrive to the background
\begin{align}\label{Fcig}
     \R^{1,d-1}\times \R_{\rho}\times U(1)\times \frac{\mathcal{N}}{U(1)}\,,
\end{align}
which corresponds to the large $\rho$ ({\it i.e.}~linear dilaton) limit of \eqref{cigargeometry10d} if we identify
\begin{align}
    \mathcal{M}=\frac{\mathcal{N}}{U(1)}\,,
\end{align}
that is the quotient $\mathcal{N}/U(1)$ in the $\mathcal{N}=2$ LG model with superpotential $W=F(z_a)$. In the following, a major role will be played by the two frames $\nc$, which is our original definition of noncritical string theory, as well as the dual frame $\ns$, which will be instrumental to relate our supergravity description of the low-energy fluctuations of the noncritical string to the critical setup discussed above. A visual representation of the three frames can be found in Figure \ref{fig:frames}, which focuses on the case $d=6$ described in the next subsection.

\subsection[The $d=6$ noncritical superstring theory]{The $\boldsymbol{d=6}$ noncritical superstring theory}
\label{sec:6dnoncriticalworldsheet}

So far, our discussion has been pretty general. From now on, we specialize to the case of $d=6$ (or $n=2)$. In this case, the central charge $c_{\mathcal{M}}$ reads
\begin{align}
    c_{\mathcal{M}}=3(1-2\bar{w})<3\,,
\end{align}
as from \eqref{barw} we find that $\bar{w}>0$. All unitary 2d $\mathcal{N}=2$ SCFTs with $c<3$ are classified and they correspond to $\mathcal{N}=2$ minimal models \cite{Zamolodchikov:1986gh}, so the only possibility is \cite{Goddard:1984vk}\footnote{Note that $\mathcal{N}=2$ minimal models are in one-to-one correspondence with ADE singularities: the CY twofold $\CY_2$ is then an ALE space with the corresponding ADE singularity, which is the description in the frame $\cy$ \cite{Gepner:1987vz,Greene:1988ut}.}
\es{Mdisk}{
\mathcal{M}=\frac{SU(2)_k}{U(1)}\,,
}
where $SU(2)_k$ denotes the $\mathcal{N}=1$ supersymmetric WZW model at level $k$ with central charge $c[SU(2)_k]=\tfrac{9}{2}-\tfrac{6}{k}$. Gauging a supersymmetric $U(1)$ gives rise to the $\mathcal{N}=2$ supersymmetric CFT \eqref{Mdisk}, which is often referred to as the parafermion disk CFT and has central charge
\es{c_SU2U1}{
c\left[\frac{SU(2)_k}{U(1)}\right]=3\,\left(1-\frac{2}{k}\right)\,,
}
in agreement with \eqref{c_M} for $d=6$. We then obtain the worldsheet CFT 
\begin{equation}
    \R^{1,5} \times \left( \frac{SL(2,\R)_k}{U(1)} \times \frac{SU(2)_k}{U(1)} \right) \Bigl/ \Z_{k} \,,
    \label{10d-ws-CFT}
\end{equation}
which is a $\mathcal{N}=2$ SCFT and as described above for general $d$, the dependence of \eqref{c_SU2U1} on $k$ cancels with that of the cigar CFT given in \eqref{c_SL2U1}, so that the model \eqref{10d-ws-CFT} has zero central charge for any value of $k$. Notice the $\Z_{k}$ orbifold in \eqref{10d-ws-CFT}. This is necessary for the $U(1)$ R-charge of the $\mathcal N=2$ on the worldsheet to be integer quantized, and hence to be used to impose the GSO projection, leading to an $\mathcal N=2$ supersymmetric spectrum in spacetime. At various places in the manuscript we will consider the worldsheet CFT \eqref{10d-ws-CFT} at level $k=2$. At this value of the level, the supercoset \eqref{Mdisk} has vanishing central charge, see eq.~\eqref{c_SU2U1}, and the worldsheet CFT \eqref{10d-ws-CFT} reduces to 
\begin{equation}
    \R^{1,5} \times \frac{SL(2,\R)_2}{U(1)} \Bigl/ \Z_{2} \,. 
    \label{8d-ws-CFT}
\end{equation} 
Notice that the string background $\R^{1,5} \times \frac{SL(2,\R)_2}{U(1)}$, where no orbifold is present, is also a consistent string background, see {\it e.g.}~\cite{Ashok:2005py}. However, as we shall see below, the presence of the $\Z_2$ orbifold is important for us: it allows in the spectrum cigar primaries which would otherwise not be present \cite{Chang:2014jta, Gadde:2009dj}, resulting in an $\mathcal N=2$ theory in spacetime.

Although, as previously discussed, the cigar resolves the strong coupling singularity of the linear dilaton CFT, we will still frequently find it useful to work in the large-$\rho$ region of the cigar geometry, where the linear dilaton provides a good approximation. In this limit, the CFT \eqref{10d-ws-CFT} becomes
\es{LDCFT}{
\R^{1,5} \times  \R_{\rho}\times \left(  U(1)_{\theta} \times \frac{SU(2)_k}{U(1)} \right) \Bigl/ \Z_{k} \,,
}
where the $U(1)$ is a circle (parametrized by a coordinate $\theta$) of radius $2/Q$, as in \eqref{lineardilatonmetricd+2}. We note that at $k=2$, we have $Q=1$: this is the free fermion radius and the $U(1)$ symmetry is enhanced to $SU(2)$.\footnote{Note that this is not enough, on its own, to guarantee the existence of an $SU(2)$ symmetry in spacetime. In this case, all the vertex operators that generate the $SU(2)$ are BRST invariant, which is what turns such worldsheet $SU(2)$ into a spacetime symmetry.} This not only implies the existence of an extra symmetry, but also extra vertex operators which can be used to build states as we shall soon see.\footnote{See Appendix E of \cite{Gadde:2009dj} for more details on the relation between this and the symmetries of the supergravity and of the dual field theory (once D-branes are added).} Moreover, using the relation \cite{Gepner:1986hr,DiFrancesco:1988xz}\footnote{See {\it e.g.}~\cite{Chang:2014jta} for the precise relation between vertex operators in the two models, including a detailed discussion of supersymmetry and the various orbifolds.}
\es{SU2_vs_SU2U1}{
SU(2)_k \simeq 
\left(  U(1)_{\theta} \times \frac{SU(2)_k}{U(1)} \right) \Bigl/ \Z_{k}\,,
}
we obtain the description of \eqref{LDCFT} in the frame $\ns$, which is just the Callan-Harvey-Strominger (CHS) background \cite{Callan:1991at}
\begin{equation}
    \R^{1,5} \times \R_{\rho} \times SU(2)_k \,, 
    \label{CHS-b}
\end{equation}
arising in the near-horizon limit of a stack of $k$ NS5 branes. In terms of the description \eqref{FNS5}, this identifies $\mathcal{N}=SU(2)_k$ in the case $d=6$, where \eqref{SU2_vs_SU2U1} (respectively \eqref{CHS-b}) is just a more precise version of \eqref{N_vs_N/U(1)} (respectively \eqref{Fcig}).

This gives a complete picture of the three T-duality frames for $d=6$, in the linear dilaton limit. In $\cy$ we have type II string theory on an ALE space, say $\R^4/\Z_k$, which is T-dual to the CHS background \eqref{CHS-b} describing the near-horizon region of a stack of $k$ localized NS5 branes, which is why we refer to this frame as $\ns$ \cite{Ooguri:1995wj}\footnote{See \cite{Gregory:1997te,Tong:2002rq,Harvey:2005ab} for a description of the role played by worldsheet instantons in the T-duality between the CY frame and that with localized NS5 branes.}. Finally, the T-duality \eqref{SU2_vs_SU2U1} relates this to the linear-dilaton geometry \eqref{LDCFT} in the frame $\nc$, where the parameter $k$ is related to the background charge $Q$ of the linear dilaton CFT by $Q=\sqrt{\tfrac{2}{k}}$ as in \eqref{c_SL2U1}. As we have seen, it is convenient to think of the linear dilaton as a limit of the cigar CFT \eqref{10d-ws-CFT}, which has the advantage that the curvature and string coupling are always finite and asymptotes the linear dilaton in a suitable limit ($\rho \to +\infty$). It is then interesting to wonder what this corresponds to in the other T-duality frames. In the frame $\ns$, the string coupling singularity of the linear dilaton can be seen as a result of the fact that the NS5 branes are coincident. A way to solve this problem is to separate the branes by a suitable distance $R$ and then take a suitable scaling limit where the distance between the NS5 branes is again taken to zero, but so is the string coupling $g_s$, in such a way that the ratio $g_{\text{eff}}=g_s/R$ is kept constant, thus defining an effective string coupling $g_{\text{eff}}$ \cite{Giveon:1999px}. The string length $\ell_s$ is kept fix in the procedure. As first observed in \cite{Sfetsos:1995ac,Sfetsos:1998xd} -- see also \cite{Israel:2005fn,Martinec:2017ztd} for interesting comments -- if the NS5 branes are separated by spreading them symmetrically  on a circle of radius $R$, then this is precisely dual to the cigar background \eqref{10d-ws-CFT}. We will be more precise about this relation in Section \ref{sec:NS5distributions}, where we will also comment on other ways to separate the NS5 branes. Finally, in the frame $\cy$ replacing the linear dilaton with the cigar amounts to a blow-up of the ADE singularity, which is replaced by a Taub-NUT geometry -- see Figure \ref{fig:frames} for a pictorial representation of the relation between the three frames. An interesting question is where the singularities of the geometry are localized in the different frames: clearly there are magnetic monopoles in $\cy$, localized at the fixed points of the circle action generated by $\frac{\partial}{\partial\psi}$ in Figure \ref{fig:frames}. After T-duality, in $\ns$ such singularities are replaced by localized NS5 branes, near which the geometry becomes singular. The situation is more subtle in $\nc$, where at a generic point in the disk the cigar geometry is completely regular. However, the target space of the parafermion disk CFT is not a regular geometry but rather a {\it T-fold}: it is only regular when transition functions between charts are combined with T-duality transformations \cite{Kiritsis:1991zt} -- see \cite{Brennan:2020bju} for a modern perspective.

\paragraph{The worldsheet CFT.} The main goal of this section is to investigate the low-lying spectrum of the string theory \eqref{10d-ws-CFT}, which we will do for arbitrary value of the level $k$. We begin by presenting more details of the worldsheet theory. Since we are ultimately interested only in the quantum numbers of the states that we analyze, we will simplify the details of the presentation by considering the limit $\rho \to +\infty$ of the cigar geometry, where the CFT \eqref{10d-ws-CFT} reduces to \eqref{LDCFT}, which is in turn equivalent to the CHS background \eqref{CHS-b}. While the spectrum of the cigar CFT contains both states localized at the tip of the cigar and a continuum of delta-function normalizable scattering states, this is possible because we are only interested in the latter, which are in one-to-one correspondence with the vertex operators of the CHS background.

Let us spell out the conventions we are going to use in the following for the CFT \eqref{CHS-b}. We work at $\alpha'=2$ and denote the six Minkowski free bosons and fermions by $\partial X_\alpha$ and $\psi_\alpha$ respectively, where $\alpha \in \{ 0, 1, \dots 5 \}$. They obey the OPEs
\begin{equation}
    \partial X_\alpha(z) \, \partial X_\beta(0) \sim -\frac{\eta_{\alpha \beta}}{z^2} \,,
\end{equation}
and
\begin{equation}
    \psi_\alpha(z) \psi_\beta(0) \sim \frac{\eta_{\alpha \beta}}{z} \,. 
\end{equation}
For the linear dilaton $\partial \rho$ and its superpartner $\psi^\rho$ we have 
\begin{equation}
\partial \rho(z) \partial \rho(0) = -\frac{1}{z^2} \,,\qquad \psi^\rho(z) \psi^\rho(0) \sim \frac{1}{z} \,,  
\end{equation}
while the three free fermions $\psi_{\pm},\psi^\vartheta$ of the supersymmetric $SU(2)_k$ WZW model in \eqref{CHS-b} (in the adjoint of $SU(2)$) obey
\begin{equation}
    \psi^\vartheta(z) \psi^\vartheta(0) \sim \frac{1}{z} \,,\qquad \psi_+(z) \psi_-(0) \sim \frac{1}{z} \,. 
\end{equation}
In the following it will prove useful to bosonize the worldsheet fermions as\footnote{Note that we are using $\vartheta$ to denote the bosonization of the two fermions $\psi_\pm$ of the $SU(2)_k$ WZW model. This is related to, but not the same as, the coordinate $\theta$ appearing in \eqref{LDCFT}, parametrizing the circular direction of the cigar, and entering the bosonization of the bosonic $\mathfrak{sl}(2,\R)_{k+2}$ Cartan $j^3 = \ii \, \partial \theta$. In particular, for $k=2$ the worldsheet CFT reduces to \eqref{8d-ws-CFT} and $\theta$ and $\vartheta$ can be identified. From the perspective of eq.~\eqref{8d-ws-CFT}, the $SU(2)_2$ fermions $\psi^\pm$ arise from the fermionization of the compact coordinate $\theta$ of the cigar, $\ii \, \partial \theta = \psi^+ \psi^-$. Similar comments apply to $\psi^\vartheta$ vs $\psi_\theta$ (superpartner of $\partial\theta$ in the cigar theory).}
\begin{align}
\tfrac{1}{\sqrt 2} (\pm \psi_0 + \psi_1) & = \ex^{\pm \phi_0} \,, &  \tfrac{1}{\sqrt 2} (\psi_2 \pm \ii \psi_3) & = \ex^{\pm \ii \phi_1} \,,\\
\tfrac{1}{\sqrt 2} (\psi_4 \pm \ii \psi_5) & = \ex^{\pm \ii \phi_2} \,, &  \tfrac{1}{\sqrt 2} (\psi^\rho \pm \ii \psi^\vartheta) & = \ex^{\pm \ii H} \  , 
\end{align}
\begin{equation}
    \psi^\pm =  \ex^{\pm \ii \vartheta} \,. \label{psipm}
\end{equation}
The holomorphic $\mathcal N=1$ superalgebra on the worldsheet is realized as
\begin{equation}
T = T^{\text m} + T^{\beta \gamma} + T^{bc} \,, \qquad G = G^{\text m} + G^{\beta \gamma} + G^{bc} \,, 
\label{T-and-G}
\end{equation}
where $G^{bc}$ and $G^{\beta \gamma}$ are the usual ghost and superghost supercurrents. The matter stress tensor and supercurrent entering eq.~\eqref{T-and-G} are given by
\begin{equation}
\begin{aligned}
    T^{\text m} & = T_{\text{flat}} + T_\rho + T_{\text{su}} \,, \\
    G^{\text m}& = G_{\text{flat}} + G_\rho + G_{\text{su}} \,, \\ 
\end{aligned}
    \label{N=1-ws}
\end{equation}
with
\begin{equation}
\begin{aligned}
   T_{\text{flat}} & = -\frac{1}{2}\left( \partial X_{\alpha} \partial X^{\alpha} +\psi^{\alpha}\partial\psi_{\alpha} \right) \,, &  \quad G_{\text{flat}} & = \ii \, \partial X^\alpha \psi_\alpha \,, \\
    T_\rho & = -\frac{1}{2} \left( (\partial \rho)^2+Q \partial^2 \rho + \psi^\rho \partial \psi^\rho \right) \,, & \quad  G_\rho & = \ii \, \partial \rho \, \psi^\rho + \ii \, Q \, \partial \psi^\rho \,,
\end{aligned}
\label{Tm-components}
\end{equation}
and 
\begin{equation}
\begin{aligned}
    T_{\text{su}} & = -\frac{1}{2} \left( (\partial \vartheta)^2 +\psi^\vartheta \partial \psi^\vartheta \right) + \frac{Q^2}{4} (K^+ K^- + K^- K^+ + 2 K^3 K^3) \,, \\
    G_{\text{su}}& = \ii \, Q \, \partial \vartheta \, \psi^\vartheta + Q \left(\frac{K^+ \psi^- + K^- \psi^+}{\sqrt 2} +  K^3 \psi^\vartheta\right)    \,. 
\end{aligned}
\end{equation}
In eqs.~\eqref{N=1-ws} and \eqref{Tm-components}, $K^a$ with $a = +, -, 3$ denote the bosonic $SU(2)$ currents at level $k-2$. We bosonize the $\beta$ and $\gamma$ superghosts as usual,
\begin{equation}
\gamma =  \ex^{\varphi} \, \eta \,, \qquad \beta = \partial \xi \, \ex^{-\varphi} \ ,
\end{equation}
with
\begin{align}
\eta(z) \, \xi(0) & \sim \frac{1}{z} \,, \label{eta-xi-OPE} \qquad \partial \varphi(z) \partial \varphi(0) \sim -\frac{1}{z^2} \,. 
\end{align}
Let us also introduce the picture charge operator 
\begin{equation}
    P = \oint \frac{\text d z}{2 \pi \ii} (-\partial \varphi + \xi \eta) \,. 
\end{equation}

\paragraph{Spacetime supercharges.} Let us construct spacetime supercharges following \cite{Kutasov:1990ua,Kutasov:1991pv}. We consider the R sector states 
\begin{equation}
    S_{\vec \epsilon \,}(z) = \ex^{-\frac{\varphi}{2}} \, \ex^{\frac{1}{2}( \ii \epsilon_\vartheta \vartheta + \ii \epsilon_H H )} V_{\hat \epsilon} \,, 
    \label{spacetime-supercharges-ansatz}
\end{equation}
where 
\begin{equation}
    \vec \epsilon = (\epsilon_\vartheta, \epsilon_H, \hat\epsilon) \in (2 \Z+1)^{\otimes 5}\,,\qquad
 \hat \epsilon = (   \epsilon_0, \epsilon_1, \epsilon_2 )\in (2 \Z+1)^{\otimes 3}\,,
\end{equation}
play the role of spinor indices in 10d ($\vec \epsilon \,$) and 6d ($\hat \epsilon$), respectively, and
\begin{equation}\label{Veps}
     V_{\hat \epsilon } = \ex^{\frac{1}{2}(\epsilon_0 \phi_0 + \ii \epsilon_1 \phi_1 + \ii \epsilon_2 \phi_2)} \,,
\end{equation}
is a vertex operator which can be thought of as a 6d spinor from the spacetime perspective. Spacetime supercharges should be physical and map physical states to physical states. We should then require that 
\begin{equation}    
    L_0 \, S_{\vec \epsilon \,}(z) = S_{\vec \epsilon}(z) \,, \qquad L_n \, S_{\vec \epsilon \, }(z) = 0 \,, \quad n > 0 \,,
\label{S-L0-physical}
\end{equation}
and 
\begin{equation}
    G_n^{\text m} \, S_{\vec \epsilon}(z) = 0 \,, \quad n \geq 0 \,.
    \label{S-G-physical}
\end{equation}
Eq.~\eqref{S-L0-physical} implies 
\begin{equation}
 \epsilon_H^2 + \epsilon_\vartheta^2 + \epsilon_0^2 + \epsilon_1^2 + \epsilon_2^2 = 5 \,,
\end{equation}
hence 
\begin{equation}
    |\epsilon_H| =|\epsilon_\vartheta| = |\epsilon_0| = |\epsilon_1| = |\epsilon_2| = 1 \,. 
    \label{all-eps=pm1}
\end{equation}
When \eqref{all-eps=pm1} is obeyed, worldsheet fermions have the following OPEs with $S_{\vec \epsilon}$,
\begin{subequations}
\begin{align}\label{gammadef}
    \psi_{\alpha}(z) \, S_{\vec \epsilon \,} (0) & \sim \frac{\ex^{-\frac{\varphi}{2}} \ex^{\frac{\ii}{2}(\epsilon_\vartheta \vartheta + \epsilon_H H)} \, [V(0)\cdot  \gamma_{\alpha}]_{\hat\epsilon}}{\sqrt 2 \, z^{\frac{1}{2}}} \,, \\
    \psi^\rho(z) \, S_{\vec \epsilon \,} (0) & \sim \epsilon_0 \epsilon_1 \epsilon_2 \frac{\ex^{-\frac{\varphi}{2}} \ex^{\frac{\ii}{2}(\epsilon_\vartheta \vartheta - \epsilon_H H)} \, V_{\hat\epsilon}(0)}{\sqrt 2 \, z^{\frac{1}{2}}} \,, \\
    \psi^\vartheta(z) \, S_{\vec \epsilon \,} (0)  & \sim \ii \, \epsilon_0 \epsilon_1 \epsilon_2 \epsilon_H \, \frac{\ex^{-\frac{\varphi}{2}} \ex^{\frac{\ii}{2}(\epsilon_\vartheta \vartheta - \epsilon_H H)} \, V_{\hat\epsilon}(0)}{\sqrt 2 \, z^{\frac{1}{2}}} \,, 
\end{align}
\label{psi-S-OPE}%
\end{subequations}
where the factors of $\epsilon_0, \epsilon_1, \epsilon_2, \epsilon_H$ in \eqref{psi-S-OPE} are guarantee that worldsheet fermions anti-commute. In \eqref{gammadef}, $\gamma_{\alpha}$ are 6d gamma matrices which appear with one of their spinor indices contracted with the vertex operator $V$.

In type II flat space string theory, eq.~\eqref{S-G-physical} would not imply any further constraint. On  the other hand, we will see momentarily that the presence of $\ii \, Q \, \psi^\vartheta \partial \vartheta + \ii \, Q \, \partial \psi^\rho$ in \eqref{N=1-ws} leads to a different result for us. In fact, from \eqref{psi-S-OPE} it follows 
\begin{equation}
    (\psi^\vartheta \partial \vartheta + \partial \psi^\rho)(z) S_{\vec \epsilon}(0) \sim \frac{\epsilon_0 \epsilon_1 \epsilon_2 (\epsilon_H \epsilon_\vartheta -1)}{2 \sqrt 2 z^{\frac{3}{2}}} \ex^{-\frac{\varphi}{2}} \ex^{\frac{\ii}{2}(\epsilon_\vartheta \vartheta - \epsilon_H H)}V_{\hat\epsilon} \,,
\end{equation}
which in turn implies 
\begin{equation}
    \epsilon_H = \epsilon_\vartheta \,. 
    \label{epsH=epstheta}
\end{equation}

A further constraint on the supercharges comes from the GSO projection, imposing mutual locality. Since the OPE of two states of the form \eqref{spacetime-supercharges-ansatz} reads
\begin{equation}
    S_{\vec \epsilon \,}(z) S_{\vec \tau \,}(0) = \mathcal O( z^{\frac{1}{4}(-1 + \vec \epsilon \cdot \vec \tau)} ) \,, 
\end{equation}
a necessary condition for mutual locality is 
\begin{equation}
    -1 + \vec \epsilon \cdot \vec \tau = 0 \quad \mod 4 \,. 
\end{equation}
It is easy to check that, when eqs.~\eqref{all-eps=pm1} and \eqref{epsH=epstheta} and their analogues for $\vec \tau$ are obeyed,  
\begin{equation}
    -1 + \vec \epsilon \cdot \vec \tau = 1 - \epsilon_0 \epsilon_1 \epsilon_2 \, \tau_0\tau_1\tau_2 \quad \mod 4 \,. 
    \label{mutual-locality}
\end{equation}
Constructing holomorphic spacetime supercharges as
\es{supercharges_physical}{
 S^+ &= \ex^{-\frac{\varphi}{2}} \, \ex^{\frac{\ii \epsilon_\vartheta}{2}(\vartheta + H)} \ex^{\frac{1}{2}(\epsilon_0 \phi_0 + \ii \epsilon_1 \phi_1 + \ii \epsilon_2 \phi_2)} \,, \qquad \text{with} \qquad \epsilon_0 \epsilon_1 \epsilon_2 = 1 \,, \\
    S^- &= \ex^{-\frac{\varphi}{2}} \, \ex^{\frac{\ii \epsilon_\vartheta}{2}(\vartheta + H)} \ex^{\frac{1}{2}(\epsilon_0 \phi_0 + \ii \epsilon_1 \phi_1 + \ii \epsilon_2 \phi_2)} \,, \qquad \text{with} \qquad \epsilon_0 \epsilon_1 \epsilon_2 = -1 \,, 
}
where the $+$ or $-$ sign denotes the chirality, eq.~\eqref{mutual-locality} shows that only supercharges of the same chirality are mutually local. 

To sum up, we started in the holomorphic sector with $2^5$ candidate supercharges. These were halved by the physical state condition and further halved by the GSO projection. Similarly, for the anti-holomorphic sector. In total, we obtain 8 supercharges in the holomorphic sector and 8 supercharges in the anti-holomorphic sector, for a total of 16 supercharges, as expected. In analogy with flat space, we will refer to the noncritical string theory as type IIA when different chiralities are chosen in the holomorphic and anti-holomorphic sectors and as type IIB when instead the same chirality is chosen in both sectors, see Table~\ref{tab:IIA-IIB-supercharges}. 

\begin{table}
    \centering
    \begin{tabular}{c|c|c}
         & Left-moving supercharges & Right-moving supercharges \\
         \hline
        Type IIA & $S^+(z)$ & $\tilde S^-(\bar z)$ \\
        Type IIB & $S^+(z)$ & $\tilde S^+(\bar z)$
    \end{tabular}
    \caption{For each choice of GSO projection, we list left-moving and right-moving mutually local spacetime supercharges.}
    \label{tab:IIA-IIB-supercharges}
\end{table}
We are now going to build physical states in the lowest level. We will impose GSO projection by requiring mutual locality with the supercharges.

\subsection{Spectrum}\label{sec:spectrumws}

We finally turn to the main point of this section: the analysis of the low-lying spectrum of the noncritical string. As anticipated, we focus on delta-function normalizable states of the cigar for generic $k$ and perform our analysis in the CHS background \eqref{CHS-b}, which is an asymptotic limit of the cigar geometry (up to T-duality). We are only interested in the states at the lowest mass-level, which we will then identify with the fields of a certain gauged supergravity in Section \ref{sec:7dsugra}. A similar analysis for the spectrum has been performed in various references on noncritical string theory \cite{Mizoguchi:2000kk,Eguchi:2000tc,Murthy:2003es,Gadde:2009dj}, compared to which we add two novel elements:
\begin{itemize}
    \item Rather than just studying the massless spectrum of the noncritical string theory \eqref{8d-ws-CFT}, which is defined for $k=2$, we perform an analysis valid more in general for the CFT \eqref{10d-ws-CFT}, at a generic value of the level $k$, adopting the asymptotic description \eqref{CHS-b}. For general $k$, vertex operators of the supersymmetric $SU(2)_k$ WZW model are obtained combining the three free fermions $(\psi_\pm,\psi^\vartheta)$ with vertex operators of spin $j$ associated with the bosonic $SU(2)_{k-2}$. In order to minimize the value of the mass, one is forced to only consider the identity operator for $SU(2)_{k-2}$ ($j=0$), so that the states we are interested in are built using only vertex operators of the linear dilaton CFT plus three free fermions. This makes the study of the low-lying spectrum for general $k$ just a simple generalization of that for $k=2$ (in which case $SU(2)_k$ reduces to the supersymmetric $U(1)_\theta$ in \eqref{SU2_vs_SU2U1}). From this perspective, our analysis can be seen as an in-depth investigation of the simplest subsector of the spectrum of double-scaled little string theory \cite{Giveon:1999px} (see also, {\it e.g.}, \cite{Giveon:1999tq,Kutasov:2000jp,Aharony:2003vk,Chang:2014jta}). 
    \item The worldsheet theory \eqref{CHS-b} only manifests 6d Poincar\'e invariance, since translations in the $\rho$ direction are broken by the presence of the linear dilaton. Nonetheless, our goal is to interpret this spectrum as associated with a 7d supergravity theory admitting the linear dilaton background as a specific solution, in the spirit of \cite{Polyakov:1997tj,Polyakov:1998ju}. With this in mind, we will adopt a 7d perspective, thinking of our results as coming from the spontaneous breaking of 7d Poincar\'e invariance to 6d, due to the choice of a specific vacuum (the linear dilaton). In fact, as we shall see, something similar holds for supersymmetry, which is also spontaneously broken by the choice of vacuum from thirty-two supercharges to the sixteen discussed in the previous subsection. Note that adopting this 7d perspective will be particularly important in the RR sector, since the type IIA and IIB theories are only distinguishable in 7d but become identical in 6d, much like critical type IIA and IIB superstring theory compactified on a circle.
\end{itemize}
Keeping in mind a 7d perspective is crucial in order to make contact with a 7d supergravity description, which we do in Section \ref{sec:7dsugra}. Moreover, we will find that such a two-derivative theory is only trustworthy in the limit of large $k$, hence the importance of performing the analysis for general $k$. 

{
At various points in our analysis it will appear natural to refer to certain states as ``massless'' or ``massive'', because of certain constraint imposed by the GSO projection and BRST invariance on the corresponding polarizations and momenta. As just discussed, however, the string is quantized in a vacuum which does not enjoy the 7d Poincaré invariance we are advocating for, and as a result the concept of mass is not well-defined. In analogy with the massless states of the ten-dimensional critical superstring, we will refer to the subsector of the lowest modes as {\it massless}. In fact, we are going to see that the holomorphic sector includes a 7d vector subject to an equivalence relation that we interpret as a gauge invariance condition in the linear dilaton vacuum. Similarly, once holomorphic and anti-holomorphic sectors are assembled, we find a graviton and a Kalb-Ramond two-form (as well as other vectors), whose polarizations also contain BRST exact states naturally interpreted as ``pure gauge''. As a result, the natural candidate for a field-theoretical description of the interactions between such states involves massless representations of the 7d Poincaré group. However, this is not the end of the story. Firstly, the supergravity theory that we are led to consider is a {\it gauged} supergravity (see {\it e.g.} \cite{Trigiante:2016mnt} for a review), and as well known the gauging introduces mass terms in the Lagrangian (namely, quadratic terms in the fields). Secondly, typical vacua for gauged supergravity are not flat and may contain vacuum expectation values for some of the scalars. Both of these facts contribute to the value of the masses of the linearized fluctuations in a certain vacuum. It is therefore possible for two fields -- whose physical polarizations satisfy the constraints that are typical of massless and massive fields respectively -- to appear at the same mass level. We anticipate here that this is exactly what happens in our case, where the states in the NSNS sector are subject to the gauge invariance conditions described above and are therefore more naturally interpreted as massless. On the other hand, in the RR sector we shall find fields whose equations of motion naively appear massive, although such states appear at the same mass level as the massless graviton in the NSNS sector. A familiar case in which something analogous happens is the 10d supergravity multiplet in AdS$_5\times S^5$, where the interactions by the fluctuations at the lowest level are described by 5d $\mathcal{N}=8$ supergravity with $SO(6)$ gauging: the various fields in the 5d supergravity multiplet have different masses in the AdS$_5$ vacuum, despite the maximal supersymmetry and the common origin from a single massless multiplet -- see {\it e.g.} \cite{Ferrara:1998ej}.
}

Before moving on to the details, we also wish to make a comment for the reader with some experience in supergravity. We anticipate here that the 7d supergravities associated with the type IIA/B version of \eqref{CHS-b} turn out to be those arising as a consistent truncation of 10d type IIA/B supergravity on $S^3$ (the target space geometry of $SU(2)_k$). Consistent truncations typically involve the identification of a subsector of states in the Kaluza-Klein (KK) decomposition which are not constrained by setting to zero all other states in the equations of motion. However, such states generally have different masses and some of the KK modes which one might naively expect to contribute to the spectrum actually turn out to be pure gauge degrees of freedom. With these comments in mind, we wish to emphasize that from the worldsheet perspective we can identify a very precise criterion to select which fields should be kept in the truncation, since they are at the same mass level and they are all built from singlet vertex operators of the bosonic $SU(2)_{k-2}$. Note that this does {\it not} imply that they are $SU(2)$ singlets, since they can sit in non-trivial $SU(2)$ representations inherited by the free fermions of the supersymmetric $SU(2)_k$.

\subsubsection{NS sector}

We begin by considering the left-moving NS sector and we build physical states by combining the usual flat space vertex operators (and ghosts) with those of the cigar (in the linear dilaton approximation). We should choose one set of spacetime supercharges from those described in \eqref{supercharges_physical} (either $S^+$ or $S^-$) and first of all require that the vertex operators are mutually local with them, which amounts to the GSO projection. In the case of the NS sector, the result is the same regardless of whether we choose to work with $S^+$ or $S^-$ (as for the 10d critical string). So far as the lowest-lying states are concerned, the first implication of this requirement is that the simplest vertex operator that one can build, the tachyon $\ex^{-\varphi}\ex^{\mathtt{j}\rho}\ex^{\ii q\cdot X}$, is excluded from the spectrum. The next-to-lightest states, which are actually compatible with the GSO projection, have the form
\es{NS-general}{
\ex^{-\varphi}\psi^A\ex^{\mathtt{j}\rho}\ex^{\ii q\cdot X}\,,
}
where $\mathtt j = -\frac{Q}{2} + \ii s$, $s \in \R$ and the indices $A,B$ run over $A, B \in \{\alpha, \rho, +, -, \vartheta \}$ and $\alpha = 0, \dots, 5$ labels $\mathbb{R}^{1,5}$ flat coordinates. In the following, it will be useful to also introduce a 7d index 
\begin{equation}
    \mu =( \alpha, \rho ) \,, 
\end{equation}
and an $SU(2)$ index 
\begin{equation}
    i = ( +,-,\vartheta ) \ ,
\end{equation}
for a total of ten candidate states. We must now require that physical states are annihilated by $G^{\mathrm{m}}_{\frac{1}{2}}$ and that they are not BRST exact. The states built using only $\psi^i$ in \eqref{NS-general} are three scalars from the 7d perspective (an $SU(2)$ triplet), and they satisfy both conditions. On the other hand, those built using $\psi^{\mu}=(\psi^{\alpha},\psi^{\rho})$ identify a 7d vector, but not all of them are physical. As in \cite{Gadde:2009dj}, this can be conveniently seen by saturating the target space index with a polarization vector $\varepsilon_{\mu}$, namely we consider the vertex operator
\es{generalNSvector}{
\varepsilon_{ \mu}\ex^{-\varphi}\psi^{ \mu}\ex^{\mathtt{j}\rho}\ex^{\ii q\cdot X}\,.
}
The requirement that this is annihilated by $G^{\mathrm{m}}_{\frac{1}{2}}$ imposes the condition
\es{NStransverse}{
\varepsilon_\alpha q^{\alpha}-\ii(\mathtt{j}+Q)\varepsilon_{\rho}=0
}
on the polarization of physical states, which we will interpret as a transversality condition in Section \ref{sec:NSNS}. Moreover, we note that the choice of polarization 
\es{}{
\varepsilon_{\alpha}=\ii\,q_{\alpha}\,\qquad
\varepsilon_{\rho}=\mathtt{j}\,,
}
leads to a BRST exact state, as it can be shown using
\es{}{
0 = G^{\text m}_{-\frac{1}{2}} \ex^{\mathtt j \rho} \ex^{\ii q \cdot X} = (\ii\,q_\alpha \psi^\alpha + \mathtt{j} \psi^\rho) \ex^{\mathtt j \rho} \ex^{\ii q \cdot X} \,. 
    \label{null-state}
}
Hence, out of the seven states generically described by \eqref{generalNSvector}, only five are physical: their polarizations are subject to the transversality condition \eqref{NStransverse} and identified under the equivalence relation
\es{NSgaugeinv}{
(\varepsilon_{\alpha}, \varepsilon_{\rho}) \sim (\varepsilon_{\alpha},\varepsilon_{\rho}) + (\ii\,q_{\alpha}, \mathtt{j}) \,, 
}
which in Section \ref{sec:NSNS} we will recognize as the condition for gauge invariance on a 7d vector.

We conclude that among the ten states described by \eqref{NS-general} only eight are physical: an $SU(2)$ triplet of 7d scalars and an $SU(2)$ singlet which is a 7d ``massless'' vector. By massless here we mean that it is a gauge field, as shown by the fact that its polarizations are identified by the equivalence relation \eqref{NSgaugeinv}. The mass-shell condition for such states can be argued from the requirement that $L_0=1$, resulting in the constraint
\begin{equation}
q^2+s^2+\frac{Q^2}{4}=0\,,\qquad
\text{or}\qquad
q^2-\mathtt{j}(\mathtt{j}+Q)=0\,.
\label{mass-shell}
\end{equation}
Equivalently, we can introduce a 7d ``momentum''
\begin{equation}
    p^\mu \equiv (q^\alpha, s) \,, \qquad  p_{\mu} \equiv (q_\alpha, s) \,,   
    \label{7d-momentum}
\end{equation}
which allows to rephrase the mass-shell condition \eqref{mass-shell} more compactly as
\es{massshell_k7d}{
p^2+\frac{Q^2}{4}=0\,.
}
Naively, this looks like the condition satisfied by the momentum of a massive field, but as described above we should not read too much into this because of the broken translation invariance in the linear dilaton direction. Rather, we consider the gauge invariance relation \eqref{NSgaugeinv} to be a fundamental principle and insist in calling this the massless level. This will be supported by the derivation of these states from a 7d gauged supergravity in Section \ref{sec:7dsugra}.

In the semiclassical limit $Q\to 0$ (or, $k\to \infty$) the slope of the linear dilaton vanishes and this is a honest 7d momentum since Poincaré invariance is restored along the $\rho$ direction. In this limit, we indeed obtain a massless momentum. Finally, let us write the physical states using the notation $\mathbf{a}_b$, where $\mathbf{a}$ denotes the dimension of the corresponding representation of the massless 7d little group $SO(5)$, while ${b}$ denotes the dimension of the corresponding $SU(2)$ representation. In the NS sector, we find
\es{}{
\mathbf{5}_1\oplus \mathbf{1}_3\,.
}

\subsubsection{R sector}

In the R sector, we consider the candidate vertex operators for physical states
\begin{equation}
 \ex^{-\frac{\varphi}{2}} \, \ex^{\frac{\ii}{2}( \epsilon_\vartheta \vartheta + \epsilon_H H)} V_{\hat \epsilon} \, \ex^{\mathtt j\rho} \ex^{\ii q \cdot X} \,, 
 \label{R-states-preliminary}
\end{equation}
where $V_{\hat \epsilon}$ is defined in \eqref{Veps} and 
\begin{equation}
    \epsilon_\vartheta, \epsilon_H, \epsilon_0, \epsilon_1, \epsilon_2 = \pm 1 \,. 
\end{equation}
The implications of the GSO projection in the R sector now {\it do} depend on which set of supercharges is chosen among those in \eqref{supercharges_physical}. Locality with $S^+$ (respectively $S^-$) requires $\epsilon_\vartheta \epsilon_H \epsilon_0 \epsilon_1 \epsilon_2 =1$ (respectively $\epsilon_\vartheta \epsilon_H \epsilon_0 \epsilon_1 \epsilon_2 =-1$). Let us begin by analyzing the states that are compatible with $S^+$, and only at the end comment on what would have changed if we had chosen $S^-$ instead. We can organize the states \eqref{R-states-preliminary} compatible with $S^+$ as 
\begin{equation}
\begin{aligned}
    & \Psi^+_{\hat \epsilon} \ex^{-\frac{\varphi}{2}} \ex^{\frac{\ii}{2}(\vartheta + H)} V_{\hat \epsilon} \, \ex^{ \mathtt j\rho} \ex^{\ii q \cdot X} \,, \\
    & \Xi^+_{\hat \epsilon} \ex^{-\frac{\varphi}{2}} \ex^{-\frac{\ii}{2}(\vartheta + H)} V_{\hat \epsilon} \, \ex^{ \mathtt j\rho} \ex^{\ii q \cdot X} \,, \\
    & \Psi^-_{\hat \epsilon} \ex^{-\frac{\varphi}{2}} \ex^{\frac{\ii}{2}(\vartheta - H)} V_{\hat \epsilon} \,  \ex^{ \mathtt j\rho} \ex^{\ii q \cdot X} \,, \\
    & \Xi^-_{\hat \epsilon} \ex^{-\frac{\varphi}{2}} \ex^{-\frac{\ii}{2}(\vartheta - H)} V_{\hat \epsilon} \,  \ex^{ \mathtt j\rho} \ex^{\ii q \cdot X} \,, 
\end{aligned}
\label{6d-all-polar}
\end{equation}
where we are summing over the repeated index $\hat \epsilon$, which plays the role of a spinor index. Here $\Psi^\pm_{\hat \epsilon}$ and $\Xi^\pm_{\hat \epsilon}$ are 6d symplectic Majorana-Weyl spinor polarizations of fixed chirality, related by the symplectic Majorana condition 
\begin{equation}\label{SMW}
    \chi^{\pm,a} = \Omega^{ab} D (\chi^{\pm,b})^*  \,, \qquad
    \chi^{\pm,a} = \begin{pmatrix} \Psi^\pm \\ \Xi^\pm \end{pmatrix}\,,
\end{equation}
in terms of the $Sp(2)$ doublets $\chi^{\pm,a}$, where $a,b=1,2$,
$D$ is the intertwiner for complex conjugation in 6d ($D^{-1}\gamma^{\alpha}_{6d}D=(\gamma^{\alpha}_{6d})^*$) and  $\Omega$ is the two-index antisymmetric tensor. One can think of the degrees of freedom as being carried by complex $\Psi^+$ and complex $\Psi^-$, with $\Xi^+$ and $\Xi^-$ fixed in terms of the latter by the symplectic Majorana condition \eqref{SMW}. For simplicity, in the following we will then neglect $\Xi^+$ and $\Xi^-$ and work in terms of $\Psi^+$ and $\Psi^-$ only. We find 16 off-shell real degrees of freedom. 

The mass-shell condition $L_0 = 1$ implies again ${p}^2+\tfrac{Q^2}{4}=0$ as in \eqref{massshell_k7d}, {\it i.e.}~the same level as the states we discussed in the NS sector. We should also require that physical states are annihilated by $G_0^{\text m}$, so we consider the action of the latter on a generic combination of the R-sector vertex operators \eqref{R-states-preliminary}. For the reasons that we just discussed, it is sufficient to consider the $\Psi^{\pm}$ polarizations (and their complex conjugates), and we find
\es{G0=0}{
    0&=G_0^{\mathrm{m}}\left[ \ex^{-\frac{\varphi}{2}} \bigl( \Psi^+_{\hat \epsilon} \ex^{ \tfrac{\ii}{2}(\vartheta+H)} + \Psi^-_{{ \hat \epsilon}} \ex^{ \tfrac{\ii}{2}(\vartheta-H)} \bigr) V_{\hat \epsilon} \, \ex^{\mathtt j\rho}\ex^{\ii q \cdot X} \right] \\[0.2cm]
    &=\ex^{-\frac{\varphi}{2}}\Bigl[ \left(\slashed{q}\Psi^+_{\hat \epsilon} + \ii\, (\mathtt j+Q) \Psi^-_{\hat \epsilon}\right) \ex^{ \tfrac{\ii}{2}(\vartheta+H)}  + \left( \slashed{q}\Psi^-_{\hat \epsilon} - \ii\, \mathtt j \, \Psi^+_{\hat \epsilon} \right)\ex^{ \tfrac{\ii}{2}(\vartheta-H)} \Bigr] V_{\hat \epsilon} \, \ex^{\mathtt j\rho} \ex^{\ii q \cdot X} \,,
}
which in turn implies the two conditions
\begin{align}
   \ii\, \slashed{q}\Psi^+ - ( \mathtt j+Q)\Psi^-=0 \,, \qquad \ii\,\slashed{q}\Psi^- +  \mathtt j\Psi^+ = 0 \,.
    \label{G0=0-system}
\end{align}
Notice that, as expected from $(G_0^{\text m})^2 = L_0^{\text m}$, the system \eqref{G0=0-system} admits solutions only when the mass-shell condition \eqref{massshell_k7d} is satisfied. Eq.~\eqref{G0=0-system} reduces the (holomorphic) degrees of freedom from the initial 16 to 8, matching the number of degrees of freedom in the NS sector. 

It is instructive to rewrite \eqref{G0=0-system} in terms of the 7d momentum $p_\mu$ introduced in \eqref{7d-momentum}. To do so, we define a 7d Dirac spinor $\Psi$ combining the two opposite chirality 6d Weyl spinors $\Psi^{\pm}$ as 
\es{}{
\Psi=\Psi^++\Psi^-\,.
}
The 7d gamma matrices $\gamma^{\mu}_{7d}$ are as usual obtained adding the chirality matrix $\gamma^{7}_{6d}$ to the 6d gamma matrices $\gamma^{\alpha}_{6d}$: 
\es{}{
\gamma^{\mu}_{7d}=(\gamma^{\alpha}_{6d},\gamma^{7}_{6d})\,,
}
so that $\gamma^{7}_{6d}\Psi^{\pm}=\pm \Psi^{\pm}$. The two equations in \eqref{G0=0-system} can then be rewritten in a more compact form as
\begin{equation}
    \bigl( \ii \, q_\alpha \gamma^{\alpha}_{6d} + \ii\,s\,\gamma^7_{6d} - \tfrac{Q}{2} \bigr) \Psi = 0 \,,
    \label{eom-spinors-6d}
\end{equation}
or equivalently
\begin{equation}
   \ii \,  {\slashed p} \Psi =  \frac{Q}{2} \Psi \,, \qquad \text{where}  \qquad  {\slashed p} \equiv p_{ \mu} \gamma^{ \mu}_{7d}\,, 
   \label{R-7d-eom-S+}
\end{equation}
in terms of 7d quantities only. Once again, the Dirac equation (and the corresponding mass-shell condition) naively appear to be massive, but we remind the reader that these states are at the same level of the NS vector, which we refer to as the massless level. 

Finally, we can go back to the question of what would change if we chose R-sector states that are mutually local with the supercharges $S^-$ in \eqref{supercharges_physical}, instead of $S^+$ as we did above. The answer is very simple: we would obtain a Dirac equation which looks very similar to \eqref{R-7d-eom-S+}, but with the opposite sign for the mass term:
\begin{equation}
   \ii \,  {\slashed p} \Psi = - \frac{Q}{2} \Psi \,.
   \label{R-7d-eom-S-}
\end{equation}
We note a significant difference between the fermions in 10d critical string theory and here. In the former case, a choice of chirality of the supercharges implies opposite chiralities for the physical massless spinors of the theory. Here, on the other hand, any choice of supercharges leads to spinors of both chiralities in 6d ($\Psi^{\pm}$). What distinguishes the two cases ($S^+$ vs $S^-$) is simply a sign in the mass term of the corresponding Dirac equations, a fact that can be expressed in a simple way when writing the equations in 7d language, for a single Dirac spinor $\Psi=\Psi^++\Psi^-$ -- see \eqref{R-7d-eom-S+} vs \eqref{R-7d-eom-S-}. While this different sign for the mass term does not affect the type of representation in the holomorphic and anti-holomorphic sector independently, we will see the importance of this sign when considering the closed string states and the difference between the type IIA and IIB GSO projections.

To summarize, we find eight physical states in the R-sector for either choice of supercharges ($S^+$ or $S^-$), like in the NS sector. Using the notation introduced for that case, they can be expressed as
\es{}{
\mathbf{4}_2\,,
}
that is an $SU(2)$ doublet of 7d Dirac spinors.\footnote{The $SU(2)$ doublet structure is given by the $\vartheta$ dependence in \eqref{R-states-preliminary}.}

\subsubsection{Closed string spectrum}

Since we are only interested in finding the $SO(5)_{LG}$ and $SO(4)\simeq [SU(2)_+\times SU(2)_-]/\Z_2$ representations of the massless closed string states, instead of considering the full vertex operators as we have done so far we can limit ourselves to studying the result of taking the tensor products between two copies of the polarizations for the massless states described above. The two copies represent of course the left- and right-moving sector of the string. We will now write the states as $\mathbf{a}_{(b_+,b_-)}$, where $\mathbf{a}$ is again the dimension of the $SO(5)_{LG}$ representation, $b_+$ (resp.~$b_-$) is the dimension of the $SU(2)_+$ (resp.~$SU(2)_-$), where $SU(2)_+$ and $SU(2)_-$ are two copies of the $SU(2)$ mentioned above, one for the left and one for the right sector.

\paragraph{NSNS sector.} In the NSNS sector, we consider
\begin{multline}
    \left(\mathbf{5}_{(1,1)}\oplus \mathbf{1}_{(3,1)}\right)\otimes 
\left(\mathbf{5}_{(1,1)}\oplus \mathbf{1}_{(1,3)}\right) \\
=\mathbf{14}_{(1,1)}\oplus \mathbf{10}_{(1,1)}\oplus \mathbf{1}_{(1,1)}\oplus \mathbf{5}_{(3,1)}\oplus \mathbf{5}_{(1,3)}\oplus \mathbf{1}_{(3,3)}\,,
\label{tensorNSNS}
\end{multline}
consisting in the usual string triplet -- the graviton, the Kalb-Ramond two-form and the dilaton -- all $SO(4)$ singlets, as well as six 7d massless vectors in the adjoint representation of $SO(4)$ and nine 7d scalars, in the rank two symmetric traceless representation of $SO(4)$. We note that a direct consequence of the transversality \eqref{NStransverse} and gauge invariance \eqref{NSgaugeinv} conditions in the left- and right-moving sectors is that the physical polarizations for the graviton, two-form and vector described above are subject to analogous conditions, which can be argued directly from taking tensor products. One can then verify that assembling the corresponding closed string vertex operators using tensor products of polarizations subject to those conditions, all the states that are obtained are indeed physical. The mass-shell condition is of course still \eqref{massshell_k7d}, and we have a total of 64 states.

\paragraph{RR sector.} The situation in the RR sector is more interesting. Naively proceeding as for the NSNS sector we find
\es{tensorRR}{
\mathbf{4}_{(2,1)}\otimes \mathbf{4}_{(1,2)}=\mathbf{10}_{(2,2)}\oplus \mathbf{5}_{(2,2)}\oplus \mathbf{1}_{(2,2)}\,,
}
corresponding to a fundamental of $SO(4)$ of massless two-forms, massless vectors and massless scalars. From this perspective, there seems to be no difference between the type IIA and IIB GSO projections. However, we can obtain more insight on the RR sector by considering explicitly the tensor product between spinor polarizations from the left ($\Psi$) and right ($\tilde{\Psi}$) moving sector. Using the 7d Fierz identity, we can write\footnote{We are neglecting the dependence on $\vartheta$. Since we have a doublet of $\Psi$ in the left sector and a doublet of $\tilde{\Psi}$ in the right sector, one should imagine using \eqref{7dFierz} (and the equations derived from it) for each of the four states forming a fundamental of $SO(4)$.}
\es{7dFierz}{
\mathcal{F}\equiv \Psi\otimes (\tilde \Psi^T C)=\frac{1}{8}\,\left[\mathcal{F}_{(0)}+[\mathcal{F}_{(1)}]_{\mu}\gamma_{7d}^{\mu}-\frac{1}{2!}[\mathcal{F}_{(2)}]_{\mu\nu}\gamma_{7d}^{\mu\nu}-\frac{1}{3!}[\mathcal{F}_{(3)}]_{\mu\nu\sigma}\gamma_{7d}^{\mu\nu\sigma}\right]\,,
}
where
\begin{align}
[\mathcal F_{(n)}]_{\mu_1\ldots \mu_n}=\tilde \Psi^T\,C\,(\gamma_{7d})_{\mu_1\ldots \mu_n}\,\Psi\,, 
\end{align}
$C$ is the intertwiner for transposition
\begin{equation}\label{C7ddef}
C^{-1} \gamma_{7d}^\mu C = - (\gamma_{7d}^\mu)^T \,, 
\end{equation}
and $(\gamma_{7d})_{\mu_1\ldots \mu_n}\equiv (\gamma_{7d})_{[\mu_1}\dots (\gamma_{7d})_{\mu_n]}$. The $\mathcal F_{(n)}$ are differential forms encoding the off-shell degrees of freedom in the RR sector. To study on-shell physical states, one should impose the Dirac equation of $\Psi$ and $\tilde{\Psi}$, working out the consequences on the differential forms $\mathcal F_{(n)}$, as for the 10d critical string. Here is when we finally observe a difference between the type IIA string, where one uses \eqref{R-7d-eom-S+} for $\Psi$ and \eqref{R-7d-eom-S-} for $\tilde{\Psi}$, and the type IIB string, where instead the same Dirac equation \eqref{R-7d-eom-S+} is used for both $\Psi$ and $\tilde{\Psi}$, according to our definition in Table \ref{tab:IIA-IIB-supercharges}. In other words, the R sector equations of motion read
\es{eomspinorsRR}{
\left(\ii\,\slashed{{p}}-\tfrac{Q}{2}\right)\Psi=0\,,\qquad
\left(\ii\,\slashed{{p}}+\mathtt{s}\,\tfrac{Q}{2}\right)\tilde{\Psi}=0\,,
}
where  $\mathtt{s}=+1$ and $\mathtt{s}=-1$ respectively denote type IIA and type IIB. This results in the following equations of motion for the bi-spinor $\mathcal{F}$ in \eqref{7dFierz}\footnote{Note the sign change in the second equation compared to \eqref{eomspinorsRR}, which is a result of taking the transpose and using \eqref{C7ddef}.}
\es{}{
\left(\ii\,\slashed{{p}}-\tfrac{Q}{2}\right)\,\mathcal{F}=0=\mathcal{F}\,\left(\ii\,\slashed{p}-\mathtt{s}\tfrac{Q}{2}\right)\,,
}
which we analyze in what follows for the two choices of sign of $\mathtt{s}$.

In the {\bf type IIA} case ($\mathtt{s}=+1$), we find that the equations of motion for the spinor polarizations in the two sectors impose
\es{IIAeom}{
\diff \mathcal{F}_{(1)}&=0\,,\quad \star \diff \star \mathcal{F}_{(2)}=0\,,\quad \diff \mathcal{F}_{(3)}=0\,,\\
        \diff \mathcal{F}_{(0)}&=\frac{Q}{2}\mathcal{F}_{(1)}\,,\quad
        \star \diff \star \mathcal{F}_{(1)}=-\frac{Q}{2}\mathcal{F}_{(0)}\,,\quad \diff \mathcal{F}_{(2)}=\frac{Q}{2}\mathcal{F}_{(3)}\,,\quad  
        \star \diff \star \mathcal{F}_{(3)}=-\frac{Q}{2}\mathcal{F}_{(2)}\,,
}
where we have moved to position space so as to be able to use a more compact differential form notation, which follows from replacing $\ii\,{p}_{\mu} \rightarrow \partial_{\mu}$. Combining the equations in \eqref{IIAeom}, we find that the sixteen on-shell degrees of freedom contained in \eqref{7dFierz} are organized, in the IIA case, into a scalar $\mathcal{F}_{(0)}$ and a two-form $\mathcal{F}_{(2)}$, with equations of motion
\begin{align}\label{eomRRIIA_ws}
    \star \diff \star\diff \mathcal{F}_{(0)}=-\frac{Q^2}{4}\mathcal{F}_{(0)}\,,\quad
    \star \diff \star\diff \mathcal{F}_{(2)}=-\frac{Q^2}{4}\mathcal{F}_{(2)}\,,
\end{align}
while $\mathcal{F}_{(1)}$ and $\mathcal{F}_{(3)}$ are related to the above via $\mathcal{F}_{(1)}=\frac{2}{Q}\diff \mathcal{F}_{(0)}$ and $\mathcal{F}_{(3)}=\frac{2}{Q}\diff \mathcal{F}_{(2)}$. As a consequence of its equations of motion, the polarization for $\mathcal{F}_{(2)}$ is transverse with respect to the 7d momentum $p_{\mu}$:
\es{}{
\star \diff \star \mathcal{F}_{(2)}=0\,.
}

On the other hand, for {\bf type IIB} ($\mathtt{s}=-1$) we impose the same equations of motion on both sides of \eqref{7dFierz} and obtain the system
\begin{align}
    \begin{split}
        \diff \mathcal{F}_{(0)}&=\mathcal{F}_{(0)}=0\,,\quad 
        \star \diff \star \mathcal{F}_{(1)}=0\,,\quad
        \diff \mathcal{F}_{(2)}=0\,,\quad 
         \star \diff \star \mathcal{F}_{(3)}=0\,,\\
          \diff \mathcal{F}_{(1)}&=-\frac{Q}{2}\mathcal{F}_{(2)}\,,\quad
          \star\diff \star \mathcal{F}_{(2)}=-\frac{Q}{2}\mathcal{F}_{(1)}\,,\quad
         \star \diff \mathcal{F}_{(3)}=-\frac{Q}{2}\mathcal{F}_{(3)}\,.
    \end{split}
\end{align}
Combining these equations, we find that now the degrees of freedom are arranged in a vector $\mathcal{F}_{(1)}$ and a self-dual three-form $\mathcal{F}_{(3)}$, subject to the equations of motion 
\begin{align}\label{eomRRIIB_ws}
    \star \diff \star \diff \mathcal{F}_{(1)}=\frac{Q^2}{4}\mathcal{F}_{(1)}\,,\quad \diff \mathcal{F}_{(3)}=\frac{Q}{2}\star \mathcal{F}_{(3)}\,,
\end{align}
while $\mathcal{F}_{(2)}=-\frac{2}{Q}\diff \mathcal{F}_{(1)}$ and $\mathcal{F}_{(0)}$ simply vanishes. As a consequence of their equations of motion, both $\mathcal{F}_{(1)}$ and $\mathcal{F}_{(3)}$ have polarizations that are transverse with respect to $p_{\mu}$, or
\es{}{
\star \diff \star \mathcal{F}_{(1)}=0\,,\quad
\star \diff \star \mathcal{F}_{(3)}=0\,.
}

Of course, the same exact analysis can be repeated for each state in the $SU(2)_+$ doublet of spinors in the left sector, tensored with each state in the $SU(2)_-$ doublet of spinors in the right sector, so we end up with a fundamental of $SO(4)$ of the fields described above. 

{
Once again, a comment about the concept of mass is in order. As apparent from \eqref{eomRRIIA_ws} and \eqref{eomRRIIB_ws}, with both GSO projections the fields in the RR sector are differential forms of various degrees, whose polarizations $\mathcal{F}_{(p)}$ are subject to constraints that are naturally interpreted as equations of motion for {\it massive} fields, where one can read off a common value for the mass of $Q/2$. Despite this fact, the states considered here satisfy the same mass-shell conditions as the graviton in the NSNS sector, which we have called massless due to the presence of a local gauge symmetry. As we explained, these two facts are not contradictory: rather, their coexistence will be explained in Section \ref{sec:7dsugra}, where we introduce two 7d gauged supergravities ($SO(4)$ and $ISO(4)$) and study their spectrum in the linear dilaton background. 

To conclude, let us comment on the relation between this discussion and the states listed in \eqref{tensorRR}, which from a spacetime perspective are classified using the massless little group $SO(5)$. In the case of type IIA, \eqref{eomRRIIA_ws} describe the degrees of freedom of a massive scalar and a massive two-form, carrying one and fifteen degrees of freedom respectively. The former is obviously not a problem, since a scalar always has one degree of freedom independently of its mass, while the fifteen degrees of freedom of the massive two-form can be thought of as arising from a description in terms of a massless vector and a massless two-form (the $\mathbf{5}$ and the $\mathbf{10}$ in \eqref{tensorRR}, respectively), coupled {\it à la} St\"uckelberg, so that they can collectively be thought of as a single massive two-form. On the other hand, in the type IIB case \eqref{eomRRIIB_ws} describe the polarizations of a massive vector and a massive self-dual three-form\footnote{\label{selfdualthreeform}Self-duality in odd dimensions for massive forms is common in gauged supergravity, as first described in \cite{Townsend:1983xs} -- see also \cite{Nastase:1999kf} for a notable instance of this. In $d=2k+1$ this corresponds to a $k-$form $S$ satisfying first order equations of motion $\diff S \sim m\,\star S$, which implies the massive $k-$form equations $\star\diff\star\diff S\sim m^2 S$. Since the latter are first-order, they only give rise to half of the usual $\binom{2k}{k}$ degrees of freedom of a massive $k-$form in $d=2k+1$ dimensions.}, carrying six and ten degrees of freedom respectively. The six from the massive vector can again be thought of in terms of a Higgs mechanism, this time between the $\mathbf{1}$ and the $\mathbf{5}$ in \eqref{tensorRR}, while the degrees of freedom of a self-dual massive three form are the same as those of a massless two form (the $\mathbf{10}$ in \eqref{tensorRR}) in seven dimensions. We note that this interpretation in terms of massless fields with St\"uckelberg couplings, which might appear arbitrary from the worldsheet perspective, is exactly what arises from the supergravity description of Section \ref{sec:7dsugra}. 

Finally, we note that insisting on a 7d interpretation of the RR states proved instrumental in distinguishing the type IIA and IIB GSO projections, at least at this mass level. Indeed, one could in principle only focus on 6d Poincaré invariance and consider the 6d Fierz identity in \eqref{7dFierz}. The result would give a set of differential forms ($\mathcal{F}^{a,b,\ell,m}$ in \eqref{RR-12-12}) with coupled equations of motion which is both hard to disentangle and very similar in the IIA and IIB case.\footnote{Indeed, the circle reduction of the corresponding 7d supergravities leads to two indistinguishable 6d theories, like in the case of 10d type IIA/B supergravity on a circle. This statement is related to T-duality and in fact an analogous symmetry was observed between the $SO(4)$ and $ISO(4)$ supergravities that are relevant here, see \cite{Malek:2015hma}.} On the other hand, grouping the various states in 7d polarizations gives much cleaner results and reveals the difference in field content between the two GSO projections, which in the language of 7d supergravity corresponds to two different choices of gauging. 
}

\paragraph{NSR and RNS sectors.} The fermionic states are contained in the RNS and NSR sectors, where the relevant tensor products of polarizations are
\es{NSRRNS}{
\mathbf{4}_{(2,1)}\otimes\left(\mathbf{5}_{(1,1)}\oplus \mathbf{1}_{(1,3)}\right) \oplus 
\left(\mathbf{5}_{(1,1)}\oplus \mathbf{1}_{(3,1)}\right)\otimes \mathbf{4}_{(1,2)}=
\mathbf{16}_{(2,1)\oplus(1,2)}\oplus
\mathbf{4}_{(2,1)\oplus(2,1)}\oplus
\mathbf{4}_{(3,2)\oplus(2,3)}\,,
}
corresponding to four gravitini and sixteen matter fermions. The physical state conditions for the matter fermions are the same as those worked out for the R sector, which can be summarized by the 7d Dirac equation \eqref{R-7d-eom-S+} (or \eqref{R-7d-eom-S-}, according to the set of supercharges that one chooses), while those for the gravitini can be obtained by combining those for the vector in the NS sector -- \eqref{NStransverse} and \eqref{NSgaugeinv} -- with the Dirac equation and projecting out the spin-1/2 degrees of freedom as usual. We have a total of 128 degrees of freedom, equal to the sum of 64 NSNS and 64 RR physical states. {Note that the counting in \eqref{NSRRNS} uses the massless 7d little group, but much like in the RR sector these degrees of freedom combine to give rise to massive-looking equations of motion. One should then consider the four matter fermions in the $(2,1)\oplus(1,2)$ as contributing to the degrees of freedom of massive $(2,1)\oplus(1,2)$ gravitini, by means of a super Higgs mechanism. As a result, one has $(2,1)\oplus(1,2)$ massive gravitini and $(3,2)\oplus(2,3)$ massive fermions. Once again, while here this might appear highly conjectural and the meaning of mass might look misleading, a precise justification of these statements will be given in Section \ref{sec:7dsugra}.}

\paragraph{Comments on supersymmetry.} Our results for the closed string spectrum are summarized in Tables \ref{tab:IIAspectrum} and \ref{tab:IIBspectrum}, respectively. We would now like to organize the states at the lowest level in supersymmetry multiplets. {As we discussed, the issue of the mass of the states is subtle due to the spontaneously broken Poincaré invariance. However, for the purpose of analyzing the supermultiplets, we can think of the states in terms of representations of the 7d little group, as we did in eqs.~\eqref{tensorNSNS}, \eqref{tensorRR} and \eqref{NSRRNS}. Let us then classify here the states in terms of massless 7d supersymmetry multiplets and postpone the discussion of how certain masses are acquired to Section \ref{sec:NSNS}. The results, which from this massless perspective are common for the IIA and IIB GSO projections are summarized in Table \ref{tab:masslessspectrum}.}
\begin{table}[!ht]
\begin{center}
\begin{tabular}{|c|c|c|}
\hline
field & sector & $SU(2)_+\times SU(2)_-$  \\
\hline
graviton & NSNS & $(1,1)$ \\
2-form & NSNS & $(1,1)$ \\
vectors & NSNS & $(1,3)\oplus (3,1)$ \\
scalars & NSNS & $(1,1)\oplus (3,3)$ \\
0-forms & RR & $(2,2)$ \\
1-forms & RR & $(2,2)$ \\
2-forms & RR & $(2,2)$\\
spinors & NSR, RNS & $(2,3)\oplus (3,2) \oplus (1,2)\oplus (2,1)$ \\
gravitini & NSR, RNS & $(1,2)\oplus (2,1)$  \\
\hline 
\end{tabular}
\caption{Lowest excited states of the worldsheet model \eqref{10d-ws-CFT}, classified according to the massless 7d little group. Note that the two GSO projections lead to two indistinguishable sets of states from this (massless) perspective, so the results summarized here hold for both the IIA and the IIB noncritical superstring.}
\label{tab:masslessspectrum}
\end{center}
\end{table}
{
As discussed earlier in this section, the worldsheet CFT \eqref{10d-ws-CFT} contains eight spacetime supercharges in each sector (holomorphic and anti-holomorphic), for a total of sixteen for the closed string. Insisting on our 7d interpretation, it would then seem appropriate to consider multiplets of 7d half-maximal supersymmetry. It is indeed possible to classify the states above in this way, with the result that one has 1 graviton + 2 gravitino + 3 vector massless multiplets of 7d half-maximal supersymmetry. However, as observed in \cite{Gadde:2009dj}, this is not the end of the story as these states can actually be thought of in terms of 7d maximal supersymmetry. As first conjectured in \cite{Gadde:2009dj} we will then proceed to interpret this spectrum as arising from a 7d maximal supergravity, which admits the linear dilaton background as one of its solutions. The full theory is then fully diffeomorphism invariant in 7d and preserves 32 supercharges, and only the choice of vacuum spontaneously breaks 7d Poincaré invariance to 6d as well as 16 of the 32 supercharges. In the standard formulation of string theory one has to choose a vacuum, and in this case this choice breaks some symmetries which we conjecture to actually exist in the supergravity theory. We will see in Section \ref{sec:NSNS} that the gauged supergravity that we associate with the $d=6$ noncritical superstring theories indeed do not admit maximally supersymmetric vacua, but only at most 1/2-BPS domain wall solutions, the simplest of which corresponds to the linear dilaton background \eqref{CHS-b}. As we discussed at the beginning of this section, a combination of this choice of vacuum and the gauging in the 7d supergravity is responsible for the fact that, while all fields appear at the same mass level, the RNS, NSR and RR fields satisfy ``massive-looking'' field equations. Precisely which fields appear to be massive and the value of the mass term is crucial in the identification of the correct gauging for the corresponding 7d supergravities and moreover distinguishes the IIA and IIB GSO projections. For this reason, we find it useful to describe again the states from this perspective in Table \ref{tab:IIAspectrum} and \ref{tab:IIBspectrum}, this time emphasizing which fields have polarizations satisfying the physical state conditions typical of massive fields.}
\begin{table}[!ht]
\begin{center}
\begin{tabular}{|c|c|c|c|}
\hline
\multicolumn{4}{|c|}{\bf Type IIA}\\
\hline
field & sector & $SU(2)_+\times SU(2)_-$ & $|\text{mass}|$ \\
\hline
graviton & NSNS & $(1,1)$ & 0\\
2-form & NSNS & $(1,1)$ & 0\\
vectors & NSNS & $(1,3)\oplus (3,1)$ & 0\\
scalars & NSNS & $(1,1)\oplus (3,3)$ & 0\\
0-forms & RR & $(2,2)$ & $Q/2$\\
2-forms & RR & $(2,2)$ & $Q/2$\\
spinors & NSR, RNS & $(2,3)\oplus (3,2)$ & $Q/2$ \\
gravitini & NSR, RNS & $(1,2)\oplus (2,1)$ & $Q/2$ \\
\hline 
\end{tabular}
\caption{Lowest excited states of the worldsheet model \eqref{10d-ws-CFT} with the type IIA GSO projection, from a 7d perspective. We describe irreducible representations of $SO(4)\simeq [SU(2)_+\times SU(2)_-]/\Z_2$ using their dimension. Recall that all fields are at the same mass level, but we are reporting the value of the mass term appearing in the equations of motion for RNS, NSR and RR fields.}
\label{tab:IIAspectrum}
\end{center}
\end{table}
\begin{table}[!ht]
\begin{center}
\begin{tabular}{|c|c|c|c|}
\hline
\multicolumn{4}{|c|}{\bf Type IIB}\\
\hline
field & sector & $SU(2)_+\times SU(2)_-$ & $|\text{mass}|$ \\
\hline
graviton & NSNS & $(1,1)$ & 0\\
2-form & NSNS & $(1,1)$ & 0\\
vectors & NSNS & $(1,3)\oplus (3,1)$ & 0\\
scalars & NSNS & $(1,1)\oplus (3,3)$ & 0\\
1-forms & RR & $(2,2)$ & $Q/2$\\
s.d. 3-forms & RR & $(2,2)$ & $Q/2$\\
spinors & NSR, RNS & $(2,3)\oplus (3,2)$ & $Q/2$ \\
gravitini & NSR, RNS & $(1,2)\oplus (2,1)$ & $Q/2$ \\
\hline 
\end{tabular}
\caption{Lowest excited states of the worldsheet model \eqref{10d-ws-CFT} with the type IIB GSO projection, from a 7d perspective. We describe irreducible representations of $SO(4)\simeq [SU(2)_+\times SU(2)_-]/\Z_2$ using their dimension. Recall that all fields are at the same mass level, but we are reporting the value of the mass term appearing in the equations of motion for RNS, NSR and RR fields.}
\label{tab:IIBspectrum}
\end{center}
\end{table}
{When comparing the results summarized here to the discussion of Section \ref{sec:7dsugra}, one should think of the states listed in Table \ref{tab:masslessspectrum} as the naive spectrum of maximal 7d {\it ungauged} supergravity. A gauging of such theory is a deformation which has the effect, among others, to introduce mass terms for some of the fields. Our results show that the $SO(4)$ and $ISO(4)$ gaugings introduce precisely the mass terms listed in Table \ref{tab:IIAspectrum} and \ref{tab:IIBspectrum} for type IIA and IIB, respectively -- see Section \ref{sec:LDspectrum}.}

\section{Boundary states for noncritical strings}
\label{sec:boundarystates}

In this section, we consider the boundary states of the $d=6$ noncritical superstring theory \eqref{8d-ws-CFT}. In particular, we study BPS D3 branes localized at the tip of the cigar and BPS D5 branes extended along the cigar in type IIB theory. As we are going to discuss in Section \ref{sec:RR}, the number of D3 branes captures the amount of colors of the dual gauge theory while the D5 branes capture the flavor of the boundary theory. Our goal here is to obtain a worldsheet computation of the backreaction generated by the inclusion of D-branes, to which we can compare the supergravity solutions with RR fields that we present in Section~\ref{sec:RR}. The comparison will necessarily be qualitative: while the computations of this section are carried out in the $k=2$ noncritical superstring \eqref{8d-ws-CFT}, the supergravity solutions of Section \ref{sec:RR} can only be trusted at large level $k$. We will comment in more detail about this in Section \ref{sec:RR}. 

In Sections~\ref{sec:D3} and \ref{sec:D5}, we construct D3 and D5 brane boundary states for the noncritical string theory~\eqref{8d-ws-CFT}. The discussion there is a simple adaptation to the $d=6$ setup of the analysis carried out in \cite{Fotopoulos:2005cn, Ashok:2005py, Murthy:2006xt} for the $d=4$ noncritical superstring $\mathbb R^{1,3} \times \frac{SL(2,\mathbb R)_1}{U(1)}$ and in \cite{DiVecchia:1997vef,Billo:1998vr} for the 10d critical superstring. We proceed in Section \ref{sec:non-vanishing-backreaction}, where we analyze which of the low-lying states identified in Section~\ref{sec:spectrum} admit non-zero overlaps. To the best of our knowledge, the analysis of Section~\ref{sec:non-vanishing-backreaction} did not appear elsewhere. In Section~\ref{sec:overlapasymptotics}, adapting the techniques developed in \cite{Ashok:2005py, Murthy:2006xt} to the \eqref{8d-ws-CFT} setup, we investigate the asymptotic behavior of the graviton backreaction in the $\rho \to \infty$ limit. While we devote most of this section to the type IIB GSO projection, we comment on the type IIA case in Section \ref{sec:D2-D4}. Finally, in Section~\ref{sec:boundary-summary} we briefly summarize our findings. 

\subsection{D3 branes}
\label{sec:D3}

Let us discuss the construction of the boundary state corresponding to a type IIB D-brane extended along four flat directions and localized at the tip of the cigar. Semiclassically, the brane configuration is
\begin{equation}
\begin{tabular}{cccc}
 IIB  & $\R^{1,3}$& $\R^2$  & Cigar  \\
   \hline
   D3  & $\times \times \times \times $ & &  
\end{tabular}
\label{D3-config}
\end{equation}
where $\times$ denotes a spacetime direction along which the brane is extended. These branes, which we will refer to as ``D3 branes'', are BPS and preserve momentum around the cigar. While the boundary state in the NSNS sector will be constructed in the canonical $(-1,-1)$ picture, in the RR sector we consider boundary states in the $(-\tfrac{1}{2}, -\tfrac{3}{2})$ picture. Analogously, when computing RR sector overlaps we will consider vertex operators in the $(-\tfrac{3}{2},-\tfrac{1}{2})$ picture. This asymmetric choice in the picture is necessary to saturate the ghost background charge independently in the holomorphic and anti-holomorphic sector and to conserve the superghost current on the brane, see \cite{Billo:1998vr, Ashok:2005py} for more details on this point. The D3-brane boundary state reads \cite{DiVecchia:1997vef, Billo:1998vr, Ashok:2005py, Fotopoulos:2005cn} 
\begin{equation}
||D3\rangle \rangle = ||D3, +\rangle \rangle_{\text{NS}} - ||D3,-\rangle \rangle_{\text{NS}} + ||D3,+\rangle \rangle_{\text{R}} + ||D3,-\rangle \rangle_{\text{R}} \,, 
\label{D3}
\end{equation}
where $\eta = \pm$ labels the choice of spin structure and 
\begin{equation}
    ||D3, \eta\rangle \rangle_{\text{NS,R}} = \frac{T}{2} ||B_X \rangle \rangle \,\, ||D0, \eta\rangle \rangle_{\text{NS,R}}^{cig} \,\, ||B_\psi, \eta \rangle \rangle_{\text{NS,R}} \,\, ||B_{gh}\rangle \rangle \,\, ||B_{sgh}, \eta \rangle \rangle_{\text{NS,R}} \,. 
    \label{D3eta}
\end{equation}
In eq.~\eqref{D3eta}, $T$ denotes the overall normalization. Its precise value will not be relevant for us. The linear combination entering \eqref{D3} is chosen by imposing the type IIB GSO projection. Before explaining the various ingredients entering eq.~\eqref{D3eta}, notice that the same choice of spin structure $\eta$ must be made in each factor of \eqref{D3eta} in order to have a well-defined periodicity for the total $\mathcal N=1$ worldsheet supercurrent \cite{Fotopoulos:2005cn, Ashok:2005py}. In eq.~\eqref{D3eta}, $||B_X \rangle \rangle \,$ is the Ishibashi state constructed out of the flat space free bosons \cite{Recknagel:2013uja},
\begin{align}
    ||B_X \rangle \rangle &= \delta(x^4) \, \delta(x^5) \, \exp \left[ - \sum_{n=1}^\infty \frac{1}{n} \alpha^\alpha_{-n} \, S_{\alpha \beta} \, \tilde \alpha^\beta_{-n} \right] \ket{0} \,, \qquad \alpha, \beta = 0, \dots , 5 \,, \\[0.2cm]
    S_{\alpha \beta} &= \begin{cases} \eta_{\alpha \beta} & \alpha, \beta = 0, \dots 3 \\ -\delta_{\alpha \beta} &  \alpha, \beta = 4,5 \end{cases} \,, 
\end{align}
implementing Neumann boundary conditions along the flat directions $\alpha, \beta = 0, \dots, 3$ and Dirichlet boundary conditions along $\alpha, \beta = 4,5$. The Ishibashi state $||B_\psi, \eta \rangle \rangle_{NS, R}$ captures the dependence on the six free fermions $\psi^\alpha$ along the flat directions, as well as the cigar free fermions $\psi^\rho$ and $\psi^\theta$. It reads \cite{DiVecchia:1997vef,Billo:1998vr,Ashok:2005py, Recknagel:2013uja}
\begin{subequations}
\begin{align}
    ||B_\psi, \eta \rangle \rangle_{NS} & = \exp \Biggl[ \ii \, \eta \sum_{m=\frac{1}{2}}^\infty \psi^{\hat \mu}_{-m} \, S_{\hat \mu \hat \nu} \, \tilde \psi^{\hat \nu}_{-m} \Biggr] \ket{0} \,,  \\
    ||B_\psi, \eta \rangle \rangle_{R} & = \exp \Biggl[ \ii \, \eta \sum_{m=1}^\infty \psi^{\hat \mu}_{-m} \, S_{\hat \mu \hat \nu} \, \tilde \psi^{\hat \nu}_{-m} \Biggr] \ket{B_\psi, \eta}_R^{(0)} \,, \\
     S_{\hat \mu \hat \nu} &= \begin{cases} \eta_{\hat \mu \hat \nu} & \hat \mu, \hat \nu = 0, 1,2, 3, \rho, \theta \\ -\delta_{\hat \mu \hat \nu} &  \hat \mu, \hat \nu = 4,5 \end{cases} \,, 
\end{align}
\label{B-psi}%
\end{subequations}
where the coordinate labels $\hat \mu, \hat \nu$ run over all eight spacetime directions $\hat \mu, \hat \nu = 0, \dots, 5, \rho, \theta$. Notice that we choose Neumann boundary conditions for the cigar free fermions $\psi^\rho$, $\psi^\theta$. We will review how this comes about momentarily. The R-sector vacuum $\ket{B_\psi, \eta}_R^{(0)}$ cannot be written as a tensor product of Ishibashi states for the flat directions and the cigar. It has to be selected by requiring the zero mode gluing conditions
\begin{subequations}
\begin{align}
    \left(\psi^\alpha_0 - i \, \eta \, \tilde \psi^\alpha_0 \right)|| B_\psi, \eta \rangle \rangle^{(0)}_R & = 0 \,, \qquad 
\text{for} \qquad \alpha = 0, 1, 2, 3 \,, \\
 \left(\psi^\alpha_0 + i \, \eta \, \tilde \psi^\alpha_0 \right)|| B_\psi, \eta \rangle \rangle^{(0)}_R & = 0 \,, \qquad 
\text{for} \qquad \alpha = 4,5 \,, \\
\left(\psi^\rho_0 - i \, \eta \, \tilde \psi^\rho_0 \right)|| B_\psi, \eta \rangle \rangle^{(0)}_R &= 0 \,, \\
\left(\psi^\theta_0 - i \, \eta \, \tilde \psi^\theta_0 \right)|| B_\psi, \eta \rangle \rangle^{(0)}_R &= 0   \,. 
\end{align}
\label{free-fermions-gluing}%
\end{subequations}
The construction of this vacuum state is standard in the literature, see \cite{DiVecchia:1997vef,Billo:1998vr, Ashok:2005py}. The state $||D0, \eta\rangle \rangle_{\text{NS,R}}^{cig}$ in \eqref{D3eta} corresponds to the D0-brane localized at the tip of the bosonic cigar\footnote{Note that the boundary states of the cigar CFT are Euclidean branes, so the D0 branes mentioned here are genuinely localized from the cigar perspective. Similarly, the D2 branes of the cigar mentioned in the next subsection are extended along two directions. This should be contrasted with Lorentzian D$p$ branes, whose worldvolume includes $p$ Euclidean directions as well as one time direction.}. For each sector and choice of spin structure, the wavefunctions read \cite{Ribault:2003ss, Eguchi:2003ik, Ahn:2003tt, Israel:2004jt, Ahn:2004qb, Fotopoulos:2004ut, Hosomichi:2004ph, Ashok:2005py, Fotopoulos:2003vc}
\begin{align}
    \Psi_{D0}(s,m) \equiv \langle \Phi^j_{m,\bar m}| \, ||D0, \pm \rangle \rangle_{\text{NS,R}}^{cig} & \sim \delta_{m, \bar m} \, \nu^{\mathtt j+\frac{1}{2}} \frac{\Gamma(-\mathtt j+m)\Gamma(-\mathtt j-m)}{\Gamma(-2 \mathtt j-1)\Gamma(1-\frac{1+2 \mathtt j}{k})} \,,
    \label{D0-overlap}
\end{align}
where $\Phi^{\mathtt j}_{m,\bar m}$ denotes a primary of the bosonic cigar coset CFT (at level $k+2$) with spin $\mathtt j$ and charge $m$ (respectively $\bar m$) under the purely bosonic current $j^3$ (respectively $\tilde j^3$). In eq.~\eqref{D0-overlap} $\sim$ denotes equality up to phases independent of $\mathtt j$ and factors of $k$ not playing a role in our analysis. In eq.~\eqref{D0-overlap}, $\nu$ denotes the worldsheet cosmological constant, see {\it e.g.}~\cite{Giveon:1998ns, Teschner:1999ug}. In the cigar theory, the bosonic charges $m$, $\bar m$ are related to momentum and winding as 
\begin{equation}
    m = \frac{1}{2}\left(n + w k \right) \,, \qquad \bar m = -\frac{1}{2}(n-w k) \,, \qquad \qquad n, w \in \Z \,, 
\end{equation}
and hence at level $k=2$ obey
\begin{equation}
    m - \bar m \in \Z \,, \qquad \text{and} \qquad m + \bar m \in 2 \Z \,. 
    \label{cigar-m-quantization}
\end{equation}
The effect of the orbifold in \eqref{8d-ws-CFT} is to relax the second condition in eq.~\eqref{cigar-m-quantization}, so that instead of \eqref{cigar-m-quantization} $m$ and $\bar m$ obey (see {\it e.g.}~\cite{Chang:2014jta})
\begin{equation}
    m - \bar m \in \Z \qquad \text{and} \qquad    m + \bar m \in \Z 
\end{equation}
in the $\Z_2$-orbifolded theory. Finally, in eq.~\eqref{D3eta} $||B_{gh}\rangle \rangle $ and $ ||B_{sgh}, \eta \rangle \rangle_{\text{NS,R}}$  denote respectively the contribution of the $b,c$ ghosts and of the $\beta, \gamma$ superghosts. Their form is as in the flat space string and explicit expressions can be found \textit{e.g.}~in \cite{Billo:1998vr}.

\paragraph{Neumann boundary conditions for the cigar free fermions.} Since this has important consequences for us, let us review why Neumann boundary conditions must be chosen for the cigar free fermions $\psi^\rho$, $\psi^\theta$. $\mathcal N=2$ superalgebras admit two distinct gluing conditions, known as A- and B-type. Different references interchange their naming, and we follow the conventions of \cite{Israel:2004jt}. A-type gluing conditions read
\begin{equation}
\left(J_{n}^{\mathcal R} - \tilde J_{-n}^{\mathcal R} \right)|| B, \eta\rangle \rangle = 0 \,, \qquad  \left(G_{r}^\pm - i \eta \, \tilde G_{-r}^{\mp} \right)|| B, \eta \rangle \rangle = 0 \,, 
\label{A-type}
\end{equation}
while B-type gluing conditions correspond to 
\begin{equation}
\left(J_{n}^{\mathcal R} + \tilde J_{-n}^{\mathcal R} \right)|| B, \eta \rangle \rangle = 0 \,, \qquad  \left(G_{r}^\pm - i \eta \, \tilde G_{-r}^{\pm} \right)|| B, \eta \rangle \rangle = 0 \,, 
\end{equation}
with $J^{\mathcal R}$ and $\tilde J^{\mathcal R}$ denoting the $\mathcal R$-currents. In both cases, $G = G^+ + G^-$ obeys the gluing conditions necessary to preserve $\mathcal N=1$ supersymmetry on the worldsheet,
\begin{equation}
\left(L_{n} - \tilde L_{-n}\right)||B, \eta\rangle \rangle  = 0 \,, \qquad \left(G_{r} - i \eta \tilde G_{-r}\right)||B, \eta \rangle \rangle  = 0 \,, \label{G=bar G}
\end{equation}
where $L_n$ denote the modes of the total stress tensor and $G_r$ the modes of the worldsheet $\mathcal N=1$ supercurrent. For the case of the $\mathcal N=2$ worldsheet cigar CFT, the D0-brane localized at the tip of the cigar corresponds to B-type boundary conditions based on the identity representation\footnote{In fact, no BPS boundary states can be constructed out of A-type identity representation Ishibashi states, see \cite{Fotopoulos:2003vc, Ribault:2003ss, Israel:2004jt, Hosomichi:2004ph, Fotopoulos:2005cn, Israel:2005fn, Benichou:2008gb}.} and $\mathcal R$-currents can be written as 
\begin{subequations}
\begin{align}
J^{\mathcal R} &= \frac{2}{k} J^3 + \psi^{++} \psi^{--} = \frac{2}{k} j^3 + \frac{k+2}{k} \psi^{++} \psi^{--} =  \frac{2}{k} j^3 - \ii \, \frac{k+2}{2k} \psi^\rho \psi^\theta \,, \\
\tilde J^{\mathcal R} &= -\frac{2}{k} \tilde J^3 + \tilde \psi^{++} \tilde \psi^{--} =- \frac{2}{k} \tilde j^3 + \frac{k+2}{k} \tilde \psi^{++} \tilde \psi^{--} =  -\frac{2}{k} \tilde j^3 + \ii \, \frac{k+2}{2k} \tilde \psi^\rho \tilde \psi^\theta \,,  
\end{align}
\label{cigaRR-currents}%
\end{subequations}
where $J^3 = j^3 + \psi^{++} \psi^{--}$ is the $\mathcal N=1$ $\mathfrak{sl}(2, \R )$ Cartan at level $k$, $j^3$ is the bosonic $\mathfrak{sl}(2, \R )$ Cartan at level $k+2$ and\footnote{We choose this notation to distinguish from the fermions in eq.~\eqref{psipm}.}
\begin{equation}
\psi^{\pm \pm} = \frac{1}{\sqrt 2}(\psi^\rho \pm \ii \psi^\theta) \,, \qquad \tilde \psi^{\pm \pm} = \frac{1}{\sqrt 2}(\tilde \psi^\rho \mp \ii \tilde \psi^\theta) \,. 
\end{equation}

The choice of B-type boundary conditions implies Neumann boundary conditions along the angular direction of the cigar \cite{Hosomichi:2004ph, Israel:2004jt, Ashok:2005py, Fotopoulos:2005cn},
\begin{equation}
\left( j^3_n - \tilde j^3_{-n} \right) || B, \eta\rangle \rangle = 0 \,, 
\label{j3 gluing}
\end{equation} 
which in turn implies Neumann boundary conditions also for the cigar free fermions $\psi^\theta$ and $\psi^\rho$, 
\begin{equation}
\left(\psi^\rho_r - \ii \, \eta \, \tilde \psi^\rho_{-r} \right)|| B, \eta \rangle \rangle = 0 \,, \qquad 
\left(\psi^\theta_r - \ii \, \eta \, \tilde \psi^\theta_{-r} \right)|| B, \eta \rangle \rangle  = 0 \,. 
\label{psi-theta-rho-Neumann}
\end{equation}
To make contact with the notation of Section~\ref{sec:spectrum}, we can re-express eqs.~\eqref{j3 gluing} and \eqref{psi-theta-rho-Neumann} respectively as\footnote{We have $j^3 = \ii \, \partial \theta$, $\tilde j^3 = - \ii \, \bar \partial \theta$ and $\partial H =  \psi^\rho \psi^\theta$, $\bar \partial \tilde H = \tilde \psi^\rho \tilde \psi^\theta$.}
\begin{equation}
\left( \partial \theta + \bar \partial \tilde \theta \right)\Bigl|_{z=\bar z} || B \rangle \rangle = 0 \,, 
\label{theta gluing}
\end{equation} 
and 
\begin{equation}
\left( \partial H - \bar \partial \tilde H \right)\Bigl|_{z=\bar z} || B \rangle \rangle = 0 \,. 
\label{H gluing}
\end{equation}

\subsection{D5 branes}
\label{sec:D5}

BPS D5 branes extended along the cigar and corresponding to the semiclassical configuration 
\begin{equation}
\begin{tabular}{cccc}
 IIB  & $\R^{1,3}$& $\R^2$  & Cigar  \\
   \hline
   D5  & $\times \times \times \times $ & & $\times \times$ 
\end{tabular}
\label{D5-config}
\end{equation}
can be constructed  out of B-type continuous representations, similarly to the D3 branes we just described. D5 branes are built as a linear combination of the Ishibashi states 
\begin{equation}
    ||D5, P, M, \eta\rangle \rangle_{\text{NS,R}} = ||B_X \rangle \rangle \,\, ||D2, P, M, \eta\rangle \rangle_{\text{NS,R}}^{cig} \,\, ||B_\psi, \eta \rangle \rangle_{\text{NS,R}} \,\, ||B_{gh}\rangle \rangle \,\, ||B_{sgh}, \eta \rangle \rangle_{\text{NS,R}} \,. 
    \label{D5eta}
\end{equation}
The precise linear combination can be determined by carrying out a modular bootstrap analysis for $k=2$, taking into account the $\Z_2$ orbifold in \eqref{8d-ws-CFT} and imposing the GSO projection. While we bypass this analysis, the form of the Ishibashi states \eqref{D5eta} will suffice for us. The Ishibashi states $||B_X \rangle \rangle$, $||B_\psi, \eta \rangle \rangle_{\text{NS,R}}$, $||B_{gh}\rangle \rangle$ and $||B_{sgh}, \eta \rangle \rangle_{\text{NS,R}}$ are defined as in the case of the D3 brane described above while for $M= 0, \frac{1}{2}$ and $P \in \R$ the D2 brane of the cigar theory has overlaps \cite{Fotopoulos:2005cn, Murthy:2006xt, Ribault:2003ss, Eguchi:2003ik, Ahn:2003tt, Israel:2004jt, Ahn:2004qb, Fotopoulos:2004ut, Hosomichi:2004ph}
\begin{align}
    \langle \Phi^{\mathtt j}_{m,\bar m}| \, ||D2, P, M, \pm \rangle \rangle_{\text{NS,R}}^{cig} & \sim \delta_{m, \bar m} \, \nu^{\mathtt j+\frac{1}{2}} \frac{\Gamma(2+2 \mathtt j)\Gamma(\frac{2 \mathtt j+1}{k})}{\Gamma(1+\mathtt j+m)\Gamma(1+\mathtt j-m)} \cos\left(\tfrac{4 \pi s P}{k}\right) \,, 
     \label{D2-overlap}
\end{align}
where we remind that $\mathtt j = -\frac{1}{2}+is$ and as above $\sim$ denotes equality up to phases that do not depend on $\mathtt j$ and factors of $k$ not playing a role in our analysis. Notice that since the cigar D2 branes entering eq.~\eqref{D5eta} are built out of B type boundary conditions --- as it was the case for D0 branes entering eq.~\eqref{D3eta} --- the cigar free fermions $\psi^\rho$ and $\psi^\theta$ obey Neumann boundary conditions. As we will see momentarily, the fact that for \emph{both} D3 and D5 branes the cigar free fermions obey Neumann boundary conditions makes it more difficult to distinguish between D3 branes and D5 branes than what experience of the flat space 10d critical string theory would naively suggest.

\subsection{Non-vanishing backreaction}
\label{sec:non-vanishing-backreaction}

In Section~\ref{sec:spectrum} we derived the low-lying spectrum of the string theory \eqref{8d-ws-CFT}. The analysis there was carried out in the linear dilaton regime, for which the worldsheet CFT effectively reduces to the CHS background \eqref{CHS-b} with $k=2$. As already discussed, vertex operators in the linear dilaton regime are in one to one correspondence with vertex operators constructed out of continuous representations in the string theory \eqref{10d-ws-CFT}. Given the analysis of the previous sections, we are now ready to understand which of the low-lying vertex operators listed in eqs.~\eqref{tensorNSNS} and \eqref{tensorRR} admit non-zero overlaps with D3 and D5 branes. We first discuss overlaps in the NSNS sector and then proceed with the RR sector. 

\paragraph{D3 branes and NSNS sector backreaction.} Adopting the 7d jargon of Section~\ref{sec:spectrum}, in the NSNS sector we identified the following states: graviton, antisymmetric Kalb-Ramond field, dilaton, six massless 7d vectors and nine scalars. See Table~\ref{tab:IIBspectrum}. From eq.~\eqref{B-psi}, it is easy to see that the off-diagonal components of the graviton, the Kalb-Ramond field and the 7d vectors all have zero overlaps. Moreover,  only three of the nine 7d scalars $V(z,\bar z)$ satisfy 
\begin{equation}
    \left( \partial \theta + \bar \partial \tilde \theta \right)\Bigl|_{z=\bar z} V(z,\bar z) \sim 0 \,, \qquad \text{and} \qquad  \left( \partial H - \bar \partial \tilde H \right)\Bigl|_{z=\bar z} V(z,\bar z) \sim 0 \,, 
\label{theta+tthetaV=0}
\end{equation}
and hence admit non-vanishing overlaps. Notice that the three scalars admitting non-zero overlaps preserve the rotational symmetry around the axis of the cigar, as one would naively expect from the semiclassical cartoon of D3 branes localized at the tip of the cigar.

\paragraph{D3 branes and RR sector backreaction.} As already anticipated, in order to compute overlaps in the RR sector, one needs to consider vertex operators in the $(-\tfrac{3}{2},-\tfrac{1}{2})$ picture. These were constructed in Appendix~\ref{app:picture changing} in the linear dilaton regime and are in one-to-one correspondence with vertex operators constructed out of continuous representations in the string theory \eqref{8d-ws-CFT}. We will then slightly abuse the notation and identify states in terms of the ``potentials'' $\mathcal A^{a,b, \ell, m}$ introduced in Appendix~\ref{app:picture changing}. In the RR sector we can repeat the argument carried out around eq.~\eqref{theta+tthetaV=0} to deduce that only 
\begin{equation}
    \mathcal A^{++, \pm, \mp} \qquad \text{and} \qquad     \mathcal A^{--, \pm, \mp} \,. 
    \label{A-neq0-overlaps}
\end{equation}
in equation \eqref{W0} admit nonzero overlap. Since $\mathcal A^{a,b,\ell,m}$ is related to $\mathcal A^{a,b,-\ell,-m}$ by complex conjugation as in \eqref{SMW}, we then have only two independent components. In addition, the free fermion gluing conditions \eqref{free-fermions-gluing} impose that only 7d five-forms have non-zero overlap. The degree of the form sourced by the brane is dictated by the number of free fermions obeying Neumann boundary conditions. The computation is analogous to the one performed for a D5 brane in type IIB critical string theory \cite{DiVecchia:1997vef}, see also \cite{Ashok:2005py} for a (4+2)d noncritical setup. Dualizing 7d five-forms as 7d two-forms and making use of the decomposition carried out in Appendix~\ref{app:picture changing}, we find that the fields with non-zero overlap are the two-forms $\tilde{\mathcal A}_{(2)}$. 

\paragraph{D5 brane backreaction.} The analysis of states admitting non-zero overlaps with D5 branes is quite similar to the one just carried out for D3 branes. In fact, exactly the same free fermion boundary state \eqref{B-psi} enters both eqs.~\eqref{D3eta} and \eqref{D5eta} and both the overlaps \eqref{D0-overlap} and \eqref{D2-overlap} vanish for $m \neq \bar m$. As a result, the vertex operators admitting non-vanishing overlaps with the D5 brane are exactly those having non-zero overlaps with the D3 brane: the diagonal component of the graviton, the dilaton, 7d two-forms, and the three 7d scalars neutral under the $U(1)_\theta$ rotational symmetry of the cigar.

\subsection{Backreaction asymptotics}
\label{sec:overlapasymptotics}

We just observed that differently from 10d flat string theory, in the noncritical string theory \eqref{8d-ws-CFT}, color D3 and flavor D5 branes source the same fields. The functional form of one-point functions is however different, as one can observe comparing eq.~\eqref{D0-overlap} with eq.~\eqref{D2-overlap}. Following the paradigm of \cite{Ashok:2005py, Murthy:2006xt}, in this section we compute the linearized backreaction sourced by D3 and D5 branes of the noncritical string theory \eqref{8d-ws-CFT} in the $\rho \to \infty$ limit. We concentrate on the graviton backreaction, as the computation for the other fields is completely analogous. 

\paragraph{D3 branes.} The metric backreaction sourced by $N_c$ parallel D3 branes is computed by inserting a closed string propagator in the overlap of the graviton with the boundary state~\eqref{D3}, 
\begin{equation}
    \frac{1}{N_c} \, \delta \tilde h_{\mu \nu }^{D3}(q, s) = \langle V_{\mu \nu } | D_{cl}||D3\rangle \rangle \sim  S_{\mu \nu } \frac{\delta(q^0) \delta(q^1) \delta(q^2) \delta(q^3)}{q^2 - \mathtt j( \mathtt j+1)} \Psi_{D0}(s,0) \,, 
\end{equation}
where $V_{\mu \nu }$ denotes the graviton vertex operator, $\Psi_{D0}(s,m)$ was introduced in eq.~\eqref{D0-overlap} and we remind that $\mu$, $\nu$ run over $0, \dots, 5, \rho$. The position space dependence is obtained by folding with the eight-dimensional Laplacian 
\begin{equation}
\ex^{\ii \, q^\alpha x_\alpha} \, \phi(\mathtt j,\rho) \,,
\label{8d-Laplacian}
\end{equation}
and by ``Fourier transforming'' as 
\begin{align}
    \frac{1}{N_c} \, \delta h_{\mu \nu }^{D3}(x, \rho) &=  \int \text dq^0 \dots \text dq^5 \int_0^\infty \text ds \, \ex^{\ii \, q^\alpha x_\alpha} \, \phi(j,\rho) \, \langle V_{\mu \nu} | D_{cl}||D3\rangle \rangle \nonumber \\
    &\sim S_{\mu \nu}  \int \text d^2 \ell \int_0^\infty \text ds \, \frac{\nu^{\ii \, s} \, \ex^{\ii \,  \vec \ell \cdot \vec y } \, \phi(\mathtt j,\rho) \, \Gamma(\frac{1}{2} - \ii \, s)^2}{(\ell^2 + s^2 + \frac{1}{4}) \, \Gamma(-2 \, \ii \, s) \, \Gamma(1-\ii \, s)} \,. 
    \label{D0-int}
\end{align}
In eqs.~\eqref{8d-Laplacian} and \eqref{D0-int}, we introduced the compact notation
\begin{equation}
    \vec y = (x_4, x_5) \,,  \qquad \vec \ell = (q^4, q^5) \,, \qquad \ell^2 = q^4 q_4 + q^5 q_5 \,, \qquad \vec \ell \cdot \vec y = q^4 x_4 + q^5 x_5 \,, 
    \label{y-def}
\end{equation}
and $\phi(\mathtt j,\rho)$ denotes the cigar theory Laplacian, which in the $k=2$, $\rho \to \infty$ limit reads \cite{Ribault:2003ss, Ashok:2005py}
\begin{equation}
    \phi(\mathtt j, \rho) \sim \ex^{\rho(2 \, \ii \, s -1)} + \frac{\Gamma(-2 \, \ii s) \, \Gamma(\frac{1}{2} + \ii s)^2 \, \Gamma(1- \ii s) \, \nu^{-\ii \, s}}{\Gamma(2 \,  \ii s) \Gamma(\frac{1}{2} - \ii s)^2 \, \Gamma(1 + \ii s) \nu^{\ii \, s}} \, \ex^{-\rho(2 \, \ii \, s + 1)} \,. 
    \label{Laplacian}
\end{equation}
Notice that this is simply the sum of incoming and outgoing wave-functions. In fact, the ratio of Gamma functions in front of $\ex^{-\rho(2 \, \ii \, s + 1)}$ in eq.~\eqref{Laplacian} is the reflection coefficient a plane wave gets when scattering off the linear dilaton potential, see {\it e.g.}~\cite{Teschner:1997ft}. Eq.~\eqref{Laplacian} is valid at large $\rho$ and corrections are exponentially suppressed. The linearized graviton backreaction \eqref{D0-int} can thus be rewritten as 
\begin{equation}
    \frac{1}{N_c} \, \delta h_{\mu \nu }^{D3}(x, \rho) \sim S_{\mu \nu} \int \text d^2 \ell \int_{-\infty}^\infty \text ds \, \frac{\nu^{\ii \, s} \, \ex^{\ii \,  \vec \ell \cdot \vec y } \, \ex^{\rho(2 \, \ii \, s -1)} \, \Gamma(\frac{1}{2} - \ii \, s)^2}{(\ell^2 + s^2 + \frac{1}{4}) \, \Gamma(-2 \, \ii \, s) \, \Gamma(1-\ii \, s)} \,. 
    \label{D0-int-2}
\end{equation}
Notice that the integrand in \eqref{D0-int-2} has poles in $s$ with non-zero residue only at 
\begin{equation}
    s = \pm \, \ii \, \sqrt{\ell^2 + \frac{1}{4}} \,. 
\end{equation}
We should emphasize that, as was noticed in \cite{Ashok:2005py}, expressions like \eqref{D0-int-2} for the graviton backreaction should be taken with caution. In \cite{Ashok:2005py} the authors carried out a similar computation at level $k=1$ in the noncritical string theory 
\begin{equation}
    \R^{3,1} \times \frac{SL(2, \R)_1}{U(1)} \,. 
\end{equation}
They noticed that the analogue of the $s$ integral in \eqref{D0-int-2} needs a precise contour prescription to be well-defined. They suggested that a good approximation to compute the graviton backreaction of D3 branes localized at the tip of the $k=1$ cigar is to deform the integration contour in such a way that only the residue of the pole in the upper half plane contributes. We believe that, when computing the backreaction of D3 branes, this prescription makes sense also for the $k=2$ noncritical string theory \eqref{8d-ws-CFT} and accordingly we find 
\begin{equation}
    \frac{1}{N_c} \, \delta h_{\mu \nu }^{D3}(x, \rho) \sim \ex^{-\rho} \, S_{\mu \nu} \int \text d^2 \ell \, \frac{\nu^{- \sqrt{\ell^2 + \frac{1}{4}}} \, \ex^{\ii \,  \vec \ell \cdot \vec y } \, \ex^{-2 \rho \sqrt{\ell^2 + \frac{1}{4}}} \, \Gamma \left(\frac{1}{2} + \sqrt{\ell^2 + \frac{1}{4}} \right)^2}{\Gamma \left(2 \sqrt{\ell^2 + \frac{1}{4}}+1\right) \Gamma\left(\sqrt{\ell^2 + \frac{1}{4}}+1\right)} \,, \label{D0-int-3}
\end{equation}
where again we neglect factors of $2, \pi$ etc. Adopting polar coordinates and making use of the identity
\begin{equation}
    \int_0^{2 \pi} \text d \phi \, \ex^{\ii \, \ell \, y \cos \phi} = 2 \pi J_0 (\ell y) \,,
\end{equation}
we obtain 
\begin{equation}
    \frac{1}{N_c} \, \delta h_{\mu \nu }^{D3}(x, \rho) \sim \ex^{-\rho} S_{\mu \nu} \int_0^\infty \text d \ell \, \frac{\ell \, \nu^{- \sqrt{\ell^2 + \frac{1}{4}}} \,  J_0 (\ell y) \, \ex^{-2 \rho \sqrt{\ell^2 + \frac{1}{4}}} \, \Gamma \left(\frac{1}{2} + \sqrt{\ell^2 + \frac{1}{4}} \right)^2}{\, \Gamma \left(2 \sqrt{\ell^2 + \frac{1}{4}}+1\right) \Gamma\left(\sqrt{\ell^2 + \frac{1}{4}}+1\right)} \,, \label{D0-int-4}
\end{equation}
where $J_0$ denotes the Bessel function of the first kind. The asymptotic large $\rho$ behavior of the integral \eqref{D0-int-4} can be computed by the Laplace approximation, see Appendix~\ref{app:Laplace-method} for a brief review. We find an asymptotic expansion of the form
\begin{equation}
\frac{1}{N_c} \, \delta h_{\mu \nu }^{D3}(x, \rho) \sim S_{\mu \nu} \, \ex^{-2\rho} \sum_{n=0}^\infty \frac{P_n(y^2)}{\rho^{n+1}} \,,
\label{D3-asymptotic-expansion}
\end{equation}
where $P_n(y^2)$ are polynomials in $y^2$ of degree $n$, {\it e.g.}
\begin{equation}
    P_0 = \frac{1}{2 \sqrt{\pi \nu}} \,, \qquad P_1 = - \frac{y^2 + 4 \log \nu + 4 \psi(\frac{3}{2})}{16 \sqrt{\pi \nu}} \,,
\end{equation}
and $\psi(z)$ is the digamma function. The sum in \eqref{D3-asymptotic-expansion} is asymptotic and is not meant to be resummed. As we observed above, the expression \eqref{Laplacian} for the cigar Laplacian is only valid at large $\rho \to \infty$ and corrections decay exponentially. As a result, also \eqref{D3-asymptotic-expansion} should only be trusted up to subleading, exponentially suppressed, corrections.

\paragraph{D5 branes.} Let us now investigate the asymptotic behavior of the D5 graviton backreaction. Proceeding along the lines of the analysis carried out above for the D3 brane, we find the backreaction
\begin{equation}
    \frac{1}{N_f} \, \delta h_{\mu \nu }^{D5}(x, \rho) \sim S_{\mu \nu} \int \text d^2 \ell \int_{-\infty}^\infty \text ds \, \frac{\nu^{\ii \, s} \, \ex^{\ii \,  \vec \ell \cdot \vec y } \, \ex^{\rho(2 \, \ii \, s -1)} \, \Gamma(1+ 2\,  \ii \, s) \, \Gamma(\ii \, s) \, \cos\left(2 \pi s P\right)}{(\ell^2 + s^2 + \frac{1}{4}) \, \Gamma(\frac{1}{2} + \ii \, s)^2}  \,. \label{D2-int-1}
\end{equation}
Let us list the poles of the integrand with non-zero residues. As in the case of the D3 brane, we find poles at 
\begin{equation}
    s = \pm \, \ii \, \sqrt{\ell^2 + \frac{1}{4}} \,. 
\end{equation}
In addition, we also find a pole at $s=0$. Following at $k=2$ the prescription formulated in \cite{Ashok:2005py, Murthy:2006xt} for $k=1$, and hence estimating the integral \eqref{D2-int-1} by its residue at $s = \ii \, \sqrt{\ell^2 + \frac{1}{4}}$, we would obtain a backreaction decaying as $\rho^{-1} \ex^{-2 \rho}$. Exactly as in the case of the D3 brane discussed above. Comparing the semiclassical configurations \eqref{D3-config} and \eqref{D5-config}, one would instead expect the D5 graviton backreaction to decay at large $\rho$ more softly than the D3 brane backreaction. We thus believe that the correct prescription to evaluate the D5 graviton backreaction is to include the contribution of the pole at $s=0$ (more precisely its principal value), which is dominant over the contribution coming from the pole at $s = \ii \, \sqrt{\ell^2 + \frac{1}{4}}$. We find 
\begin{equation}
    \frac{1}{N_f} \, \delta h_{\mu \nu }^{D5}(x, \rho) \sim  S_{\mu \nu} \, \ex^{-\rho}  \int \text d^2 \ell \, \frac{\ex^{\ii \,  \vec \ell \cdot \vec y } }{\ell^2 + \frac{1}{4}} \sim S_{\mu \nu} \, \ex^{-\rho}  K_0 \left(\frac{y}{2} \right)  \,. 
    \label{D2-int-2}
\end{equation}
where $K_0$ is the modified Bessel function of the second kind. As it was the case above, eq.~\eqref{D2-int-2} should only be trusted up to subleading, exponentially suppressed corrections.

\subsection{D2 and D4 branes in Type IIA}
\label{sec:D2-D4}

It is not difficult to extend the analysis of the previous sections to describe in the type IIA noncritical string theory \eqref{8d-ws-CFT} D2 and D4 branes, corresponding to the semiclassical configurations 
\begin{equation}
\begin{tabular}{cccc}
 IIB  & $\R^{1,2}$& $\R^3$  & Cigar  \\
   \hline
   D2  & $\times \times \times $ & &  
\end{tabular}
\label{D2-config}
\end{equation}
and 
\begin{equation}
\begin{tabular}{cccc}
 IIB  & $\R^{1,2}$ & $\R^3$  & Cigar  \\
   \hline
   D4  & $\times \times \times $ &  & $\times \times$ 
\end{tabular}
\label{D4-config}
\end{equation}
respectively. Let us briefly mention the few differences with respect to the analysis carried out above for D3 and D5 branes. 

\paragraph{D2 branes.} The boundary state for D2 branes corresponding to the semiclassical configuration \eqref{D2-config} can be constructed similarly to the D3 brane boundary state \eqref{D3}, by defining Dirichlet instead of Neumann boundary conditions for the free boson $X^3$ and its superpartner $\psi^3$. For each choice of spin structure, eq.~\eqref{D3eta} is replaced by 
\begin{equation}
    ||D2, \eta\rangle \rangle_{\text{NS,R}} = \frac{T'}{2} ||B_X \rangle \rangle \,\, ||D0, \eta\rangle \rangle_{\text{NS,R}}^{cig} \,\, ||B_\psi, \eta \rangle \rangle_{\text{NS,R}} \,\, ||B_{gh}\rangle \rangle \,\, ||B_{sgh}, \eta \rangle \rangle_{\text{NS,R}} \,, 
    \label{D2eta}
\end{equation}
where now $||B_X \rangle \rangle$ is 
\begin{align}
    ||B_X \rangle \rangle &= \delta(x^3) \, \delta(x^4) \, \delta(x^5) \, \exp \left[ - \sum_{n=1}^\infty \frac{1}{n} \alpha^\alpha_{-n} \, S_{\alpha \beta} \, \tilde \alpha^\beta_{-n} \right] \ket{0} \,, \qquad \alpha, \beta = 0, \dots , 5 \,, \\[0.2cm]
    S_{\alpha \beta} &= \begin{cases} \eta_{\alpha \beta} & \alpha, \beta = 0, 1, 2 \\ -\delta_{\alpha \beta} &  \alpha, \beta = 3,4,5 \end{cases} \,, 
\end{align}
and $||B_\psi, \eta \rangle \rangle_{\text{NS,R}} $ reads
\begin{subequations}
\begin{align}
    ||B_\psi, \eta \rangle \rangle_{NS} & = \exp \Biggl[ \ii \, \eta \sum_{m=\frac{1}{2}}^\infty \psi^{\hat \mu}_{-m} \, S_{\hat \mu \hat \nu} \, \tilde \psi^{\hat \nu}_{-m} \Biggr] \ket{0} \,,  \\
    ||B_\psi, \eta \rangle \rangle_{R} & = \exp \Biggl[ \ii \, \eta \sum_{m=1}^\infty \psi^{\hat \mu}_{-m} \, S_{\hat \mu \hat \nu} \, \tilde \psi^{\hat \nu}_{-m} \Biggr] \ket{B_\psi, \eta}_R^{(0)} \,, \\
     S_{\hat \mu \hat \nu} &= \begin{cases} \eta_{\hat \mu \hat \nu} & \hat \mu, \hat \nu = 0, 1,2, \rho, \vartheta \\ -\delta_{\hat \mu \hat \nu} &  \hat \mu, \hat \nu = 3,4,5 \end{cases} \,.  
\end{align}
\end{subequations}
In the R sector, eq.~\eqref{free-fermions-gluing} for $\alpha =3$ is now replaced by 
\begin{equation}
     \left(\psi^3_0 + i \, \eta \, \tilde \psi^3_0 \right)|| B_\psi, \eta \rangle \rangle^{(0)}_R = 0 \,.
\end{equation}
Expressions for the ghosts contribution can again be found in \cite{Billo:1998vr}, while $||D0, \eta\rangle \rangle_{\text{NS,R}}^{cig}$ is still constructed as in eq.~\eqref{D0-overlap}. The analysis of overlaps with the  ``massless'' spectrum proceeds as in the previous sections and in the NSNS sector we find again that only the diagonal component of the graviton, the dilaton, and the three scalars preserving the $U(1)_\theta$ symmetry of the cigar admit non-vanishing overlaps. In the R-sector, only 7d three-forms (dual to 7d four-forms) of the form \eqref{A-neq0-overlaps} feature a non-zero backreaction. The asymptotic behavior of the graviton can be estimated by the integral
\begin{equation}
    \frac{1}{N_c} \, \delta h_{\mu \nu }^{D2}(x, \rho) \sim S_{\mu \nu}  \int \text d^3 \ell \int_0^\infty \text ds \, \frac{\nu^{\ii \, s} \, \ex^{\ii \,  \vec \ell \cdot \vec y } \, \phi(\mathtt j,\rho) \, \Gamma(\frac{1}{2} - \ii \, s)^2}{(\ell^2 + s^2 + \frac{1}{4}) \, \Gamma(-2 \, \ii \, s) \, \Gamma(1-\ii \, s)} \,, 
    \label{D0-int-IIA}
\end{equation}
where now $\ell$ and $y$ are defined according to 
\begin{equation}
\begin{aligned}
    \vec y & = (x_3, x_4, x_5) \,,  \qquad &  \vec \ell & = (q^3, q^4, q^5) \,, \\
    \ell^2 & = q^3 q_3 + q^4 q_4 + q^5 q_5 \,, & \qquad \vec \ell \cdot \vec y & = q^3 x_3 + q^4 x_4 + q^5 x_5 \,.  
\end{aligned}
    \label{y-def-IIA}
\end{equation}
Proceeding as in the previous sections, we find the asymptotic behavior
\begin{equation}
\frac{1}{N_c} \, \delta h_{\mu \nu }^{D2}(x, \rho) \sim S_{\mu \nu} \, \ex^{-2\rho} \sum_{n=0}^\infty \frac{P_n'(y^2)}{\rho^{n+\frac{3}{2}}} \,,
\label{D2-asymptotic-expansion}
\end{equation}
where $P_n'$ are polynomials of degree $n$. 

\paragraph{D4 branes.} Similarly to what we just discussed for D2 branes, boundary states for D4 branes corresponding to the semi-classical configuration \eqref{D4-config} can be constructed by replacing, in the expressions for the D5 brane discussed above, the Neumann boundary conditions for $X^3$ and $\psi^3$ with Dirichlet boundary conditions. The ``massless'' fields with non-zero backreaction are the same that we found in the analysis of the D2 brane. The backreaction of the graviton can be computed by the integral
\begin{equation}
    \frac{1}{N_f} \, \delta h_{\mu \nu }^{D4}(x, \rho) \sim S_{\mu \nu} \int \text d^3 \ell \int_{-\infty}^\infty \text ds \, \frac{\nu^{\ii \, s} \, \ex^{\ii \,  \vec \ell \cdot \vec y } \, \ex^{\rho(2 \, \ii \, s -1)} \, \Gamma(1+ 2\,  \ii \, s) \, \Gamma(\ii \, s) \, \cos\left(2 \pi s P\right)}{(\ell^2 + s^2 + \frac{1}{4}) \, \Gamma(\frac{1}{2} + \ii \, s)^2}  \,, \label{D2-int-1-IIA}
\end{equation}
resulting in the asymptotic $\rho \to \infty$ behavior
\begin{equation}
    \frac{1}{N_f} \, \delta h_{\mu \nu }^{D4}(x, \rho) \sim S_{\mu \nu} \, \frac{\ex^{-\rho} \, \ex^{-\frac{|y|}{2}}}{|y|} \,. 
    \label{D2-int-2-IIA}
\end{equation}

\subsection{Summary}
\label{sec:boundary-summary}

Let us take stock and summarize the results of this section. We analyzed the worldsheet construction of boundary states for D3 color branes and D5 flavor branes in the noncritical type IIB string theory \eqref{sec:RR_IIBNC} and of D2 and D4 branes in type IIA. 

\paragraph{D3 and D5 branes in type IIB.} We observed that D3 and D5 branes source the same background fields. Adopting the 7d perspective described in Section \ref{sec:spectrum}, in the NSNS sector the fields with non-zero backreaction are the diagonal component of the graviton, the dilaton, and three scalars. We observe that the three scalars with non-zero backreaction preserve the $U(1)_\theta$ symmetry of the cigar. In the RR sector, the only non-vanishing backreaction is given by the 7d two-form $\tilde{\mathcal A}_{(2)}$. While D3 and D5 branes cannot be distinguished according to which fields they source, the asymptotic behavior of their backreaction for $\rho \to \infty$ is different: compare eqs.~\eqref{D3-asymptotic-expansion} and \eqref{D2-int-2}.

\paragraph{D2 and D4 branes in type IIA.} Similarly to what happens in type IIB for D3 and D5 branes, in type IIA the D2 and D4 branes \eqref{D2-config} and \eqref{D4-config} source the same fields. The fields backreacting are in the NSNS sector again the diagonal component of the graviton, the dilaton, and three scalars preserving the $U(1)_\theta$ symmetry. Instead, in the RR sector D2 and D4 branes source 7d three-forms. The asymptotic behavior of the graviton backreaction for D2 and D4 branes is given by eqs.~\eqref{D2-asymptotic-expansion} and \eqref{D2-int-2-IIA} respectively. 

\section{The 7d supergravity}
\label{sec:7dsugra}

We would now like to identify a two-derivative gauged supergravity that captures the interactions between the massless\footnote{In the sense described in Section \ref{sec:spectrumws}.} degrees of freedom of the 6d noncritical string at low energy. The simplest vacuum of such theory should be the linear-dilaton background, hence why our requirement that it should be a {\it gauged} supergravity: this is only possible if a scalar potential is present. What is less clear is whether it should be a 7d theory (thus incorporating the Minkowski directions as well as the linear dilaton direction) or an 8d one (including also the circle direction parametrized by $\theta$ in \eqref{lineardilatonmetricd+2}). We suggest that a description with manifest supersymmetry is only possible in 7d supergravity, which can be justified in terms of the embedding of the noncritical string in ten-dimensional string theory. As we reviewed in Section \ref{sec:spectrum}, there are three equivalent descriptions: let us focus in particular on the frames that we called $\ns$ (corresponding to a distribution of NS5 branes) and the frame $\nc$ (the cigar). The two frames correspond to exact solutions of the 10d equations of motion, that in the asymptotic $\rho \to +\infty$ region are 
\begin{align}
    \ns:\quad \R^{1,5}\times \R_{\rho}\times S^3\,,\qquad
    \nc:\quad \R^{1,5}\times \R_{\rho}\times U(1)\times \frac{SU(2)}{U(1)}\,,
\end{align}
and are related by a T-duality. It is then natural to guess that one can obtain a 7d theory by reducing type II supergravity on $S^3$, or an 8d theory reducing on $SU(2)/U(1)$. Naively, since the two backgrounds are T-dual the two procedures should be somehow equivalent from the point of view of string theory. However, at the level of supergravity, it turns out that only the geometry $\ns$ is supersymmetric (preserving 16 supercharges), but supersymmetry is completely broken by the T-duality which relates this frame to $\nc$. The crucial point, as already observed by various authors \cite{Kiritsis:1993pb,Sfetsos:1995ac,Alvarez:1995np,Hassan:1999bv,Kelekci:2014ima}, is roughly speaking that the Killing spinors in the supersymmetric geometry $\ns$ depend on the direction of the T-duality (see the discussion in \cite{Plauschinn:2017ism} for more details). While at the level of the worldsheet this leads to a non-locally realized supersymmetry \cite{Hassan:1995je}, from the perspective of the supergravity the background after T-duality does not preserve any Killing spinors. The bottom line is that the disk $SU(2)/U(1)$ is a non-supersymmetric geometry and there is little hope to obtain a {\it super}gravity by performing this dimensional reduction. On the other hand, all of the supersymmetry is manifest in the frame $\ns$, and reducing 10d supergravity on $S^3$ does indeed give rise to a supergravity in 7d \cite{Cvetic:2000ah}.

Another indication for the fact that the correct description is in terms of a 7d theory is to actively look for a putative 8d gauged supergravity that serves our purposes. As discussed in the previous sections it should be a maximally supersymmetric theory (see the discussion at the end of Section~\ref{sec:spectrum}), so the spectrum is fixed, and all gaugings of maximal 8d gauged supergravity have been classified \cite{Salam:1984ft,AlonsoAlberca:2003jq,Bergshoeff:2003ri,deRoo:2011fa}. A basic requirement on the putative 8d supergravity is that it admits a supersymmetric linear dilaton solution as a vacuum, that is a background of the form \eqref{LDCFT}, excluding the $SU(2)/U(1)$ part. Supersymmetric domain walls of maximal 8d supergravity with general gauging are classified in \cite{AlonsoAlberca:2003jq}, and none of them is compatible with a linear dilaton solution. One can also check this by direct inspection making an ansatz, even without requiring supersymmetry.

For these reasons, the rest of the paper will be devoted to the study of certain 7d gauged supergravities and their supersymmetric solutions. We start in this section with a general review of maximal 7d gauged supergravity, mostly based on \cite{Samtleben:2005bp}. We will then identify two gaugings that are particularly interesting for our purposes, and discuss the corresponding supergravities more in detail. Both gaugings admit a linear dilaton solution and in the final part of the section we present one of our main results: a precise matching between the spectrum of the $SO(4)$ and $ISO(4)$ gauged supergravitiy in the linear dilaton vacuum, and the low-lying spectrum of the $d=6$ noncritical superstring presented in Section \ref{sec:spectrumws}, with type IIA and IIB GSO projection respectively.

\subsection{Review of maximal 7d gauged supergravity}

Let us now briefly review the maximal 7d supergravities. It is convenient to think of a {\it gauging} as a certain deformation of the ungauged theory, by means of a parameter $g$, the gauge coupling. All gaugings of maximal 7d supergravity were classified in \cite{Samtleben:2005bp}, which we review here. Let us start with the case of maximal 7d ungauged supergravity, whose field content is given by the vielbein $e_{\mu}^{\,\,\,m}$, 4 gravitini $\psi_{\mu}^a$, 5 two-forms $B_{M,\mu\nu}$, 10 vectors $A^{MN}_{\mu}=A^{[MN]}_{\mu}$, 16 matter fermions $\chi^{abc}$ and 14 scalars parametrizing the coset space $SL(5)/SO(5)$ via the matrix $\Vfive_M^{\,\,\,ab}$. Here $\mu,\nu,\ldots=0,\ldots,6$ are spacetime indices, $m,n,\ldots=0,\ldots,6$ are frame indices, $M,N,\ldots=1,\ldots,5$ are indices of the $SL(5)$ global symmetry and finally $a,b,\ldots=1,\ldots,4$ are indices of the $SO(5)_R\simeq USp(4)_R/\Z_2$ R symmetry. We follow the same group theory conventions as \cite{Samtleben:2005bp} (see also Appendix \ref{app:grouptheory}) and in Table \ref{tab:7dfields} we summarize the field content with the respective representations under $SL(5)$, $USp(4)_R$ and the 7d massless little group $SO(5)_{LG}$.
\begin{table}[!ht]
\begin{center}
\begin{tabular}{|c||c|c|c|c|c|c|}
\hline
fields & $e_{\mu}^{\,\,\,m}$ & $\psi_{\mu}^a$ & $B_{M,\mu\nu}$ & $A^{MN}_{\mu}$ & $\chi^{abc}$ & $\Vfive_M^{\,\,\,ab}$ \\
\hline \hline
$SO(5)_{LG}$ & $\mathbf{14}$ & $\mathbf{16}$ & $\mathbf{10}$ & $\mathbf{5}$ & $\mathbf{4}$ & $\mathbf{1}$ \\
\hline
$USp(4)_R$ & $\mathbf{1}$ & $\mathbf{4}$ & $\mathbf{1}$ & $\mathbf{1}$ & $\mathbf{16}$ & $\mathbf{5}$ \\
\hline
$SL(5)$ & $\mathbf{1}$ & $\mathbf{1}$ & $\mathbf{5}$ & $\mathbf{\overline{10}}$ & $\mathbf{1}$ & $\mathbf{5}$ \\
\hline \hline
\# d.o.f.& 14 & 64 & 50 & 50 & 64 & 14 \\
\hline 
\end{tabular}
\caption{Field content of the 7d maximal ungauged supergravity, organized in representations of the little group $SO(5)_{LG}$, R-symmetry group $USp(4)_R$ and global symmetry group $SL(5)$.}
\label{tab:7dfields}
\end{center}
\end{table}
Ungauged supergravities do not have a potential for the scalar fields, only have abelian gauge groups (under which none of the fermions are charged) and do not require a specific choice of duality frame for the $p$-form gauge fields of the theory. A gauging of the theory, on the other hand, is a deformation that changes these facts as follows:
\begin{itemize}
    \item A subset of the 10 vector fields $A^{MN}_{\mu}$ is used to gauge a subgroup $G_0$ of the global symmetry group $SL(5)$, with ordinary derivatives replaced by $G_0$-covariant derivatives\footnote{Note that the $SO(5)\subset SL(5)$ appearing in the $SL(5)/SO(5)$ scalar coset is identified with the R-symmetry $SO(5)_R$. Hence, even fields that are singlets under $SL(5)$ can become charged under the gauge symmetry if they belong to non-trivial representations of $SO(5)_R$. This is the case, for instance, for the gravitini.}. Given a set of generators $t^M_{\,\,N}$ of $SL(5)$, the generators $X_{MN}$ of $G_0$ are conveniently identified using the so-called embedding tensor $\Theta_{MN,P}{}^Q$ as
\begin{align}\label{X_from_Theta}
    X_{MN}=\Theta_{MN,P}{}^Q\,t^P_{\,\,Q}\,.
\end{align}
The embedding tensor is subject to a set of linear constraints, which restrict its form to\footnote{The tensors $Y$ and $Z$ parametrize the $\mathbf{15}$ and $\mathbf{\overline{40}}$ representations of $SL(5)$, respectively. As such, they satisfy $Y_{MN}=Y_{(MN)}$ and $Z^{MN,P}=Z^{[MN],P}=Z^{[MN,P]}$.}
\begin{align}\label{Theta_from_YZ}
   \Theta_{MN,P}{}^Q=\delta^{Q}_{[M}Y_{N]P}-2\epsilon_{MNPRS}Z^{RS,Q}\,, 
\end{align}
as well as non-linear constraints which ensure the closure of the Lie algebra $\mathfrak{g}_0$ of $G_0$.

\item The gauged theory has a potential for the $SL(5)/SO(5)$ scalars $\Vfive_M^{\,\,\,N}$, which is proportional to the gauge coupling squared $g^2$ and is completely specified by the embedding tensor (or, equivalently, by $Y_{MN}$ and $Z^{MN,P}$).

\item A choice of gauging also fixes a certain distribution of degrees of freedom among the $p$-form gauge fields. The gauge fields involved in the gauging remain massless, while the others can be either rendered massive with a Higgs mechanism that involves some of the scalars, or they can in turn provide the longitudinal degrees of freedom that are necessary to make some of the two-forms massive. Another option that is realized in seven dimensions is that the massless two-forms $B_{\mu\nu}$ are dualized to massive self-dual three-forms $S_{\mu\nu\rho}$ -- see footnote \ref{selfdualthreeform} for more details.

More details on maximal 7d gauged supergravity can be found in \cite{Samtleben:2005bp}, including the general form of the action and supersymmetry transformations for arbitrary choice of gauging. 

\end{itemize}

\subsection{The relevant gaugings}

As we discussed in Section \ref{sec:6dnoncriticalworldsheet}, the embedding of the 6d noncritical string in 10d type II theories can be discussed in terms of distributions of NS5 branes, whose near-horizon geometry is the CHS background \eqref{CHS-b}. As it turns out, both type IIA and type IIB supergravity admit a consistent truncation on $S^3$, as first observed in \cite{Cvetic:2000ah,Samtleben:2005bp}. Such truncation leads to two maximal 7d gauged supergravities with $ISO(4)$ and $SO(4)$ gauging for the IIA and IIB reduction, respectively. As we shall discuss in greater detail in the rest of this section, both theories admit a linear dilaton solution that breaks half of the supersymmetry which uplifts to the CHS background, only distinguished by the 6d chirality of the respective Killing spinors. Moreover, we shall soon provide a detailed analysis of the spectrum of linearized fluctuations on the linear dilaton background for the two 7d supergravities, which turns out to match precisely our findings from the worldsheet perspective. Note that because of the T-duality that relates the two frames $\nc$ and $\ns$, our proposal is that the $ISO(4)$ theory obtained reducing 10d type IIA supergravity should provide the low energy description of type IIB noncritical string theory, while the $SO(4)$ theory arising from 10d type IIB should describe type IIA noncritical strings. Finally, the gauge groups of the two supergravities of interest have a common maximal compact subgroup $SO(4)$, which naturally can be thought of
\es{so4su2su2}{
SO(4)\simeq [SU(2)_+\otimes SU(2)_-]/\Z_2\,,
}
so the 7d fields can be arranged into irreps of the groups $SU(2)_{\pm}$, which can be identified with the $SU(2)$ symmetry of the $SU(2)_k$ WZW model appearing in the worldsheet description of 6d noncritical strings in the left and right sector.

In the next two paragraphs we describe the bosonic field content of the $ISO(4)$ and $SO(4)$ gauged theories. 

\paragraph{$\boldsymbol{ISO(4)}$ gauged supergravity: IIB noncritical strings.}

We find that the theory describing the massless fluctuations of type IIB noncritical superstrings in 6d is maximal 7d $ISO(4)$-gauged supergravity, which can be obtained from a consistent truncation of type IIA supergravity in ten dimension on $S^3$ -- for details see \cite{Cvetic:2000ah}, where uplift formulas for the bosonic sector are also given. The Lagrangian and supersymmetry transformation for this theory can be extrapolated from the results of \cite{Samtleben:2005bp} once the embedding tensor is fixed using \eqref{Theta_from_YZ} with
\es{YZ_ISO(4)}{
Y_{MN}=\text{diag}(1,1,1,1,0)\,,\quad Z^{MN,P}=0\,.
}
Note that the choice of gauging breaks $SL(5)$ covariance and it is therefore convenient to split $M,N,\ldots$ indices as 
\es{sl5toso4}{
M=(i,\,0)\,,\quad i=1,\ldots,4\,,
}
where now $i,j,\ldots =1,\ldots,4$ are fundamental indices of $SO(4)$, the compact part of the gauge group $ISO(4)$. The technical details on the Lagrangian and Killing spinor equations can be found in \cite{Samtleben:2005bp}, while here we describe some general features. We begin by discussing the effect of the gauging on the distribution of degrees of freedom, while also labelling the fields of the theory as NSNS or RR, according to their origin from the fields of type IIA supergravity. As we shall see, such labeling also corresponds to the one obtained by comparing the spectrum in the linear dilaton background to that of type IIB noncritical superstrings.

To describe the field content, it is convenient to break $SL(5)$ covariance to the $SO(4)$ subgroup which represents the maximal compact part of the two gaugings. Focusing on bosonic fields, we write
\es{fields_so5toso4}{
\Vfive \to (\Vfour,\,b_i,\,\phi_0)\,,\quad
A^{MN}_{\mu} \to (A^{[ij]}_{\mu},\,A^i_{\mu})\,,\quad
B_{M,\mu\nu}\to (B_{i,\mu\nu},\,B^0_{\mu\nu})\,,
}
where $\Vfour\in SL(4)/SO(4)$ is a new scalar coset representative, $b_i$ is a $\mathbf{4}$ of $SO(4)$ of scalars and $\phi_0$ is a singlet. For the one-forms we have used $\mathbf{10}\to \mathbf{6}\oplus \mathbf{4}$ and for the two-forms $\mathbf{5}\to \mathbf{4}\oplus\mathbf{1}$ under $SO(5)\to SO(4)$. The gauge fields $A^{[ij]}_{\mu}$ are NSNS and gauge the $SO(4)\subset ISO(4)$, while the $A^i_{\mu}$ are RR and gauge the $\R^4\subset ISO(4)$. The RR two-forms $B_{i,\mu\nu}$ are dualized to massive self-dual three-forms $S_{i,\mu\nu\rho}$, while $B^0_{\mu\nu}$ is NSNS and remains massless. Finally, the 9+1 scalars $(\Vfour,\phi_0)$ are NSNS, while the $b_i$ are RR and do not appear in the scalar potential. Moreover, they have a St\"uckelberg coupling to the gauge fields $A^i_{\mu}$, thus effectively making the latter massive, at least at the linearized level.

\paragraph{$\boldsymbol{SO(4)}$ gauged supergravity: IIA noncritical strings.}

We find that the supergravity that is relevant for type IIA noncritical superstring theory is maximal 7d $SO(4)$-gauged supergravity, which can be obtained from a consistent truncation of type IIB supergravity in ten dimension on $S^3$, as first observed in \cite{Samtleben:2005bp} -- see \cite{Malek:2015hma} for the uplift formulas. Again all details can be worked out directly from the general results of \cite{Samtleben:2005bp} when the embedding tensor is specified. The choice that leads to $SO(4)$ gauging is
\es{Zso4}{
Y_{MN}=0\,,\quad 
Z^{MN,P}=\delta^{5[M}\delta^{N]P}\,.
}
This suggests again a split of $SL(5)$ indices as in \eqref{sl5toso4}, so once again we relabel the fields as in \eqref{fields_so5toso4}. The theory with $SO(4)$ gauging is completely equivalent to $ISO(4)$ gauged supergravity in the NSNS sector, while the two differ in the RR sector, as one could have anticipated given the respective ten-dimensional origins of the two theories. The $b_i$ in this case have a massive scalar action, while the $A^i_{\mu}$ have a St\"uckelberg  to the two-forms $B_{i,\mu\nu}$ so that the two sets of fields can be viewed as four massive two-forms. 

The bosonic field content of the two theories is summarized in Table \ref{tab:NSNSsector} for the NSNS sector and Table \ref{tab:RRsector} for the RR sector: note that they reproduce the states obtained from the worldsheet analysis of Section \ref{sec:spectrum}, which are summarized in Tables \ref{tab:IIAspectrum} and \ref{tab:IIBspectrum} for type IIA ($SO(4)$ gauged supergravity) and IIB ($ISO(4)$ gauged supergravity) noncritical strings, respectively.
\begin{table}[!ht]
\begin{center}
\begin{tabular}{|c||c|c|c|}
\hline
NSNS fields (massless) & two-forms & one-forms & scalars \\
\hline\hline
$ISO(4)$ & 1 & 6 & 10  \\
\hline 
$SO(4)$ & 1 & 6 & 10  \\
\hline 
\end{tabular}
\caption{NSNS fields of the $ISO(4)$ and $SO(4)$ gauged supergravities. All fields have a massless action at the linearized level.}
\label{tab:NSNSsector}
\end{center}
\end{table}
\begin{table}[!ht]
\begin{center}
\begin{tabular}{|c||c|c|c|c|}
\hline
RR fields (massive) & three-forms & two-forms & one-forms & scalars \\
\hline\hline
$ISO(4)$ & 4 & 0 & 4 & 0 \\
\hline 
$SO(4)$ & 0 & 4 & 0 & 4 \\
\hline 
\end{tabular}
\caption{RR fields of the $ISO(4)$ and $SO(4)$ gauged supergravities. All fields have a massive action at the linearized level.}
\label{tab:RRsector}
\end{center}
\end{table}

\subsection{Spectrum in the linear dilaton background}\label{sec:LDspectrum}

As we just discussed, the NSNS sector is common to both theories. Here we consider the simplest truncation, where only the metric tensor and the 7d dilaton $\phi_0$ are non-trivial. The Lagrangian in this sector is 
\es{Llineardilaton}{
\mathcal{L}=R-20(\partial\phi_0)^2+\frac{g^2}{4}\ex^{4\phi_0}\,.
}
This is enough to describe the simplest solution of our 7d supergravities: the linear dilaton vacuum
\es{lineardilaton7d}{
\ds_7&=\ex^{-4\phi_0}\,\left[\ds(\R^{1,5})+\diff\rho^2\right]\,,\\
\phi_0&=-\frac{Q}{10}\rho\,,
}
which uplifts to the linear dilaton vacuum of type II supergravity (in the string frame)
\es{lineardilaton10d}{
\ds_{10}&=\ds(\R^{1,5})+\diff\rho^2+\frac{4}{Q^2}\ds(S^3)\,,\\
H_3&=\frac{8}{Q^2}\,\vol(S^3)\,,\\
\Phi&=-\frac{Q}{2}\rho\,,
}
describing the near-horizon region of a stack of $k$ NS5 branes, with
\es{Qtok}{
Q=\sqrt{\frac{2}{k}}\,,
}
as in Section \ref{sec:6dnoncriticalworldsheet}, and we have identified the gauge coupling with the number of NS5 branes as 
\es{Qtog}{
g^2=\frac{8}{k}\,.
}
The only difference between the solution in the $ISO(4)$ and $SO(4)$ gauging is in the structure of the 16 Killing spinors that the solution preserves, as we shall review in the next subsection in the context of more general solutions. Here we consider the linearized equations of motion in this background for the various fields appearing in the two 7d supergravities of interest, showing that they reproduce the spectrum of the associated noncritical superstring theories, which we have reviewed in Section \ref{sec:6dnoncriticalworldsheet}. To this end, it is useful to introduce a notion of 7d string frame, in which the worldsheet results have been obtained. We set
\es{strinframe7d}{
g_{\mu\nu}^S=\ex^{4\phi_0}g_{\mu\nu}\,,
}
where $g_{\mu\nu}$ is the 7d Einstein frame metric. Apart from this section, we will always work in the Einstein frame in 7d, so we only specify the superscript $S$ for the string frame but do not add a superscript $E$ since it is understood that we are working in the Einstein frame unless otherwise specified.

Now, given a generic field $\mathbb{F}$ in 7d supergravity we consider its linearized action in the linear dilaton background \eqref{lineardilaton7d} in the string frame, and write it in terms of a polarization tensor $\bar{\mathbb{F}}$ times a plane-wave dependence on the 7d coordinates ${x}^{{\mu}}=(x^{\alpha},\,\rho)$ which mimics that of the vertex operators on the worldsheet\footnote{Note that by an abuse of notation we denote with the same symbol $x$ both the 6d and the 7d coordinates. Since these appear mostly contracted with momenta, it should be clear that $k\cdot x$ refers to a 6d contraction (since we denote the 6d momentum with $k_\alpha$) while $p\cdot x$ refers to a 7d contraction (since we denote the 7d momentum with $p_\mu$)}. For example, for each NSNS field $\mathbb{F}$ we set
\es{fieldansatz}{
\mathbb{F}_{\text{NSNS}}=\bar{\mathbb{F}}_{\text{NSNS}}\,\ex^{\ii\,q\cdot x+\mathtt{j}\, \rho}\equiv \ex^{-\tfrac{Q}{2}\rho}\,\bar{\mathbb{F}}_{\text{NSNS}}\,\ex^{\ii\,{p}\cdot {x}}\,,
}
where as inSection \ref{sec:spectrumws} we set $\mathtt{j}=-\tfrac{Q}{2}+\ii s$, $s\in \R$, $k_{\mu}$ denotes the 6d momentum and we have introduced a 7d momentum $p_{{\mu}}=(q_{\alpha},s)$. The story for RR fields is slightly different, as we shall see. We will now consider the quadratic action for all (bosonic) fields $\mathbb{F}$, as dictated by 7d gauged supergravity, and show that the equations of motion that follow from it reproduce the massless spectrum of 6d noncritical superstrings. The story is slightly different for the NSNS and RR sector, so we consider the two separately.

\paragraph{NSNS sector.} Let us start from the NSNS sector, which contains scalars, vectors, a two-form and the metric. We observe that in the 7d string frame (which is defined by \eqref{strinframe7d}), the action for the NSNS fields takes the form
\es{LNSNS}{
\mathcal{L}_{NSNS}=\ex^{-10\phi_0}\,\mathcal{L}'_{NSNS}\,,
}
where $\mathcal{L}'_{NSNS}$ does not depend on the 7d dilaton $\phi_0$, except for the kinetic term of the dilaton itself. To study the spectrum in the background \eqref{lineardilaton7d}, it suffices to expand the Lagrangian to the quadratic level. Moreover, since via a standard construction one can build the graviton and Kalb-Ramond two-form states from the tensor product of two vectors, we shall consider for simplicity only the Lagrangian for NSNS scalars and vectors in the background \eqref{lineardilaton7d}. We begin by observing that any NSNS scalar field $\varphi$ has an action which, once expanded to quadratic order, reads
\es{}{
S_{\varphi}=\int \diff^7x\,\ex^{Q\rho}\partial^{{\mu}}\varphi\,\partial_{{\mu}}\varphi\,,
}
so that imposing the equations of motion 
\es{}{
\partial^{{\mu}}\left[\ex^{Q\rho}\partial_{{\mu}}\varphi\right]=0\,,
}
on an ansatz of the form \eqref{fieldansatz} with $\mathbb{F}\equiv \varphi$ gives the mass-shell condition
\es{massshellLagrangian}{
-p^2\equiv -q^2-s^2=\frac{Q^2}{4}\,,\quad
\text{or}\quad
-q^2+\mathtt{j}(\mathtt{j}+Q)=0\,,
}
which reproduces precisely the result \eqref{massshell_k7d}. The quadratic action for any of the NSNS massless vectors $A$, on the other hand, is
\es{S_A_NSNS}{
S_{A}=\int \diff^7 {x}\,\ex^{Q\rho}F^{ \mu \nu}F_{ \mu  \nu}\,,\quad F=\diff A\,,
}
which gives the equations of motion
\es{eomApre}{
\partial^{ \mu}\left[\ex^{Q\,\rho}F_{ \mu  \nu}\right]=0\,.
}
Using the ansatz \eqref{fieldansatz} with $\mathbb{F}\equiv A$, \eqref{eomApre} can be expressed as 
\es{eomA}{
\left[-p^2-\tfrac{Q^2}{4}\right]A_{ \mu}+(q,-\ii\,\mathtt{j})_{\mu}\,(q,-\ii(\mathtt{j}+Q))^{ \nu}A_{ \nu}=0\,.
}
To interpret this, we note that since $A$ is a massless vector its action \eqref{S_A_NSNS} is of course invariant under gauge transformations $A\to A+\diff \lambda$, which writing $\lambda$ as in \eqref{fieldansatz} with $\mathbb{F}\equiv \lambda$ can be expressed as the condition on polarizations
\es{gaugeinvNSNS}{
\bar{A}_{ \mu}\sim \bar{A}_{ \mu}+\ii\,\bar{\lambda}(q,-\ii\,\mathtt{j})_{ \mu}\,,
}
which is equivalent to \eqref{NSgaugeinv}. We can then use gauge transformations to fix the Lorenz gauge, which sets the divergence of $A$ to zero. The covariant divergence of $A$ in the linear dilaton background gives
\es{transvANSNS}{
\partial^{ \mu}\left[\ex^{Q\rho}A_{ \mu}\right]=0\,,\quad \Rightarrow\quad
(q,-\ii(\mathtt{j}+Q))^{ \nu}A_{ \nu}=0\,,
}
which sets to zero the second term in \eqref{eomA}, thus reproducing the same mass-shell condition as for the scalar. Moreover, we note that \eqref{transvANSNS} is equivalent to the tranversality condition \eqref{NStransverse} found from the worldsheet analysis.

The analysis for the other fields in the NSNS sector, the graviton and the Kalb-Ramond two-form, works in a similar way, with the usual transversality and gauge invariance conditions replaced by suitable generalizations of \eqref{gaugeinvNSNS} and \eqref{transvANSNS}. After unphysical states are projected out, all NSNS fields $\mathbb{F}$ have equations of motion that can be expressed as
\es{}{
\partial^{{\mu}}\left[\ex^{Q\rho}\partial_{{\mu}}\mathbb{F}\right]=0\,,
}
which appears massless but still gives rise to the massive-looking mass-shell condition \eqref{massshellLagrangian} due to the presence of the warping factor $\ex^{Q\rho}$ sitting in front of the NSNS Lagrangian in the string frame, see \eqref{LNSNS}.

\paragraph{RR sector.} For RR fields, we note two important differences. First, in the 7d string frame the RR Lagrangian is independent of the 7d dilaton $\phi_0$. Second, the quadratic Lagrangians for all RR fields in both 7d supergravities appear to be massive (after all St\"uckelberg mechanisms have been suitably taken into account). From the worldsheet perspective, we can see this as a consequence of the fact that in each open R sector we have massive-looking Dirac equations \eqref{R-7d-eom-S+}. Nonetheless, the mass-shell condition is still the same as the one for NSNS fields \eqref{massshellLagrangian}, since we are considering states at the same level.

In order to reproduce the equations of motion for RR states derived from the worldsheet, it turns out to be necessary to use a slightly modified version of \eqref{fieldansatz}, where the dependence on the $\rho$ direction only includes the momentum $s$ and not the overall warping $\ex^{-\tfrac{Q}{2}{\rho}}$. This is due to fact that for NSNS fields this term is necessary to compensate the dilaton-dependent warping in \eqref{LNSNS}, while the RR Lagrangian does not contain powers of the dilaton. Then, for RR fields we set
\es{}{
\mathbb{F}_{\text{RR}}=\bar{\mathbb{F}}_{\text{RR}}\ex^{\ii (q\cdot x +s\rho)}= \bar{\mathbb{F}}_{\text{RR}}\ex^{\ii \,p\cdot x}\,.
}

Let us begin from the RR sector of the $ISO(4)$-gauged theory, where the fields are four pairs ($A_{\mu},b$) of vectors and scalars with a St\"uckelberg coupling (so that effectively we have four massive vectors, at the quadratic level), as well as four self-dual massive three-forms $S_{\mu\nu\rho}$. In a gauge which sets the scalars $b=0$, we find the quadratic Lagrangian 
\es{LRR_7d_ISO4}{
\mathcal{L}^{ISO(4)}_{RR}\sim \star \left[\diff A\wedge\star \diff A+\frac{Q^2}{4}A\wedge\star A\right]+\star \left[S\wedge\star S+\frac{Q}{2}S\wedge\diff S\right]\,,
}
where the contractions and the Hodge dual are computed with respect to a flat 7d metric. Note that the mass terms in the original action of 7d supergravity appear with powers of the gauge coupling $g$, which here we have replaced using \eqref{Qtog}. The corresponding equations of motion
\es{eomRR_7d_ISO4}{
\star \diff \star \diff A=\frac{Q^2}{4}A=0\,,\quad  \diff S=\frac{Q}{2}\star S\,,
}
which are the same as \eqref{eomRRIIB_ws} upon identifying $A\leftrightarrow \mathcal{F}_{(1)}$ and $S\leftrightarrow \mathcal{F}_{(3)}$. From \eqref{eomRR_7d_ISO4} it also follows that the polarization tensors associated with $A$ and $S$ are tranvserse with respect to the 7d momentum \eqref{7d-momentum}, which is the same condition arising from the worldsheet analysis. 

The story is very similar for the $SO(4)$-gauged theory, where the RR fields are four pairs of two-forms and vectors $(B_{\mu\nu},A_{\mu})$ with St\"uckelberg couplings at the quadratic level, as well as four scalars $b$ with a massive quadratic action. Gauging away the vectors yields the Lagrangian 
\es{LRR_7d_SO4}{
\mathcal{L}^{SO(4)}_{RR}\sim  \star \left[\diff B\wedge\star \diff B+\frac{Q^2}{4}B\wedge\star B\right]+\star \left[\diff b \wedge\star \diff b+\frac{Q^2}{4}b\wedge\star b\right]\,,
}
which gives the equations of motion
\es{eomRR_7d_SO4}{
\star \diff \star \diff B=-\frac{Q^2}{4}B\,,\quad  \star \diff \star \diff b=-\frac{Q^2}{4}b\,.
}
These can be now identified with the physical state conditions \eqref{eomRRIIA_ws} found from the worldsheet in the RR sector of the noncritical IIA theory, upon identifying $b\leftrightarrow \mathcal{F}_{(0)}$ and $B \leftrightarrow \mathcal{F}_{(2)}$. Again, the mass-shell condition is the usual one \eqref{massshellLagrangian} and the polarizations are transverse with respect to the 7d momentum \eqref{7d-momentum}. 

To summarize, the difference between the NSNS and the RR sector is that in the former case we have massless quadratic Lagrangians, but with a warping factor (in the string frame) due to the dilaton $\phi_0$, which contributes to the mass term in the mass-shell condition \eqref{massshellLagrangian} in the linear dilaton background. On the other hand, in the RR sector there is no warping, but after taking into account the various St\"uckelberg mechanisms determined by the gauging, one obtains a massive Lagrangian which leads to the same condition \eqref{massshellLagrangian} as in the NSNS sector. 

We shall not discuss the case of fermions in detail.

\section{NSNS backgrounds}\label{sec:NSNS}

Let us now turn to study solutions of the 7d gauged supergravities introduced in the previous section where the only non-trivial fields are associated with states in the NSNS sector of the noncritical string. Since this sector is common to the two GSO projections, we do not need to distinguish between the $ISO(4)$ and the $SO(4)$ theory. While this section is mostly a review of domain-wall solutions already discussed in the supergravity literature, we present an interpretation of those results with our perspective on noncritical string theory in mind. Moreover, in subsection \ref{sec:8dsugra} we discuss the relation between the 7d guaged supergravities introduced in the previous section and an 8d effective action for the noncritical string that has been used by various authors \cite{Klebanov:2004ya,Kuperstein:2004yf,Bigazzi:2005md}, showing a connection between the 8d ``fake superpotential'' approach \cite{Townsend:1984iu,Skenderis:1999mm,Celi:2004st} and the genuine 7d BPS equations arising from supergravity.

\subsection{Distributions of NS5 branes from 7d}\label{sec:NS5distributions}

Here we consider solutions of 7d gauged supergravity that describe distributions of flat NS5 branes on 10d. This is analogous to the setup originally considered in \cite{Bakas:1999ax} for D3 branes from 5d gauged supergravity and in \cite{Bakas:1999fa} for M2 and M5 branes from 4d and 7d gauged supergravity respectively. The case of general sphere reductions was considered in \cite{Cvetic:2000zu}, while comments on the specific case of NS5 branes were presented in \cite{Bakas:2000nt}. In this subsection we essentially review the ideas discussed in those works, specializing them to the present setup. We consider the common NSNS sector of the two 7d supergravities discussed in this paper and further truncate to only the metric and the ten NSNS scalars. We can then use the $SO(4)$ symmetry to diagonalize the scalar coset representatives $\Vfive$ and $\Vfour$. It is convenient to write
\es{cosetrepNSNS}{
\mathcal{M}_{MN}\equiv \Vfive\,\Vfive^T=
\begin{pmatrix}
    \ex^{-2\phi_0}M_{ij} & 0\\
    0 & \ex^{8\phi_0}
\end{pmatrix}\,,
\quad
M_{ij}\equiv \Vfour\,\Vfour^T=\ex^{10\phi_0}\,\text{diag}(\ex^{2f_1},\,\ex^{2f_2},\,\ex^{2f_3},\,\ex^{2f_4})\,,
}
where the dilaton $\phi_0$ is given in terms of the four scalars $f_i$ as
\es{phi0fromfi}{
\phi_0=-\frac{1}{20}\sum_{i=1}^4f_i\,.
}
The action for the scalars $f_i$ is
\es{7dNSNSaction}{
S=\int \diff^7x\left[R-\frac{1}{4}\text{tr}\left(\mathcal{M}^{-1}\de\mathcal{M}\right)^2+\frac{g^2}{32}\ex^{4\phi_0}\,\left(\text{tr}(M^{-1})^2-2\text{tr}(M^{-2})\right)\right]
}
We are interested in solutions describing flat NS5 branes, so we make an ansatz
\es{ansatzNS5distribution}{
\ds_7=\ex^{-4g}\left[\ds(\R^{1,5})+\ex^{2h}\diff s^2\right]\,,
}
with $g$, $h$ and $f_i$ functions of $s$ only. We find a family of $\tfrac{1}{2}$-BPS solutions where the scalars are given by
\es{elli}{
\ex^{2f_i}=s^2+\ell^2_i\,,
}
and $\ell_i$ are four parameters whose interpretation we shall soon discuss. It is worth noting that 
\es{phi0NSNSall}{
\ex^{-40\phi_0}=\prod_{i=1}^4(s^2+\ell_i^2)\,,
}
which is the only combination of the scalars appearing in the final expression of the metric
\es{7dmetricNSNSall}{
\ds_7=\ex^{-4\phi_0}\ds(\R^{1,5})+\ex^{16\phi_0}\frac{16\,s^2}{g^2}\,\diff s^2\,.
}
Several comments are in order. First, we note that the 16 Killing spinors $\epsilon$ preserved by this solution satisfy the projection
\es{projectionDW}{
\gamma_s\,\epsilon=
\begin{cases}
\epsilon\,,\quad & ISO(4)\,,\\
\gammaflavor_5\,\epsilon\,,\quad & SO(4)\,,
\end{cases}
}
where we remind that $\epsilon$ is a $(\mathbf{8},\mathbf{4})$ of $Spin(1,6)\otimes USp(4)_R$ and $\gammaflavor_5$ is a $SO(5)\simeq USp(4)/\Z_2$ gamma matrix, as reviewed in Appendix \ref{app:so5usp4}. From the point of view of $\R^{1,5}$, the spacetime gamma matrix $\gamma_s$ plays the role of the chirality matrix, and of course $\gammaflavor_5$ has eigenvalues $\pm 1$. So we can interpret the projection \eqref{projectionDW} as the statement that the supercharges preserved by the solution all have the same chirality in the $ISO(4)$-gauged theory, while they come in pairs with opposite chirality in the $SO(4)$-gauged supergravity. This is of course in agreement that the former comes from the reduction of type IIA, while the latter arises from a consistent truncation of type IIB ten-dimensional supergravity.

Second, we note that in the limit $s\to +\infty$ (or equivalently setting all $\ell_i=0$), after the change of coordinates
\es{}{
s\to \ex^{\tfrac{g}{4}\rho}\,,
}
the metric \eqref{7dmetricNSNSall} reduces to that of the linear dilaton solution which we presented in \eqref{lineardilaton7d}, and moreover all scalars $f_i$ are identical in this limit and given by $f_i=-5\phi_0$, where $\phi_0$ is also the same as that appearing in \eqref{lineardilaton7d}, up to the identification \eqref{Qtog} between the gauge coupling $g$ and the parameter $Q$.

Finally, we can obtain an interpretation for this solution and for the parameters $\ell_i$ studying the uplifted version of the solution in ten dimensions. Using the uplift formulas of \cite{Cvetic:2000ah} one finds that the ten-dimensional solution in the string frame can be written as
\es{10dsolNSNS}{
\ds_{10}&=\ds(\R^{1,5})+\ex^{2\Phi}\left[\sum_{i=1}^4\left(\ex^{-2f_i}(\mu_i)^2\right)s^2\,\diff s^2+\sum_{i=1}^4\left(\ex^{2f_i}(\diff\mu_i)^2\right)\right]\,,\\
H_3&=-2\star \left[\diff\Phi\wedge\vol(\R^{1,5})\right]\,,\\
\ex^{2\Phi}&=\frac{16}{g^2}\frac{\prod_{i=1}^4\ex^{-f_i}}{\sum_{i=1}^4\left(\ex^{-2f_i}(\mu_i)^2\right)}\,,
}
where $\sum_i(\mu_i)^2=1$ hence the $\mu_i$ parametrize $S^3\subset \R^4$. The interpretation of this becomes clear after introducing four coordinates $z_i$ which parametrize $\R^4$ via
\es{}{
z_i=\ex^{f_i}\,\mu_i\,,
}
which allow to rewrite the solution as
\es{10dsolNS5branes}{
\ds_{10}&=\ds(\R^{1,5})+H\,\ds(\R^4)\,,\\
H_3&=-2\star \left[\diff\Phi\wedge\vol(\R^{1,5})\right]\,.\\
\ex^{2\Phi}&=H\,,
}
In eq.~\eqref{10dsolNS5branes}, $H$ is fixed by comparing the expression of the dilaton between \eqref{10dsolNSNS} and \eqref{10dsolNS5branes} and it satisfies $\Delta_4 H=0$, where $\Delta_4$ is the Laplacian in $\R^4$ parametrized by Cartesian coordinates $z_i$. This is simply the $\tfrac{1}{2}$-BPS solution of type II supergravity describing a distribution of NS5 branes extended along $\R^{1,5}$ and localized in $\R^4$, where their location is identified by the singularities of the function $H$, whose expression can be given as
\es{Hfromdensity}{
H=\frac{4}{Q^2}\int_{\R^4}\diff^4z' \frac{\rho(z')}{|z-z'|^4}\,. 
}
In eq.~\eqref{Hfromdensity}, $\rho$ is the normalized density of NS5 branes in $\R^4$ and we have ``dropped the 1'' in the expression of the harmonic function $H$ since 7d supergravity only captures the near-horizon region of solutions. Note that while every choice of $H$ that is harmonic in $\R^4$ is such that \eqref{10dsolNS5branes} solves the equations of motion and Killing spinor equations of type II supergravity, only functions $H$ (hence, distributions of branes) of the form fixed by the dilaton in \eqref{10dsolNSNS} can be obtained from the uplift of solutions in 7d supergravity. This constraint on the form of $H$ was understood in \cite{Bakas:1999ax,Bakas:1999fa} (see also \cite{Cvetic:1999xx,Cvetic:2000zu}) as arising from the requirement that the spectrum of fluctuations around these vacua should be computable equivalently from the 10d and from the 7d metric, as a consequence of the fact that the truncation is consistent. The relation between the parameters $\ell_i$ (see eq.~\eqref{elli}) determine the isometries of the distribution of NS5 branes in $\R^4$. In all cases we have a continuous distribution, and the cases which preserve a continuous group of symmetries were analyzed in \cite{MariosPetropoulos:2005rtq}, which we now briefly review. One of the $\ell_i$ can be set to zero with a shift of $s^2$ and a redefinition of the other parameters, so without loss of generality we can set $\ell_4=0$. Then, the minimal choice that preserves a continuous subgroup of $SO(4)$ is $\ell_3=0$, which in \cite{MariosPetropoulos:2005rtq} was proved to correspond to the normalized density
\es{}{
\rho(z)=\frac{1}{\pi\,\ell_1\,\ell_2}\delta \left(1-\frac{z_1^2}{\ell_1^2}-\frac{z_2^2}{\ell_2^2}\right)\delta(z_3)\delta(z_4)\,,
}
describing a continuous distribution of NS5 branes on an ellipsis with axes $\ell_1$ and $\ell_2$ lying in the $(z_1,z_2)$ plane. The symmetry of this distribution is $U(1)\otimes \Z_2$, corresponding to rotations in the transverse plane to the ellipsis and to swapping the two axes, respectively. It was argued in \cite{MariosPetropoulos:2005rtq} that this model should admit an exact worldsheet CFT description, and that it corresponds to a marginal deformation of the CHS background. Interesting observation on the dual worldsheet description can be found in \cite{Fotopoulos:2007rm,Prezas:2008ua,Fotopoulos:2010wc}, which have a holographic spirit similar to the one advocated in our work. Moreover, there is a number of limiting cases, which likewise are expected to (or actually already do) have a worldsheet description. We list them below as they play a role in this paper.
\begin{itemize}
    \item $\ell_1=\ell_2\equiv \ell$: in this limit the ellipsis degenerates into a circle of radius $\ell$, along which the branes are uniformly distributed. The preserved symmetry is $U(1)\otimes U(1)$. The worldsheet description of this supergravity background is the CFT \eqref{10d-ws-CFT}, given by a product of the Euclidean cigar and the parafermion disk CFT, as first observed in \cite{Sfetsos:1998xd}. Given its importance in this context, we discuss this case in detail in the next subsection.
    
    \item $\ell_1\equiv \ell,\ell_2\to 0$: this case corresponds to a denegerate limit in which one of the two axes of the ellipsis collapses and the branes are distributed on a segment of length $2\ell$ along the $z_1$ axis. The density of branes in this case is {\it not} uniform and is given by
    \es{}{
    \rho(z)=\frac{1}{\pi\,\ell}\left(1-\frac{z_1^2}{\ell^2}\right)^{-1/2}\delta\left(1-\frac{z_1^2}{\ell^2}\right)\delta(z_2)\delta(z_3)\delta(z_4)\,.
    }
    Introducing spherical coordinates for $S^3$ via 
    \es{}{
    \mu_1=\cos\psi\,,\quad \mu_2=\sin\psi\,\cos\theta\,,\quad
    \mu_3+\ii\,\mu_4=\sin\psi\,\sin\theta\,\ex^{\ii\alpha}\,, 
    }
    we can express the metric in the $\R^4$ transverse to the branes as
    \es{NS5bar}{
    \frac{Q^2}{4}\,H\,\ds(\R^4)=\sqrt{1+\frac{\ell^2}{s^2}}\,\left[\frac{\diff s^2}{s^2+\ell^2}+\diff\psi^2+\frac{s^2\,\sin^2\psi}{s^2+\ell^2\,\sin^2\psi}(\diff\theta^2+\sin^2\theta\,\diff\alpha^2)\right]\,.
    }
    \item $\ell_1\equiv \ell,\ell_2\to\infty$: this is a different degenerate limit in which the branes are uniformly distributed along two infinitely extended bars located at $z_1=\pm \ell$. The corresponding density is
    \es{}{
    \rho(z)=\lim_{L\to \infty}\Theta(L-|z_2|)\,[\delta(z_1-\ell)+\delta(z_1+\ell)]\,\delta(z_3)\delta(z_4)\,,
    }
    and it turns out that this solution is T-dual to the Eguchi-Hanson metric with two nuts at $z_1=\pm \ell$, as detailed in \cite{MariosPetropoulos:2005rtq}.

    \item $\ell_1=\ell_2=0$: this is the simplest case, in which the branes are localized at the origin of $\R^4$ and the coordinate transformation $r\to \ex^{g\rho/4}$ shows that the solution in this case reduces to the CHS background \eqref{CHS-b}.
\end{itemize}

\subsection{Branes on a circle, cigar and T-duality}

In this subsection we consider more in detail the case in which the NS5 branes are uniformly distributed on a circle, since it is particularly relevant for the worldsheet description of noncritical strings. The solution presented here was originally obtained in \cite{Sfetsos:1995ac} in the study of the interplay between supersymmetry and T-duality, which plays a crucial role in this work. In \cite{Sfetsos:1998xd} it was then obtained from a distribution of $k$ localized branes on a circle of radius $R$, in a limit of large $k$ and in the decoupling limit \cite{Giveon:1999px} -- see also \cite{Israel:2005fn}
\es{}{
g_s,\,R\to 0\,, \quad
\ell_s,\, \frac{R}{g_s} = \text{fixed}\,,
}
where $g_s$ is the string coupling and $\ell_s$ the string length. The result can be obtained directly from \eqref{10dsolNSNS} by setting $\ell_1=\ell_2\equiv \ell$ and $\ell_3=\ell_4=0$, changing coordinates using
\es{}{
s\to \ell\,\sinh\rho\,,
}
and choosing angular coordinates for $S^3$ via
\es{}{
\mu_1+\ii\,\mu_2=\sin\eta\,\ex^{\ii\xi_1}\,,\qquad 
\mu_3+\ii\,\mu_4=\cos\eta\,\ex^{\ii\xi_2}\,,
}
which allow to express the full 10d solution as 
\es{10dbranescircle}{
\ds_{10}&=\ds(\R^{1,5})+\frac{4}{Q^2}\,\left[\diff\rho^2+\diff\eta^2+\frac{\tan^2\eta\,\diff\xi_1^2+\tanh^2\rho\,\diff\xi_2^2}{1+\tan^2\eta\,\tanh^2\rho}\right]\,\\
B_2&=\frac{4}{Q^2}\frac{\diff\xi_1\wedge\diff\xi_2}{1+\tan^2\eta\,\tanh^2\rho}\,,\\
\ex^{2\Phi}&=\frac{8}{Q^2(\cos2\eta+\cosh2\rho)}\,.
}
Note that this solution, where the NS5 branes are uniformly distributed along a circle (parametrized by $\xi_1$), preserves $U(1)\times U(1)$ isometries in the $\R^4$ transverse to the worldvolume of the branes. One should think of this space as $\R^4=\C\oplus\C$, where the branes lie on a circle in the first copy of $\C$, while they are localized at the origin of the second. The Killing vectors $\partial_{\xi_1}$ and $\partial_{\xi_2}$ generate $U(1)$ rotations around the origin of the first and second copy, respectively. Note that they are both Killing vectors only in the limit of a uniform distribution of branes, while for a finite number $k$ of NS5 branes localized at points along a circle (which is the starting point of \cite{Sfetsos:1995ac,Sfetsos:1998xd}) the isometry generated by $\partial_{\xi_1}$ is broken to $\Z_k$. In terms of the $SO(4)$ compact part of the gauge group of the two supergravities that we consider, which as in \eqref{so4su2su2} we think of as $[SU(2)_+\otimes SU(2)_-]/\Z_2$, $\partial_{\xi_1}$ and $\partial_{\xi_2}$ correspond to the sum and difference of the Cartan generators of the two copies of $SU(2)$, not the Cartan generators themselves.

Using the standard rules of T-duality, it is straightforward to show that T-dualizing along $\de_{\xi_1}$ and changing coordinates with
\es{}{
\xi_1\to \frac{4}{Q^2}\psi_1\,,\quad
\xi_2\to \psi_1+\psi_2\,,
}
gives rise to
\es{cigardisk1}{
\ds_{10}&=\ds(\R^{1,5})+\frac{4}{Q^2}\,\left[\diff\rho^2+\tanh^2\rho\,\diff\psi_1^2+\diff\eta^2+\cot^2\eta\,\diff\psi_2^2\right]\,\\
B_2&=0\,,\\
\ex^{2\Phi}&=\frac{1}{\sin^2\eta\,\cosh^2\rho}\,,
}
while a T-duality along $\de_{\xi_2}$ together with the change of coordinates
\es{}{
\xi_2\to \frac{4}{Q^2}\psi_2\,,\quad
\xi_1\to \psi_1-\psi_2\,,
}
leads to
\es{cigardisk2}{
\ds_{10}&=\ds(\R^{1,5})+\frac{4}{Q^2}\,\left[\diff\rho^2+\coth^2\rho\,\diff\psi_1^2+\diff\eta^2+\tan^2\eta\,\diff\psi_2^2\right]\,\\
B_2&=0\,,\\
\ex^{2\Phi}&=\frac{1}{\cos^2\eta\,\sinh^2\rho}\,.
}
Both \eqref{cigardisk1} and \eqref{cigardisk2} are equivalent target space descriptions of the worldsheet CFT
\es{cigardisk_sugra}{
\R^{1,5}\times \frac{SL(2)}{U(1)}\times \frac{SU(2)}{U(1)}\,,
}
where the ``cigar'' $\frac{SL(2)}{U(1)}$ is parametrized by coordiantes $(\rho,\psi_1)$ and the ``disk'' $\frac{SU(2)}{U(1)}$ is spanned by $(\eta,\psi_2)$ -- see also \cite{Israel:2005fn} for a description of the geometry \eqref{10dbranescircle} from the point of view of the cigar CFT and a study of its boundary states. The two descriptions are related by further T-dualities along $\de_{\psi_1}$ and $\de_{\psi_2}$, which leave the structure of the solution unaltered except for replacing $(\cosh\rho,\sinh\rho)\to (\sinh\rho,\cosh\rho)$ and $(\cos\eta,\sin\eta)\to (\sin\eta,\cos\eta)$, respectively. Clearly, in the limit $\rho\to+\infty$ we recover the linear dilaton CFT
\es{}{
\R^{1,5}\times \R_\rho\times U(1)_{\psi_1}\times \frac{SU(2)}{U(1)}\,.
}
We note here that while the application of a T-duality to the geometry \eqref{10dbranescircle} does indeed reproduce the target space geometry \eqref{cigardisk1} of the cigar CFT \eqref{10d-ws-CFT}, this only happens in the limit $k\to \infty$ where the $k$ NS5 branes are uniformly distributed along a circle. On the other hand, the worldsheet CFT \eqref{10d-ws-CFT} is well-defined for any finite integer value of $k \geq 2$ and in particular as evident from \eqref{c_SU2U1} (see also the comments below \eqref{c_M}) the disk CFT decouples for $k=2$ since its central charge vanishes, thus defining an 8d target space geometry.\footnote{See Section 3 of \cite{Martinec:2017ztd} for some interesting comments on why the smearing is necessary from the supergravity perspective.} This fact emphasizes one of the shortcomings of our attempt to provide a supergravity description for the interaction between the massless modes of the noncritical string: the latter is defined for $k=2$, but the supergravity provides a good description only in the limit of large $k$.

An interesting feature of the relationship between the distribution of branes on a circle and the background \eqref{cigardisk_sugra} is that the specific direction that one has to choose to perform T-duality completely breaks supersymmetry, as already observed in \cite{Sfetsos:1998xd}. This justifies the statements made at the beginning of this section and explains why we are forced to work in 7d to have a manifestly supersymmetric description. It is also tempting to generalize the lesson learned from the distribution of NS5 branes on a circle to the more general distributions discussed earlier in this section, since they can all be thought as arising from a similar decoupling limit to that described in \cite{Sfetsos:1998xd}. This is particularly interesting for the distribution of branes on a segment considered in \eqref{NS5bar}, since that model preserves an $SU(2)$ symmetry geometrically, which corresponds to the diagonal $SU(2)_D$ subgroup of $SU(2)_+\otimes SU(2)_-$, see Appendix \ref{app:so4}. When considering configurations with D-branes in Section \ref{sec:RR}, this $SU(2)$ corresponds to (part of) the R-symmetry of the dual field theory living on the worldvolume of the branes, so having a model which explicitly preserves it geometrically\footnote{In the cigar model the $SU(2)$ would only be preserved for $k=2$, so it arises as a stringy effect rather than a geometric one.} is particularly convenient. The $SU(2)$ model \eqref{NS5bar} is similar to the distribution of branes on a circle in the sense that applying a T-duality along the Cartan of $SU(2)$ to \eqref{NS5bar} one goes to a frame with no Kalb-Ramond field and completely broken supersymmetry (at least from the supergravity perspective). However, while this procedure applied to \eqref{10dbranescircle} gives rise to \eqref{cigardisk1}, which is the target space of a well-known and studied CFT, the result of the same T-duality applied to \eqref{NS5bar} naively gives rise to a complicated metric, which we were not able to identify with the target space of any known 2d CFT. It would be very interesting to further investigate this model and gain a better understanding of the worldsheet CFT in the frame with no Kalb-Ramond field as described above.

We end by considering a version of \eqref{7dNSNSaction} which is tailored to the problem of obtaining the solution \eqref{10dbranescircle} upon uplift to 10d, since this will be useful for the next subsection. If in \eqref{7dNSNSaction} we set
\es{}{
f_1=f_2\equiv -5\phi_0-\phi_1\,,\quad
f_3=f_4\equiv -5\phi_0+\phi_1\,,
}
we obtain the action
\es{U(1)^2action}{
S=\int \diff^7 x\,\sqrt{-g}\,\left[R-20(\de\phi_0)^2-4(\de\phi_1)^2+\frac{g^2}{4}\ex^{4\phi_0}\right]\,,
}
which has a supersymmetric solution of the form
\es{7dcigarsol}{
\ds_7&=\ex^{-4\phi_0}\left[\ds(\R^{1,5})+\frac{16}{g^2}\diff\rho^2\right]\,,\\
\ex^{-10\phi_0}&=\frac{\ell^2}{2}\sinh(2\rho)\,,\quad\ex^{2\phi_1}=\tanh\rho\,,
}
whose uplift is precisely \eqref{10dbranescircle}.

\subsection{The cigar from an 8d effective action}\label{sec:8dsugra}

Previous approaches to the problem of finding a supergravity description of noncritical string theories are mostly based on the so-called {\it fake superpotential} approach \cite{Townsend:1984iu,Skenderis:1999mm,Celi:2004st}. After formulating a guess for a putative effective action, where the supersymmetry of the string model is not manifest, in some cases one is able to derive the corresponding (second order) equations of motion from some first order equations. The latter involve an object -- the {\it fake superpotential} -- which mimics the behavior of the superpotential of a supersymmetric theory, hence the name. This is the approach adopted, for instance, in \cite{Klebanov:2004ya,Kuperstein:2004yf,Bigazzi:2005md}. The shortcoming of this method is that it is not systematic and it is not based on an actual understanding of the underlying supersymmetry (if any). Moreover, the effective action is simply guessed, rather than derived from a microscopic description as in our work. Despite these shortcomings, some progress has nonetheless been made using this approach and here we would like to show that, at least for certain NSNS solutions, what makes it possible to obtain a fake superpotential at all is the supersymmetry of the theory after dimensional reduction.

\begin{figure}[!ht]
\begin{center}
\begin{tikzpicture}
    \node[draw, rectangle] (node1) at (0,0) {10d type II A/B};
    \node[]            (node1R) at (2,0) {};
    \node[]            (node1D) at (0,-0.5) {};
    \node[]                     at (-1,-1.5) {$S^3$};
    \node[]                     at (-1,-2) {reduction};
    \node[draw, rectangle] (node2) at (10,0) {10d type II B/A};
    \node[]            (node2L) at (8,0) {};
    \node[]            (node2D) at (10,-0.5) {};
    \node[]                     at (11,-0.9) {Disk};
    \node[]                     at (11,-1.4) {reduction};
    \node[draw, rectangle] (node3) at (0,-4) {7d $ISO(4)/SO(4)$ sugra};
    \node[]            (node3U) at (0,-3.5) {};
    \node[]            (node3R) at (2.5,-4) {};
    \node[draw, rectangle] (node4) at (10,-2.5) {8d effective action};
    \node[]            (node4U) at (10,-2) {};
    \node[]            (node4L) at (8,-2.5) {};
    \node[]                     at (6,-3.4) {$S^1$};
    \node[]                     at (6,-3.9) {reduction};
    
    \draw[<->] (node1R) -- (node2L) node[midway, above] {T-duality};
    \draw[->] (node1D) -- node[above, pos=0.6] {} (node3U);
    \draw[->] (node4L) -- node[above, pos=0.55] {}(node3R);
    \draw[->] (node2D) -- node[above, pos=0.65] {} (node4U);
\end{tikzpicture}
\end{center}
\caption{Chain of reductions and T-dualities that relate type IIA/B supergravity in 10d to 7d $ISO(4)/SO(4)$ gauged supergravity and to the 8d effective action \eqref{8dactionNSNSstring}, at least in the subsector of interest.}
\label{fig:108d7d}
\end{figure}
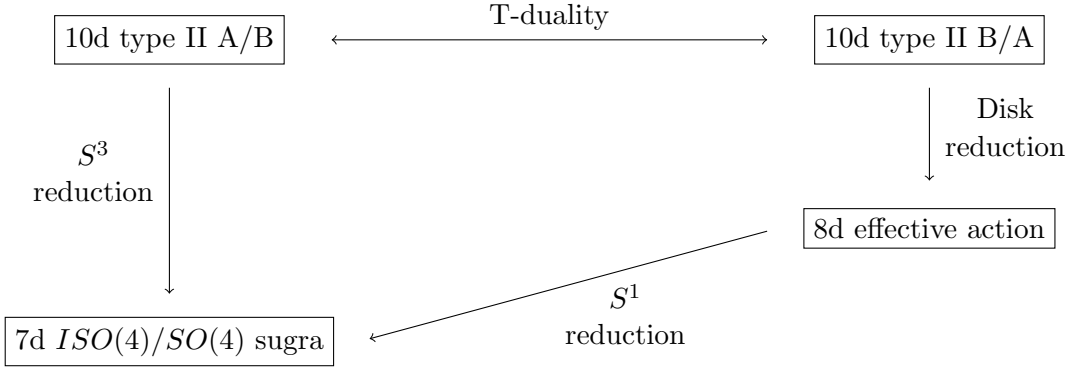

To be concrete, consider the 8d effective action
\es{8dactionNSNSstring}{
S_{\text{8d}}=\int \diff^8\hat{x}\,\sqrt{-\hat{g}^{(S)}}\,\ex^{-2\hat{\varphi}}\left[\hat{R}^{(S)}+4(\partial\hat{\varphi})^2+Q^2\right]\,,
}
where the superscript ``$S$'' is meant as a remark of the fact that we are in the string frame, while the hat notation is used for 8d quantities (which will be soon reduced to 7d ones, without hats). We can move to the Einstein frame with the Weyl rescaling (we omit the ``$E$'' superscript)
\es{}{
\hat{g}^{(S)}_{\hat \mu \hat \nu}=\ex^{\tfrac{2}{3}\hat{\varphi}}\hat{g}_{\hat \mu\hat\nu}\,,
}
leading to
\es{}{
S_{\text{8d}}=\int \diff^8\hat{x}\,\sqrt{-\hat{g}^{(S)}}\,\left[\hat{R}-\frac{2}{3}(\partial\hat{\varphi})^2+Q^2\,\ex^{\tfrac{2}{3}\hat{\varphi}}\right]\,.
}
Any solution of this model which admits a Killing vector $\de_\psi$ can be reduced to 7d by means of the standard formulas for Kaluza-Klein circle reduction (see, {\it e.g.}, Section 3.1 of \cite{deWit:2003gqa})
\es{uplift7d8d}{
\ds_8&=\ex^{2\alpha\phi}\,\ds_7+\ex^{2\beta\phi}\,(\diff\psi+A)^2\,,\\
\hat{\varphi}&=\varphi\,,
}
where the fields without hats ($g_{\mu\nu}$, $A_{\mu}$, $\varphi$ and $\phi$) are now 7d fields, and the choice
\es{}{
\beta=-5\alpha\,,\quad \alpha^2=\frac{1}{60}\,,
}
gives an Einstein frame action in 7d with canonical normalization for $\phi$. The further change of variables
\es{}{
\varphi=5\phi_0-\phi_1\,,\quad
\phi=\frac{1}{3\alpha}(\phi_0+\phi_1)\,,
}
leads to the action
\es{7dfrom8d}{
S=\int \diff^7 x \left[R-20(\partial\phi_0)^2-4(\partial\phi_1)^2-\frac{1}{4}\ex^{-4(\phi_0+\phi_1)}F^2+Q^2\,\ex^{4\phi_0}\right]\,,
}
where $F=\diff A$, which matches precisely \eqref{U(1)^2action} after identifying $g=2Q$ using \eqref{Qtog} and setting the gauge field $A$ to zero. Hence, even without an understanding of the putative Killing spinor equations underlying the 8d action proposed in \eqref{8dactionNSNSstring}, since its dimensional reduction gives rise to a subsector of the 7d gauged supergravities that we consider in this section we have a very precise notion of supersymmetry. One simply has to solve the Killing spinor equations in 7d and then uplift back in 8d, where the uplift of \eqref{7dcigarsol} with the formulas \eqref{uplift7d8d}, once translated back to the string frame, is (recall that $A=0$ for this solution)
\es{}{
\ds_8&=\ds(\R^{1,5})+\frac{4}{Q^2}\left[\diff\rho^2+\coth^2\rho\,\diff\psi_1^2\right]\,,\\
\ex^{2\hat{\varphi}}&=\frac{1}{\ell^2\,\sinh^2\rho}\,,
}
which is the cigar solution in 8d and we have redefined $\psi\to\tfrac{2}{Q}\psi_1$. This same solution can also be obtained from the reduction of \eqref{cigardisk2} on the disk $SU(2)/U(1)$. The situation is exemplified in Figure \ref{fig:108d7d}.

The same chain of dualities and reduction holds, of course, for the linear dilaton solution, since this arises as the $\rho\to \infty$ limit of the cigar. We can illustrate the above mechanism in another example, namely the solution found in \cite{Bigazzi:2001aj,Gauntlett:2001ps} describing NS5 branes wrapped on $S^2$ inside a K3 surface: this is the top-left corner of Figure \ref{fig:108d7d}. Such solution was originally obtained from the uplift of a solution to the 7d model \eqref{7dfrom8d}, which in turn is a subsector of both relevant gaugings of 7d supergravity. The 7d solution and its uplift are of the form
\es{wrappedNS5}{
\R^{1,3}\times \R_{\rho}\times S^2\qquad\longrightarrow \qquad
\R^{1,3}\times S^2\times \R^2\times \tilde{\R}^2\,,
}
where we write the 10d solution as in eq.~(2.26) of \cite{DiVecchia:2002ks}, emphasizing the fact that $\tilde{\R}^2$ is fibered over $S^2$ and together they give the backreacted metric of the K3 surface. It was already observed in \cite{Hori:2002cd} that a suitable T-duality of the 10d solution in \eqref{wrappedNS5} leads to the background
\es{kkl410d}{
\R^{1,3}\times \text{KKL}_4\times \frac{SU(2)}{U(1)}\,,
}
where $\text{KKL}_4$ is a representative of the family of exact string theory backgrounds $\text{KKL}_{2m}$ found in \cite{Kiritsis:1993pb}, which admit a worldsheet description in terms of $\mathcal{N}=2$ Landau-Ginzburg models. The case $m=1$ corresponds to the parafermion disk CFT: $\text{KKL}_{2}\equiv SU(2)/U(1)$. The target space of the $\text{KKL}_4$ CFT is of the form
\es{kkl4}{
\text{KKL}_4=\R_\rho\times \tilde{S}^3=\R_\rho\times S^2\tilde{\times}U(1)\,,
}
where $\tilde{S}^3$ is known as the Berger sphere: a certain squashing of $S^3$ preserving $SU(2)\times U(1)$ isometries (for a review see, {\it e.g.}, \cite{Zoubos:2004qm}), and $\tilde{\times}$ represents a fibration of the $U(1)$ over $S^2$. This background sits on the top-right corner of Figure \ref{fig:108d7d} and the only non-trivial fields are metric and dilaton, but no Kalb-Ramond field. Reducing on the disk $SU(2)/U(1)$ is trivial since the solution is factorized, and it gives rise to a solution of the 8d model \eqref{8dactionNSNSstring}, again with only metric and dilaton (the bottom-right corner of Figure \ref{fig:108d7d}). Further reducing along the $U(1)$ appearing in \eqref{kkl4} using the formulas \eqref{uplift7d8d} gives back the 7d solution found in \cite{Bigazzi:2001aj,Gauntlett:2001ps}, sitting in the bottom-left corner of the diagram in Figure \ref{fig:108d7d}. Note that the fibration in \eqref{kkl4} is responsible for a non-zero gauge field in 7d, which in turn allows the NS5 branes to wrap $S^2$ in a supersymmetric way, via a partial topological twist.

It would be interesting to extend this relation to other fields or solutions of the 7d supergravity, and in particular to the RR sector.

\section{RR backgrounds}\label{sec:RR}

We now proceed to consider solutions of the gauged supergravities discussed in the previous section which include RR fields. This section contains almost exclusively original results, although unfortunately our conclusions are less clear-cut than in the rest of the paper.

Our long-term goal is to understand how supergravity can be used to obtain a worldsheet formulation of 4d gauge theories such as pure $\mathcal{N}=2$ SYM or $\mathcal{N}=2$ SQCD, using noncritical string theory. While this is certainly an ambitious program, steps in this direction (or for the similar case of 4d $\mathcal{N}=1$ gauge theories) have been made by various authors, both from a worldsheet perspective \cite{Aharony:1998ub,Fotopoulos:2005cn,Ashok:2005py,Israel:2005fn,Murthy:2006xt,Benichou:2008gb} and using a supergravity approach \cite{Polyakov:1998ju,Polyakov:2000fk,Kuperstein:2004yk,Klebanov:2004ya,Alishahiha:2004yv,Kuperstein:2004yf,Bigazzi:2005md,Casero:2005se,Bertoldi:2007sf,Nunez:2010sf}. In this paper we were able to obtain a supergravity formulation of the noncritical superstring in which curvature corrections are actually suppressed for large $k$ (as opposed to previous approaches where they were simply ignored, although non-negligible), while also having a formal derivation of first-order BPS equations based on the underlying supersymmetry of the theory. 

It is also worth pointing out that the very same 4d $\mathcal{N}=2$ gauge theories that we wish to describe with the noncritical superstring have been investigated from a supergravity perspective in various papers, where they are realized with different brane setups from the one considered here. For instance in \cite{Gauntlett:2001ps,Bigazzi:2001aj,DiVecchia:2002ks,DiVecchia:2002gw} 4d $\mathcal{N}=2$ SYM is realized from D5 branes on a two-sphere\footnote{See also \cite{Boisvert:2024jrl} for the case of D5 wrapped on a spindle and a recent discussion of wrapped fivebrane solutions.} (flavor was also included in \cite{Paredes:2006wb}), while an approach with fractional D3 branes is put forward in \cite{DiVecchia:2002gw,Bertolini:2000dk,Petrini:2001fk} (with flavor included in \cite{Bertolini:2001qa}). See also \cite{Maldacena:2000yy,Apreda:2001qb,Bertolini:2002yr,Marotta:2002gc,Nunez:2003cf,Barranco:2011vt,Burrington:2004id,Casero:2006pt,Casero:2007pz,Casero:2007jj,Hoyos-Badajoz:2008znk} for the case of $\mathcal{N}=1$ supersymmetry as well as \cite{Maldacena:2001pb,DiVecchia:2001uc,Edelstein:2002xx,Ramallo:2008ew,Canoura:2008at} for analogous investigations in the case of 3d gauge theories. While these are interesting ways of addressing this problem, we stress that they all focus on the Calabi-Yau frame $\cy$ discussed in Section \ref{sec:spectrum}, which is related to noncritical strings on the cigar backgrounds by a series of two T-dualities. T-duality is of course a symmetry of the full string theory, but some information is lost when only focusing on the supergravity approximation, which makes it non-trivial to explicitly relate these approaches to the solutions described in this paper. Understanding such relation remains however an important problem for the future, since it might also help clarify some of the shortcomings of the solutions discussed in this section.

\subsection{Gauge theories from noncritical strings}

Let us begin by discussing the brane setup that one would hope to describe. As we shall soon discuss, the actual solutions that we present here are not expected to reproduce this precise setup, but it is still interesting to discuss it and keep it in mind as a reference. Let us consider the case of 4d $\mathcal{N}=2$ gauge theories first, from which the 3d case can be obtained in a trivial way. Our initial ambition was to obtain a supergravity description for theories defined by $k=2$ NS5 branes plus D-branes, such as pure $\mathcal{N}=2$ SYM or $\mathcal{N}=2$ SQCD. The distribution of branes corresponding to these theories is represented in Table \ref{tab:ciagarbranes4d}, where pure SYM corresponds to $N_f=0$ while SCQCD is the case $N_f=2N_c$.
\begin{table}[!ht]
\begin{center}
\begin{tabular}{|c||c|c|c|c|c|c|c|c||c|c|}
\hline
 & \multicolumn{4}{c|}{$\R^{1,3}_x$} & \multicolumn{2}{c|}{$\R^{2}_y$} & \multicolumn{2}{c||}{$SL(2)/U(1)$} & \multicolumn{2}{c|}{$SU(2)/U(1)$}\\
\hline
$N_c$ D3 & $\times$ & $\times$ & $\times$ & $\times$ & \hspace{0.3cm} &  \hspace{0.3cm} & \hspace{0.75cm} & \hspace{0.75cm} & $\times$ & $\times$\\
\hline
$N_f$ D5 & $\times$ & $\times$ & $\times$ & $\times$ & \hspace{0.3cm} & \hspace{0.3cm}& $\times$ & $\times$& \hspace{0.75cm} & \hspace{0.75cm} \\
\hline 
\end{tabular}
\caption{Distribution of D-branes in type IIB noncritical string theory describing a 4d $\mathcal{N}=2$ $SU(N_c)$ gauge theory with $N_f$ hypermultiplets.}
\label{tab:ciagarbranes4d}
\end{center}
\end{table}
Note that we have included the $SU(2)/U(1)$ disk CFT in the picture, but recall that this is only present for general $k$, while its central charge vanishes for the intereseting case of $k=2$, which is why the color branes are referred to as D3-branes: one should have in mind an eight-dimensional spacetime. These are precisely the boundary states described in Section \ref{sec:boundarystates}: the D3-branes are D0-branes of $SL(2)/U(1)$, localized at the tip of the cigar, tensored with a suitable boundary state of the Minkowski part (and of the disk, for $k>2$), while the D5-branes are D2-branes of $SL(2)/U(1)$, extended along the cigar, tensored with the same boundary state in $\R^{1,3}_x\times \R^{2}_y$. From the perspective of the 4d field theory living on the common worldvolume directions ($\R^{1,3}_x$), $N_c$ is the number of colors while $N_f$ is the number of flavors. 

As described in the previous section, a supergravity description for noncritical string theories is only possible in the T-dual frame where the cigar is replaced by a distribution of NS5 branes of type IIA. Similarly, the same T-duality replaces the D3 with D4 branes and the D5 with D6 branes, thus realizing a Hanany-Witten (HW) setup in type IIA \cite{Hanany:1996ie} -- see \cite{Karch:1998yv} for a discussion of the equivalence between the two frames and \cite{Elitzur:1997hc,Giveon:1998sr,Elitzur:2000pq} for more details on the relation between configurations of branes of this type and the dual gauge theories. The corresponding configuration of branes is described in Table \ref{tab:IIAbranes4d}.
\begin{table}[!ht]
\begin{center}
\begin{tabular}{|c||c|c|c|c|c|c|c|c|c|c|}
\hline
 & \multicolumn{4}{c|}{$\R^{1,3}_x$} & \multicolumn{2}{c|}{$\R^{2}_y$} & $\R_u$ & \multicolumn{3}{c|}{$\R^3_v$}\\
\hline
$k$ NS5 & $\times$ & $\times$ & $\times$ & $\times$ & $\times$ &  $\times$ & \hspace{0.3cm} & \hspace{0.3cm} & \hspace{0.3cm} & \hspace{0.3cm}\\
\hline
$N_c$ D4 & $\times$ & $\times$ & $\times$ & $\times$ & \hspace{0.3cm} &  \hspace{0.3cm} & $\times$ & \hspace{0.3cm} & \hspace{0.3cm} & \hspace{0.3cm}\\
\hline
$N_f$ D6 & $\times$ & $\times$ & $\times$ & $\times$ & \hspace{0.3cm} & \hspace{0.3cm}& \hspace{0.3cm} & $\times$& $\times$ & $\times$ \\
\hline 
\end{tabular}
\caption{HW setup in type IIA critical string theory that is T-dual to the configuration of Table \ref{tab:ciagarbranes4d}.}
\label{tab:IIAbranes4d}
\end{center}
\end{table}
For $k=2$ this should be interpreted as follows: there are two NS5 branes localized in $\R^4=\R_u\times \R^3_v$, where $u$ is the direction of a line connecting the two. The D4 branes are suspended between the two NS5 branes, with a worldvolume of finite extent along the $u$-direction, while the D6 branes are localized along $u$ and extended along the transverse $\R^3_v$. Due to the HW effect, one could trade the D6 branes for semi-finite D4 branes which extend along $u$ between the NS5 branes and infinity. From the point of view of the dual 4d gauge theory, $\R^2_y$ represents the Coulomb branch. If written in polar coordinates,
\es{}{
\ds(\R^2_y)=\diff y^2+y^2\,\diff\alpha^2\,,
}
solutions where $\partial_\alpha$ is an isometry realize the $U(1)_r$ symmetry of a 4d $\mathcal{N}=2$ QFT geometrically. As well known, for $\mathcal{N}=2$ asymptotically free $SU(N_c)$ gauge theory with $N_f$ hypermultiplets ($N_f < 2N_c$), the chiral anomaly breaks the $U(1)_r$ to a $\mathbb{Z}_{2(2N_c-N_f)}$ subgroup~\cite{Seiberg:1994rs}, which moreover is completely spontaneously broken at a generic point on the Coulomb branch. It is an interesting question how this picture should be reproduced in supergravity. So far as the spontaneous breaking is concerned, as discussed the $\R^2_y$ plane is identified with the Coulomb branch and at this level we can choose whether to consider solutions for which $\partial_\alpha$ is a Killing vector (that is, theories at the origin at the Coulomb branch) or not. A different question is how the breaking due to the anomaly can be captured by the supergravity description. In a related context, the gravity counterpart of the chiral anomaly was identified in solutions where the geometry does possess a $U(1)$ isometry, which however is broken by the RR fields to a discrete subgroup \cite{DiVecchia:2002ks, Gauntlett:2001ps}. While we shall not be able to explicitly demonstrate that the same happens here, in our solutions we will also find RR fields extended along the $\R^2_y$ plane, so it is plausible that an analogous mechanism can be invoked here.

Aside from this comment, we also note that writing
\es{R3v_S2R}{
\ds(\R^3_v)=\diff v^2+v^2\,\ds(S^2_R)\,,
}
solutions that preserve the isometries of $S^2_R$ realize the $SU(2)_R$ symmetry geometrically, and here we shall work under this assumption. 

A completely analogous story (except for the chiral anomaly, of course) holds for D branes of type IIA noncritical string theory, which are T-dual to a NS5-D3-D5 HW setup of type IIB critical string theory describing 3d $\mathcal{N}=4$ gauge theories. This can be obtained from a trivial T-duality of the configurations described above, performed along one of the space-like directions in $\R^{1,3}_x$. The resulting setup in type IIA noncritical string theory is described in Table \ref{tab:ciagarbranes3d}, while its T-dual description as a HW setup in critical type IIB is represented in Table \ref{tab:IIBbranes3d}.
\begin{table}[!ht]
\begin{center}
\begin{tabular}{|c||c|c|c|c|c|c|c|c||c|c|}
\hline
 & \multicolumn{3}{c|}{$\R^{1,2}_x$} & \multicolumn{3}{c|}{$\R^{3}_y$} & \multicolumn{2}{c||}{$SL(2)/U(1)$} & \multicolumn{2}{c|}{$SU(2)/U(1)$}\\
\hline
$N_c$ D3 & $\times$ & $\times$ & $\times$ & \hspace{0.3cm} & \hspace{0.3cm} &  \hspace{0.3cm} & \hspace{0.75cm} & \hspace{0.75cm} & $\times$ & $\times$\\
\hline
$N_f$ D5 & $\times$ & $\times$ & $\times$ & \hspace{0.3cm} & \hspace{0.3cm} & \hspace{0.3cm}& $\times$ & $\times$& \hspace{0.75cm} & \hspace{0.75cm} \\
\hline 
\end{tabular}
\caption{Distribution of D-branes in type IIA noncritical string theory describing a 3d $\mathcal{N}=4$ $SU(N_c)$ gauge theory with $N_f$ hypermultiplets.}
\label{tab:ciagarbranes3d}
\end{center}
\end{table}
\begin{table}[!ht]
\begin{center}
\begin{tabular}{|c||c|c|c|c|c|c|c|c|c|c|}
\hline
 & \multicolumn{3}{c|}{$\R^{1,2}_x$} & \multicolumn{3}{c|}{$\R^{3}_y$} & $\R_u$ & \multicolumn{3}{c|}{$\R^3_v$}\\
\hline
$k$ NS5 & $\times$ & $\times$ & $\times$ & $\times$ & $\times$ &  $\times$ & \hspace{0.3cm} & \hspace{0.3cm} & \hspace{0.3cm} & \hspace{0.3cm}\\
\hline
$N_c$ D3 & $\times$ & $\times$ & $\times$ & \hspace{0.3cm} & \hspace{0.3cm} &  \hspace{0.3cm} & $\times$ & \hspace{0.3cm} & \hspace{0.3cm} & \hspace{0.3cm}\\
\hline
$N_f$ D5 & $\times$ & $\times$ & $\times$ & \hspace{0.3cm} & \hspace{0.3cm} & \hspace{0.3cm}& \hspace{0.3cm} & $\times$& $\times$ & $\times$ \\
\hline 
\end{tabular}
\caption{HW setup in type IIB critical string theory that is T-dual to the configuration of Table \ref{tab:ciagarbranes3d}.}
\label{tab:IIBbranes3d}
\end{center}
\end{table}
Note that for 3d $\mathcal{N}=4$ gauge theories the Coulomb branch is now $\R^3_y$, which we can write in polar coordinates as
\es{}{
\ds(\R^3_y)=\diff y^2+y^2\,\ds(S^2_L)\,,
}
where the isometries of $S^2_L$ realize geometrically the $SU(2)_L$ part of the $SO(4)\simeq SU(2)_L\times SU(2)_R$ R-symmetry, while the $SU(2)_R$ part is realized by the isometries of a two-sphere $S^2_R$ in $\R^3_v$ as in \eqref{R3v_S2R}.

While this represents our original ambition, one has to deal with the fact that in the context of two-derivative supergravity it is not possible to obtain a T-dual description of the cigar background with localized NS5 branes, as we have reviewed in the previous section. Rather, the only geometrical description available in the frame $\ns$ includes a large number $k\to \infty$ of NS5 branes arranged on a circle. This is closely related to the fact that the curvature of the solutions scales with $k^{-1}$, so that supergravity is only trustworthy for large $k$. When adding boundary states, the gauge theories that we end up describing are complicated by the fact that the D4-branes are now stretched between an infinite number of smeared NS5 branes and are therefore also smeared. We can imagine the existence of a family of 4d gauge theories parametrized by an integer $k$ (in the spirit of, {\it e.g.}, \cite{Aharony:2012tz}), with our supergravity description valid for $k\to \infty$ while theories such as $\mathcal{N}=2$ pure SYM and SQCD are defined for $k=2$, where the supergravity description fails.

\subsection{Backgrounds with RR fields: noncritical type IIB}
\label{sec:RR_IIBNC}

We would now like to consider a solution of 7d $ISO(4)$ gauged supergravity which realizes the HW setup of Table \ref{tab:IIAbranes4d} after uplift to 10d type IIA supergravity. While in Table \ref{tab:ciagarbranes4d} we have provided the T-dual description in noncritical type IIB superstring theory in terms of boundary states of the cigar, here we encounter a problem. As discussed in the previous section, the supergravity realization of the cigar involves an arrangement of an infinite number of NS5 branes on a circle, which only preserves a $U(1)\times U(1)$ isometry in the space transverse to the branes. If, on the other hand, we wish to realize the $SU(2)_R$ symmetry of the dual 4d $\mathcal{N}=2$ gauge theory geometrically, we are forced to choose a different arrangement, namely the branes on a segment described in Section \ref{sec:NS5distributions}. This has the advantage that we are dealing with a larger group of isometries, which simplifies the ansatz in gauged supergravity, but the drawback that the worldsheet CFT corresponding to this configuration of NS5 branes is not understood from the perspective of noncritical string theory and therefore we do not have a good understanding of its boundary states. The two descriptions of course agree in the limit $k=2$, where there are only two NS5 branes and the $SU(2)_R$ in the $SL(2)_2/U(1)$ CFT arises at the worldsheet level because the cigar is at the free fermion radius for this value of $k$: unfortunately, though, this aspect cannot be captured by the supergravity description.

\subsubsection[A truncation of $ISO(4)$ gauged supergravity]{A truncation of $\boldsymbol{ISO(4)}$ gauged supergravity}

We would like to find a truncation of the gauged supergravity that preserves Poincaré invariance on $\R^{1,3}_x$ as well as the $SU(2)_R$ isometry of $S^2_R$ upon uplift to 10d type IIA supergravity\footnote{This $SU(2)_R$ is the group called $SU(2)_D$ in Appendix \ref{app:so4}. In terms of the $SO(4)$ index $i=1,\ldots,4$ introduced in \eqref{sl5toso4}, $SU(2)_R$ generates rotations in the space $i=1,2,3$. In other words, it is the $SU(2)\subset SO(4)$ such that under $SO(4)\to SU(2)$, the branching rule for the fundamental representation is $\mathbf{4}\to \mathbf{1}+\mathbf{3}$.}. Out of the 14=10(NSNS)+4(RR) scalars of the theory, there are 2 NSNS singlets of $SU(2)_R$ ($f_1$ and $f_4$ in \eqref{cosetrepNSNS}, with $f_2=f_3=f_1$) and 1 RR singlet $b$. None of the NSNS vectors preserves $SU(2)_R$, while there is a single RR vector which is invariant, which we call $A_\mu$. The requirement that our solution should preserve Poincaré invariance on $\R^{1,3}_x$ implies that all self-dual massive three-forms should vanish, while in principle there is still room for the NSNS two-form $B^0_{\mu\nu}$ appearing in \eqref{fields_so5toso4}, since it is a singlet of $SO(4)$. However, a non-trivial $B^0_{\mu\nu}$ would be associated with the existence of NS5 branes in type IIA that are extended along some directions in $\R^4=\R_u\times \R^3_v$, on top of the distribution of NS5 branes extended along $\R^{1,3}_x\times \R^2_y$ which is present just by virtue of considering solutions of this specific gauged supergravity. For this reason, we set $B^0_{\mu\nu}=0$. So we end up with a trunctation of our 7d guaged supergravity containing two NSNS scalars $\phi_0,\phi_1$\footnote{In terms of the four scalars $f_i$ of \eqref{cosetrepNSNS}, $f_1=f_2=f_3=\phi_1-5\phi_0$ and $f_4=-3\phi_1-5\phi_0$.}, one RR scalar $b$ and one RR vector $A_{\mu}$, where we would like to interpret the latter two fields as those sourced by the D5 and D3 branes in Table \ref{tab:ciagarbranes4d}, respectively, or the D6 and D4 branes in the dual picture of Table \ref{tab:IIAbranes4d}. As we shall soon see, however, this is not quite accurate. The action for this model, which can be obtained from either \cite{Cvetic:2000ah} or \cite{Samtleben:2005bp}, is
\es{actionSU2RRISO4}{
S=\int d^7x\,\sqrt{-g}\,\left[R+\mathcal{V}-20(\partial\phi_0)^2-12(\partial\phi_1)^2-4\ex^{6(\phi_0-\phi_1)}F^2-\frac{1}{2}\ex^{10\phi_0+6\phi_1}(\diff b+g\,A)^2\right]\,,
}
where $g$ is the 7d gauge coupling, $F=\diff A$ and the potential is
\es{}{
\mathcal{V}=\frac{g^2}{32}\ex^{4(\phi_0-\phi_1)}\,\left(3+6\ex^{8\phi_1}-\ex^{16\phi_1}\right)\,.
}
We note that, as anticipated in Section \ref{sec:LDspectrum}, the RR scalar $b$ does not appear in the potential and moreover it is coupled {\it à la} St\"uckelberg to the gauge field $A$: we can set $b=0$ with a gauge transformation and work with a massive gauge field $A$. Solutions of \eqref{actionSU2RRISO4} preserve supersymmetry if the following Killing spinor equations (KSEs) are satisfied\footnote{We only display the independent components of the more general KSEs which can be obtained setting to zero the fermions in the supersymmetry transformations given in \cite{Samtleben:2005bp}. The two algebraic KSEs are obtained from $\delta\chi^{abc}$ of \cite{Samtleben:2005bp} as two independent linear combinations of $(\gammaflavor_4)^d_{\,\,\,c}(\gammaflavor_4)_{ab}\delta\chi^{abc}$ and $(\gammaflavor_5)^d_{\,\,\,c}(\gammaflavor_5)_{ab}\delta\chi^{abc}$.}
\es{}{
0=&\left[\nabla_{\mu}-\frac{g}{80}\ex^{2(\phi_0-\phi_1)}(3+\ex^{8\phi_1})\gamma_{\mu}\right.\\
&\left. +\left(\frac{g}{4}\ex^{5\phi_0+3\phi_1}\,A_{\mu}+\frac{1}{10}\ex^{3(\phi_0-\phi_1)}(\gamma_{\mu}^{\,\,\,\nu \lambda}+8\gamma^{\nu}\delta_\mu^\lambda)F_{\nu\lambda}\right)\gammaflavor_{45}\right]\epsilon\,,\\
0=&\left[\slashed{\partial}\phi_0+\frac{g}{80}\ex^{2(\phi_0-\phi_1)}(3+\ex^{8\phi_1})-\left(\frac{g}{8}\ex^{5\phi_0+3\phi_1}\slashed{A}-\frac{3}{20}\ex^{3(\phi_0-\phi_1)}\slashed{F}\right)\gammaflavor_{45}\right]\epsilon\,,\\
0=&\left[\slashed{\partial}(\phi_0-\phi_1)+\frac{g}{20}\ex^{2(\phi_0-\phi_1)}(2-\ex^{8\phi_1})+\frac{2}{5}\ex^{3(\phi_0-\phi_1)}\slashed{F}\,\gammaflavor_{45}\right]\epsilon\,,
}
where $\epsilon$ is a spinor of $Spin(1,6)$ as well as of $USp(4)\simeq SO(5)$, while $\gammaflavor_{45}=\gammaflavor_{[4}\gammaflavor_{5]}$, where an explicit representation for the flavor gamma matrices $\gammaflavor_M$, $M=1,\ldots,5$, is given in Appendix \ref{app:so5usp4}.

\subsubsection{Solutions with eight supercharges}

We are now ready to make an ansatz for our solution. The metric should be of the form
\es{}{
\ds_{10}=\ex^{-4\phi_0}\,\left[\ex^{g_1}\,\ds(\R^{1,3}_x)+\frac{4\,\ex^{g_2}}{g^2}\frac{\diff r^2}{r^2}+\ex^{g_3}\,\ds(\R^2_y)\right]\,,
}
where $g_i$ ($i=1,2,3$) are functions of $r$ and $\vec{y}$, with $r$ playing the role of a linear-dilaton direction (at least asymptotically). The two scalars $\phi_0$ and $\phi_1$ are also arbitrary functions of $r$ and $\vec{y}$, and we take 
\es{}{
A=A_1\,\diff y_1+A_2\,\diff y_2+A_r\,\diff r\,,
}
for arbitrary $A_1$, $A_2$ and $A_r$. We look for solutions preserving eight supercharges, which we think of as realizing the 4d $\mathcal{N}=2$ super-Poincaré algebra on $\R^{1,3}_x$. We find that this requires the spinor $\epsilon$ to satisfy the projections
\es{}{
\gamma_r\epsilon=\epsilon\,,\quad
\gammaflavor_{45}\epsilon=\gamma_{y_1y_2}\epsilon\,,
}
where the first is the same projection satisfied by the domain walls presented in Section \ref{sec:NS5distributions}, while the second further reduces the supersymmetry by a half and is dictated by the structure of the KSEs. The corresponding BPS equations allow to express all the functions appearing in the ansatz in terms of a unique function (up to some convenient gauge choices) which we call $K\equiv K(r,\vec{y})$, as we now describe. The BPS equations set $A_r=0$ and fix the massive vector $A$ to be of the form
\es{A7d}{
A=\frac{1}{\sqrt{r}}\star_y\diff_y K\equiv 
\frac{-\partial_{y_2}K\,\diff y_1+\partial_{y_1}K\,\diff y_2}{\sqrt{r}}\,,
}
where $\star_y$ and $\diff_y$ refer to the hodge star and exterior derivative taken in $\R^2_y$ with a flat metric $\ds(\R^2_y)=\diff y_1^2+\diff y_2^2$. The metric functions are fixed in terms of $K$ by
\es{}{
\ex^{-g_1-g_2}=g\,K\,,\quad
\ex^{-g_1+g_2}=g\,(K-2r\,\partial_r K)\,,\quad
g_3=-g_1\,,
}
while the scalars $\phi_0$ and $\phi_1$ are conveniently expressed in terms of $g_1$ and $g_2$ as
\es{}{
\phi_0&=\frac{1}{5}g_1+\frac{1}{20}g_2+\frac{1}{10}\log r\,,\quad 
\phi_1=\frac{1}{4}g_2\,.
}
The function $K$ is then constrained by a second-order non-linear partial differential equation (PDE) which reads
\es{PDE4d}{
\Delta_y K+\frac{g^3}{8}r^2\,\partial^2_rK^2=0\,.
}
Similar PDEs have appeared in related contexts, see {\it e.g.}~\cite{Imamura:2001cr,Macpherson:2016xwk,Legramandi:2018itv,Lozano:2022ouq}. This equation appears to be complicated to solve in general, but let us make a comment on the simplest case, where explicit solutions reduce to backgrounds that we have already studied. If we drop the dependence on $\vec{y}$, that is we assume that $K(r,\vec{y})\equiv f(r)$, we find the general solution
\es{}{
f(r)=\sqrt{c_1+c_2\,r}\,,
}
for some integration constants $a$ and $b$. The RR field $A_{\mu}$ is set to zero by this choice, so we end up with purely NSNS solutions. In particular, we note that setting $c_2=0$ and $c_1=g^{-1}$, or $K=g^{-1}$, reproduces the linear dilaton solution \eqref{lineardilaton7d}, provided that we set $r=\ex^{-\tfrac{g}{2}\rho}$. On the other hand, keeping both $a$ and $b$ we find that the choice $K=g^{-1}\sqrt{1+\ell^2\,r}$ gives the distribution of NS5 branes on a segment discussed around \eqref{NS5bar}, after setting $s=1/\sqrt{r}$.

Even without solving the PDE \eqref{PDE4d} explicitly, it is clear that AdS$_5$ solutions do not exist -- we shall comment more on this at the end of this section.

\subsubsection{Uplift to 10d type IIA supergravity}

To interpret the 7d solution above it is convenient to consider the uplift to ten dimensions, for which we can use the expressions in \cite{Cvetic:2000ah}. The resulting string frame solution reads (see Appendix \ref{app:typeII} for our conventions)
\es{IIAsol_rpsi}{
    \diff s^2_{10}=&\ex^{g_1}\,\diff s^2(\R^{1,3}_x)+\ex^{-g_1}\diff s^2(\R^2_y)+\ex^{g_2}\frac{4}{g^2}\frac{\diff r^2}{r^2}+\frac{16}{g^2}\left[\ex^{-g_2}\diff\psi^2+\frac{1}{\Omega}\sin^2\psi\,\diff s^2(S^2_R)\right]\,,\\
    C_1=&4\cos\psi\,A\,,\\
    B_2=&\frac{16}{g^2}\left(\psi-\frac{\ex^{g_2}\sin 2\psi}{2\Omega}\right)\vol(S^2_R)\,,\\
    C_3=&\frac{64\ex^{-g_2}\sin^3\psi}{g^2\,\Omega}A\wedge\vol(S^2_R)\,,\\
    \ex^{2\Phi}=&\frac{\ex^{2g_1}}{\Omega}\,r\,,
}
with $A$ is the 7d massive vector and
\begin{align}
    \Omega=\ex^{g_2}\,\cos^2\psi +\ex^{-g_2}\,\sin^2\psi\,,
\end{align}
where the $S^3$ transverse to the NS5 branes is parametrized by $\psi\in [0,\pi)$ and polar coordinates $(\theta_R,\phi_R)$ for the $S^2_R$.\footnote{Here there is only one $S^2$, so the subscript $R$ is irrelevant. We keep it nonetheless since in the 10d type IIB solution dual to 3d gauge theories there is also an additional $S^2$ in the internal space, which we refer to as $S^2_L$.} To discuss the structure of the solution, it is useful to consider an alternative set of coordinates, namely we introduce\footnote{Another, similar, change of coordinates is also interesting: setting
\es{}{
x_1=\frac{4\sqrt{r}}{g}\frac{\psi}{\sin\psi}\,,\quad
x_2=\frac{4}{g\,\sqrt{r}}\sin\psi\,,\quad
x_3=y_1\,,\quad
x_4=y_2\,,
}
one obtains the coordinates $x_1,\ldots,x_4$ used in \cite{Macpherson:2016xwk} for the classification of $\R^{1,3}$ solutions with eight supercharges and geometric $SU(2)_R$ symmetry in type IIA. In particular, our solutions belong to the class analyzed in Section 4.3, see also Appendix C.
}
\es{uvcoords}{
u=\frac{1}{\sqrt{r}}\cos\psi\,,\quad
v=\frac{1}{\sqrt{r}}\sin\psi\,,
}
with $u\in \R$ while $v\in \R_+$ as a consequence of $\psi\in [0,\pi)$. In these coordinates, the metric and dilaton of \eqref{IIAsol_rpsi} can be rewritten as
\es{IIAsol_uv}{
\diff s^2_{10}=&\ex^{g_1}\,\diff s^2(\R^{1,3}_x)+\ex^{-g_1}\diff s^2(\R^2_y)+\frac{16}{g^2}\frac{1}{u^2+v^2}\left[\Omega^{-1}\,\diff s^2(\R^3_v)+\Omega\,\mathrm{D}u^2\right]\,,\\
        \ex^{2\Phi}=&\frac{\ex^{2g_1}}{u^2+v^2}\,\Omega^{-1}\,,
}
where now
\es{Deltauv}{
\Omega=\frac{\ex^{g_2}u^2+\ex^{-g_2}v^2}{u^2+v^2}\,,
}
and
\es{Du_R3v}{
 \mathrm{D}u=\diff u+\frac{\ex^{g_2}-\ex^{-g_2}}{\Omega}\frac{u\,v}{u^2+v^2}\diff v\,,\quad
    \diff s^2(\R^3_v)=\diff v^2+v^2\,\diff s^2(S^2_R)\,.
}
To understand the meaning of these coordinates, it is useful to consider solutions which asymptote the linear dilaton, that is 
\es{Klimit}{
K=g^{-1}+\ldots\,,
}
where the ellipsis denote corrections that come in (possibly non-integer) powers of $r$, with $\vec{y}$-dependent coefficients. In this case, since asymptotically 
\es{LDasympt}{
\ex^{g_1}\simeq \ex^{g_2}\simeq 1\,,\quad \Omega\simeq 1,\quad\mathrm{D}u \simeq \diff u\,,
}
we see that in this limit the part of the metric in square brackets in \eqref{IIAsol_uv} reduces to that of $\R^4$ written as
\es{}{
\ds(\R^4)=\diff u^2+\ds(\R^3_v)\,.
}

The main advantage of the coordinates $u,v$ introduced in \eqref{uvcoords} is that they allow for a clearer understanding of the configuration of branes. In particular, following the conventions outlined in the last paragraph of Appendix \ref{app:typeIIAconventions} for the introduction of electric potentials, we note that a suitable choice of gauge exists such that
\es{electric_potentials_4d}{
B_6&=\ex^{g_1-2\Phi}\vol(\R^{1,3}_x)\wedge\vol(\R^2_y)\,,\\
C_5&=\frac{4}{g}\ex^{g_1}\,\Omega\,\vol(\R^{1,3}_x)\wedge\mathrm{D} u\,,\\
C_7&=\frac{64}{g^3}\ex^{2\Phi}\vol(\R^{1,3}_x)\wedge\vol(\R^3_v)\,.
}
From the expressions above we can read off the fact that, if the solution is asymptotic to the linear dilaton according to \eqref{LDasympt}, then the brane setup is precisely that described in Table \ref{tab:IIAbranes4d}.

We can now consider the problem of quantizing the fluxes. For this purpose, it would be interesting to have explicit solutions describing the intersection between localized branes, which would allow one to study explicitly the compact cycles in the geometry. However, it is rather challenging to find solutions to the PDE \eqref{PDE4d} and we have not found a solution that realizes this setup. However, we can still make some interesting comments by working under the conditions \eqref{LDasympt}, which correspond to the assumption that we can study the geometry ``far enough'' from the D-branes, so that their backreaction is negligible. Note that this is the opposite regime from the near-horizon limit that one would like to take to explore holographically the dual QFTs, but it is still sufficient to determine the number of branes involved in the solution. Let us then proceed by assuming that the explicit solution for $K$ is of the type \eqref{Klimit}, implying the behavior \eqref{LDasympt}, so that the geometry ``at infinity'' looks like
\es{asymptotic_geometry}{
\ds_{10}\simeq \ds(\R^{1,3}_x)+\ds(\R^2_y)+\frac{16}{g^2}\frac{1}{u^2+v^2}\left[\diff u^2+\ds(\R^3_v)\right]\,.
}
The simplest analysis is for the three-form flux $H_3$, which is sourced magnetically by NS5 branes. As clear from \eqref{electric_potentials_4d}, the latter are extended along $\R^{1,5}_{\parallel}\equiv \R^{1,3}_x\times \R^2_y$, so they are localized at points in the transverse $\R^4_{\perp}\equiv \R_u\times \R^3_v$. We can then identify a three-cycle setting $x=y=0$ (the origin of $\R^{1,5}_{\parallel}$) and surrounding the NS5 branes with a three-sphere in $\R^4_{\perp}$, which can be done by setting $u=s\,\cos\psi$ and $v=s\,\sin\psi$, with $s\in \R_+$ and $\psi\in(0,\pi)$\footnote{Note that change of coordinates is the same as in \eqref{uvcoords}, with $r=s^{-2}$.}: the resulting three-sphere is the same $S^3$ used for the uplift, parametrized by $\psi$ and $S^2_R$. Integrating $H_3$ on this cycle gives
\es{}{
k\equiv \frac{1}{(2\pi)^2\alpha'}\int_{S^3} H_3=\frac{8}{g^2}\,,
}
where we have set $\alpha'=2$ and consistently with the notation adopted in the rest of the paper we call $k$ the number of NS5 branes. This fixes the relation between the gauge coupling of the supergravity and the number of NS5 branes as we have already discussed in \eqref{Qtok} in terms of the parameter $Q=g/2$. 

The quantization of D-brane charges is a bit more subtle. Let us begin with D6 branes, which are extended along $\R^{1,3}\times \R^3_v$ according to \eqref{electric_potentials_4d}, so to find a two-cycle where to integrate $F_2$ we set $x=v=0$ and we need to identify a two-sphere in the transverse $\R^3_{\text{D6}}\equiv \R^2_y\times \R_u$, which from \eqref{asymptotic_geometry} has a metric
\es{}{
\ds(\R^3_{\text{D6}})=\ds(\R^2_y)+\frac{16}{g^2}\frac{\diff u^2}{u^2}\,.
}
We can then set 
\es{}{
u=\ex^{\tfrac{g}{4}r_6\,\cos\eta_6}\,,\quad
\vec{y}=y\,
\begin{pmatrix}
    \cos \alpha\\
    \sin \alpha
\end{pmatrix}=r_6\,\sin\eta_6\,
\begin{pmatrix}
    \cos \alpha\\
    \sin \alpha
\end{pmatrix}\,,
}
which identifies a two-sphere $S^2_{\text{D6}}$ surrounding the origin $r_6=0$ of $\R^3_{\text{D6}}$, parametrized by $\eta_6\in(0,\pi)$ and $\alpha\in (0,2\pi)$. We are from now on going to assume that $\frac{\partial}{\partial \alpha}$ is a Killing vector, which implies that the $U(1)_r$ part of the R-symmetry of the dual 4d QFT is unbroken. We find that setting $v=0$ one has
\es{}{
\left.F_2\right|_{v=0}=\left.\diff A\right|_{v=0}\,,
}
where $A$ is the 7d massive vector introduced in \eqref{A7d}. We can then compute the total number of D6 branes via
\es{ND6}{
N_{\mathrm{D6}}\equiv \frac{1}{(2\pi\sqrt{\alpha'})g_s}\int_{S^2_{\text{D6}}}F_2=\frac{\sqrt{2}}{\pi\,g_s}\mathcal{I}_{\text{D6}}[A]\,,\quad
\mathcal{I}_{\text{D6}}[A]\equiv \int_{S^2_{\text{D6}}}\left.\diff A\right|_{v=0}\,.
}
The logic is similar for D4 branes, which according to \eqref{electric_potentials_4d} are extened along $\R^{1,3}_x\times \R_u$, so we need to identify a four-cycle inside $\R^5_{\text{D4}}\equiv \R^{1,3}_x\times \R^3_v$. The metric on this space, following from \eqref{asymptotic_geometry}, is
\es{R5D4}{
\ds(\R^5_{\text{D4}})=\ds(\R^2_y)+\frac{16}{g^2}\left[\frac{\diff v^2}{v^2}+\ds(S^2_R)\right]\,,
}
which is actually of the form $\R^3\times S^2_R$, with the $\R^3$ parametrized by $\vec{y}$ and $v$. The four-cycle $\Sigma_{\text{D4}}$ needed to integrate the four-form RR flux is then given by the product of a two-sphere $S^2_{\text{D4}}$ surrounding the origin of this $\R^3$ and the two-sphere $S^2_R$ dual to $SU(2)_R$: $\Sigma_{\text{D4}}=S^2_{\text{D4}}\times S^2_R$. Here $S^2_{\text{D4}}$ is identified in a similar way to $S^2_{\text{D6}}$, that is after the change of coordinates
\es{}{
v=\ex^{\tfrac{g}{4}r_4\,\cos\eta_4}\,,\quad
\vec{y}=y\,
\begin{pmatrix}
    \cos \alpha\\
    \sin \alpha
\end{pmatrix}=r_4\,\sin\eta_4\,
\begin{pmatrix}
    \cos \alpha\\
    \sin \alpha
\end{pmatrix}\,,
}
it is the sphere surrounding the origin $r_4=0$ and parametrized by $\eta_4\in(0,\pi)$ and $\alpha\in(0,2\pi)$. We note that generally speaking the number of D4 branes in this context is a Page charge, which can be obtained integrating either $\tilde{F}_4$ (which is gauge-invariant, but not closed) or $F_4$ (which is closed, but not gauge invariant) -- see Appendix \ref{app:typeII} for our conventions. In particular, in principle one could perform a large gauge transformation of $B_2$ which changes the D4 branes charge. However, since this would affect $F_4$ in the form of a term $F_2\wedge B_2$ which vanishes in the limit \eqref{LDasympt} and after setting $u=0$, we find that such ambiguity does not affect the computation of the charge in this case. Indeed, we find
\es{}{
\left. \tilde{F}_4\right|_{u=0}=
\left. \diff C_3\right|_{u=0}=
\frac{64}{g^2}\vol(S^2_R)\wedge \left.\diff A\right|_{u=0}\,,
}
from which we can finally compute the flux
\es{ND4}{
N_{\mathrm{D4}}\equiv \frac{1}{(2\pi\sqrt{\alpha'})^3 g_s}\int_{\Sigma_{\text{D4}}}\tilde{F}_4=\frac{\sqrt{2}k}{\pi^2\,g_s}\mathcal{I}_{\mathrm{D4}}[A]\,, \quad 
\mathcal{I}_{\mathrm{D4}}[A]\equiv 
\int_{S^2_{\text{D4}}}\left.\diff A\right|_{u=0}\,.
}

Note that in \eqref{ND6} and \eqref{ND4} we have introduced two integrals $\mathcal{I}_{\mathrm{D6}}[A]$ and $\mathcal{I}_{\mathrm{D4}}[A]$, whose evaluation determines the final result for the fluxes. We cannot evaluate those integrals directly because that would require an explicit solution to the PDE \eqref{PDE4d}, but some comments are in order. First, we note that although we have given two distinct definitions for these integrals, they actually evaluate to the same result for any solution arising from the uplift of a 7d gauged supergravity solution. This is because they are only distinguished by setting either $u=0$ or $v=0$, but the massive 7d vector $A$ only depends on the combination $r=(u^2+v^2)^{-1}$ -- see \eqref{A7d} -- so for the solutions discussed here we actually have $\mathcal{I}_{\mathrm{D6}}[A]=\mathcal{I}_{\mathrm{D4}}[A]\equiv \mathcal{I}[A]$. This fact also has another implication: our solutions do not allow to have only color (D4) branes without flavor (D6) branes. This is related to the fact that the massive 7d gauge field $A$ sources both $F_2$ and $F_4$ in ten dimensions, but we do not have an interpretation for this fact from the perspective of noncritical string theory. Another comment is related to the fact that the ratio between $N_{\mathrm{D4}}$ and $N_{\mathrm{D6}}$ is irrational and more precisely given by $k/\pi$, so that only one of the two types of D-brane charges can actually be quantized. We believe that the reason for this is that there is a smearing in the distribution of D4 branes along $S^2_R$, so that they are not fully localized. This fact is manifest in the metric on the space transverse to the D4 branes, \eqref{R5D4}, which is not $\R^5$ but rather $\R^3\times S^2_R$.

\subsubsection{Solving the PDE}\label{sec:solvingPDE4d}

To conclude this subsection, let us comment on possible solutions to the PDE \eqref{PDE4d}. As we already discussed, the simplest choice $K(r,\vec{y})\equiv f(r)$ gives the distribution of NS5 branes on a segment found in \eqref{NS5bar} and as such has vanishing RR fields. Thus, here we focus on solutions with non-trivial dependence on both $r$ and $\vec{y}$, which we would like to interpret as distributions of D4 and D6 branes on top of the arrangement \eqref{NS5bar} of NS5 branes. In doing so, we remind the reader of the analysis carried out in Section \ref{sec:boundarystates} for the backreaction of the boundary states of the cigar on the metric. While that analysis strictly speaking only holds for the cigar at level $k=2$ (that is, two NS5 branes), and as such does not apply to the solutions considered here (that is, branes on a segment for large $k$), we can still take away two qualitative lessons that we shall compare our results with. The first is that the backreaction of color branes in \eqref{D3-asymptotic-expansion} is subleading compared to that of flavor branes \eqref{D2-int-2} in an expansion for large $\rho$. The second is the type of functions of $\vec{y}$ that appear in this expansion around the linear dilaton solution differs qualitatively for the two types of branes: we find polynomials of $\vec{y}$ for color branes and Bessel functions for flavor branes.

Since all the information that we were able to extract from the analysis of the boundary states is in the form of an asymptotic expansion around the linear dilaton region, recalling that in such regime the coordinate $r$ used in this section maps to the usual $\rho$ used for linear dilaton solution and in the worldsheet analysis via $r=\ex^{-g\rho/2}$, so that it is natural to consider an expansion in (possibly fractional) powers of $r$. One can then try an ansatz of the form
\es{ansatzK_4d}{
K=\sum_{n=0}^{\infty}r^{n/a}f_n(\vec{y})\,,
}
for {\it a priori} arbitrary values of $a>0$. The PDE \eqref{PDE4d} then constrains the functions $f_n(\vec{y})$ via
\es{seriesexp_K_4d}{
\Delta_y f_n(\vec{y})+\frac{g^3}{8a^2}n(n-a)\sum_{j=0}^n f_j(\vec{y})f_{n-j}(\vec{y})=0\,,
}
that can be solved order by order. While this does not seem particularly instructive, and indeed we are not generally able to find closed-form expressions for general solutions or even just a specific criterion to fix integration constants arising at each order, we do find that for a specific value of $a$ a truncated solution exists. Setting $a=2$, which corresponds to an expansion in powers of $\sqrt{r}$, we find that\footnote{Note that while any constant at leading order would work simply from the perspective of solving \eqref{PDE4d}, we set it to $g^{-1}$ to reproduce the linear dilaton asymptotics. Interestingly, as we comment in what follows this specific value gives a Yukawa-type potential with the same mass as the RR fields.}
\es{truncated_4d}{
K=g^{-1}+\sqrt{r}\,h(\vec{y})\,,\quad \text{with} \quad
\Delta_y h(\vec{y})=\frac{g^2}{16}\,h(\vec{y})\,,
}
solves \eqref{PDE4d} exactly. If, for simplicity, we assume rotational invariance in $\R^2_y$ so that $h$ only depends on $y\equiv||\vec{y}||$, solutions to the resulting ODE for $h$ in \eqref{truncated_4d} that are regular as $y\to \infty$ are of the form  
\es{yukawa_4d}{
h(y)=K_0\big(\tfrac{g}{4}y\big)\,,
}
where $K_0$ is the same modified Bessel function of the second kind appearing in \eqref{D2-int-2}, which we note corresponds to the expression of the Yukawa potential in 2d with mass $Q/2$. Note that this is exactly the mass of the RR particles of the theory, as found from the worldsheet in Table \ref{tab:IIBspectrum} and from the 7d supergravity in \eqref{eomRR_7d_ISO4}. One could of course continue the asymptotic expansion including higher powers of $\sqrt{r}$, but it is pretty remarkable that a truncated solution exists at all and, to the best of our knowledge, such solution has not previously appeared in the literature. We have not been able to carry out a detailed analysis for the quantization of the fluxes in this case, probably due to the (at least partial) smearing of the D-branes, but we certainly think that this solution is worth of a more careful investigation in the future.

Another interesting type of solutions arising from an expansion of the type \eqref{ansatzK_4d} is of polynomial type, which remind of the type of functions appearing in the backreaction of D3 branes in \eqref{D3-asymptotic-expansion}, although here we are only including an expansion in exponentials of $\rho$, without any power-law behavior. We find that solutions exist where $f_n(y)$ (still assuming rotational invariance for simplicity) is a polynomial in $y$ of degree $2n-2$, and it is even possible to fix the integration constants consistently in such a way that 
\es{}{
f_n(y)=c_n\,y^{2n-2}\,,\quad \text{with}\quad
c_n=-\frac{g^3}{32a^2}\frac{n(n-a)}{(n-1)^2}\sum_{k=1}^{n-1}c_kc_{n-k}\,.
}
This is a non-linear recursion relation that unfortunately appears hard to solve for any $a$, although note that its expression slightly simplifies for $a=1$. We note that $c_0=0$, so that in this case the leading order is {\it not} the one leading to a linear dilaton asymptotics. The meaning of this solution is unclear, although given its particularly simple structure it would be interesting to further investigate possible interpretations and look into a closed form expression for the coefficients $c_n$, at least for $a=1$.

Finally, we would like to mention possibly the simplest non-trivial solution to \eqref{PDE4d}, which arises from an ansatz with separation of variables. Namely, setting 
\begin{equation}
K(r,\vec{y})=f(r)g(\vec{y})
\end{equation}
solves the PDE if 
\begin{equation}
f(r)=g^{-1}\sqrt{1+\ell^2 r} \,,
\end{equation} 
as in the solution \eqref{NS5bar} describing NS5 branes on a segment, for any harmonic function $g(\vec{y})$. As opposed to the ones listed so far, the interpretation of this solution is quite clear and is discussed in Appendix \ref{app:smearedsol}: it describes a configuration where the distribution of D-branes is fully smeared in such a way as to respect the symmetries of the distribution of NS5 branes on a segment. As a consequence of the smearing, it is not possible to quantize the associated fluxes.

\subsection{Backgrounds with RR fields: noncritical type IIA}\label{sec:RR_IIANC}

We now consider the analogue of the problem considered in Section \ref{sec:RR_IIBNC}, but in the 7d $SO(4)$ gauged supergravity associated with type IIA noncritical string theory. The spirit is precisely the same as that of the previous section, and so are the considerations regarding the NSNS fields. On the other hand, in the RR sector the one $SU(2)_R$-invariant massive vector considered in Section \ref{sec:RR_IIANC} is now replaced by a massive two-form $B_{\mu\nu}$. In principle, one could then work out the action and Killing spinor equations for these fields and classify supersymmetric solutions as done for the $ISO(4)$ gauged supergravity. However, since at the end of the day most of the insights on the solution in that case were gained from the uplift to ten dimensions, we can take a shortcut. Namely, we start from the solution \eqref{IIAsol_rpsi} in 10d type IIA supergravity and perform a T-duality along one of the spacelike $\R^{1,3}$ directions. This gives solution in 10d type IIB supergravity which is smeared along the T-duality direction, but its localized version (along that direction) is straightforward to guess. For this reason we shall work directly in ten dimensions, but we will give the expression of the corresponding seven-dimensional massive two-form $B_{\mu\nu}$.

In the string frame, the ten-dimensional solution in type IIB supergravity corresponding to the procedure outlined above can be expressed as\footnote{We follow the conventions outlined in the last paragraph of Appendix \ref{app:typeIIBconventions} and denote with $C_4^{\ma}$ the magnetic four-form potential. The five-form flux entering the equations of motion and Killing spinor equations in \ref{app:typeIIBconventions} is then $\tilde{F}_5=(1+\star)(\diff C_4^{\ma}-H_3\wedge C_2)$.}
\es{IIBsol_rpsi}{
    \diff s^2_{10}=&\ex^{g_1}\,\diff s^2(\R^{1,2}_x)+\ex^{-g_1}\diff s^2(\R^3_y)+\ex^{g_2}\frac{4}{g^2}\frac{\diff r^2}{r^2}+\frac{16}{g^2}\left[\ex^{-g_2}\diff\psi^2+\frac{1}{\Omega}\sin^2\psi\,\diff s^2(S^2_R)\right]\,,\\
    C_2=&4\cos\psi\,B\,,\\
    B_2=&\frac{16}{g^2}\left(\psi-\frac{\ex^{g_2}\sin 2\psi}{2\Omega}\right)\vol(S^2)\,,\\
    C_4^{\ma}=&\frac{64\,\ex^{-g_2}\,\sin^3\psi}{g^2\,\Omega}B\wedge\vol(S^2)\,,\\
    \ex^{2\Phi}=&\frac{\ex^{g_1}}{\Omega}\,r\,,
}
where
\es{}{
    \Omega=\ex^{g_2}\,\cos^2\psi+\ex^{-g_2}\,\sin^2\psi\,,
}
while the 7d massive two-form $B$ is given by
\begin{align}
    B=\frac{1}{\sqrt{r}}\star_y\diff_{y}K\,,
\end{align}
in terms of a function $K\equiv K(r,\vec{y})$. Now we have a three-dimensional Coulomb branch parametrized by three coordinates $\vec{y}$. The R-symmetry of the dual 3d gauge theory is $SO(4)\simeq (SU(2)_R\times SU(2)_L)/\Z_2$, where the $SU(2)_R$ part is always realized geometrically in \eqref{IIBsol_rpsi}, while the $SU(2)_L$ part is visible in the metric if we write
\es{}{
\ds(\R^3_y)=\diff y^2+y^2\,\ds(S^2_L)\,,
}
and assume that the solution preserves the $SU(2)_L$ isometry of $S^2_L$, {\it i.e.}~the functions only depend on the radial direction $y$. With the definitions above, \eqref{IIBsol_rpsi} is a solution to the equations of motion of type IIB supergravity preserving eight supercharges (thus realizing 3d $\mathcal{N}=4$ supersymmetry of $\R^{1,2}_x$) if and only if the function $K$ satisfies the PDE
\es{PDE3d}{
\Delta_y K+\frac{g^3}{8}r^2\,\partial^2_rK^2=0\,,
}
which is completely analogous to \eqref{PDE4d}, with the only difference that now $\Delta_y$ is the 3d Laplacian. As a consequence, the structure of the solutions is also closely related to that discussed for \eqref{PDE4d} in Section \ref{sec:solvingPDE4d} and we shall not repeat it here (note that again we do not find AdS$_4$ solutions). The only comment that we wish to make is that a truncated solution (in the small $r$ expansion) also exists here, which has the same structure as \eqref{truncated_4d} but with $\vec{y}$ coordinates on $\R^3_y$ and $\Delta_y$ a 3d Laplacian. As a consequence, the analogue of the solution \eqref{yukawa_4d} is now (assuming rotational invariance as in that case)
\es{}{
h(y)=\frac{1}{y}\ex^{-\tfrac{g}{4}y}\,,
}
which is the 3d Yukawa potential with mass $g/4=Q/2$. Once again, this is the same mass as that of the RR fields and moreover $h(y)$ above is the same function appearing in the study of the asymptotic backreaction of flavor branes on the cigar geometry, see \eqref{D2-int-1-IIA}. It would be worth studying this exact solution more in detail in the future, as to the best of our knowledge it has not previously appeared in the literature.

Much like in the IIA case, under the assumption that there is an asymptotic linear dilaton limit the geometry is conveniently analyzed introducing coordinates $u$ and $v$ as in \eqref{uvcoords}, in terms of which the metric and dilaton are
\es{IIBsol_uv}{
\diff s^2_{10}=&\ex^{g_1}\,\diff s^2(\R^{1,2}_x)+\ex^{-g_1}\diff s^2(\R^3_y)+\frac{16}{g^2}\frac{1}{u^2+v^2}\left[\Omega^{-1}\,\diff s^2(\R^3_v)+\Omega\,\mathrm{D}u^2\right]\,,\\
        \ex^{2\Phi}=&\frac{\ex^{2g_1}}{u^2+v^2}\,\Omega^{-1}\,,
}
with $\Omega$, $\mathrm{D}u$ and $\ds(\R^3_v)$ as in (\ref{Deltauv}-\ref{Du_R3v}). Using the conventions of Appendix \ref{app:typeIIBconventions}, we introduce the electric RR potentials in a suitable gauge
\es{}{
B_6&=\ex^{-2\Phi}\vol(\R^{1,2}_x)\wedge\vol(\R^3_y)\,,\\
C_4^{\el}&=\frac{4}{g}\ex^{g_1}\Omega \vol(\R^{1,2}_x)\wedge\mathrm{D}u\,,\\
C_6&=\frac{64}{g^3}\ex^{2\Phi}\vol(\R^{1,3}_x)\wedge\vol(\R^3_v)\,,
}
showing that the configuration of branes is precisely the one described in Table \ref{tab:IIBbranes3d}.

Much like everything else in this solution, the quantization of the charges associated with the various branes proceeds in a completely analogous way to the type IIA solution. The quantization of the flux of $H_3$ follows exactly the same reasoning and also leads to $g=\sqrt{8/k}$. The fluxes of D3 and D5 branes are computed in the same way as those for D4 and D6 branes in the previous subsection, with the only difference that the massive vector $A$ is replaced by the massive two-form $B$ and $\R^2_y$ with $\R^3_y$. The flux $F_3$ associated with D5 branes is then integrated on a three-sphere $S^3_{\text{D5}}$ surrounding the origin of $\R^3_y\times \R_u$ located at $v=0$ via
\es{ND5}{
N_{\text{D5}}\equiv \frac{1}{(2\pi\sqrt{\alpha'})^2g_s}\int_{S^3_{\text{D5}}}F_3=\frac{1}{2\pi^2g_s}\mathcal{I}_{\text{D5}}[B]\,,\quad
\mathcal{I}_{\text{D5}}[B]\equiv \int_{S^3_{\text{D5}}}\left.\diff B\right|_{v=0}\,.
}
On the other hand, the number of D3 branes can be obtained integrating equivalently $\tilde{F}_5$ of $F_5$ over a five-cycle $\Sigma_{\text{D3}}=S^3_{\text{D3}}\times S^2_R$, where $S^3_{\text{D3}}$ is identified as a three-sphere surrounding the origin of $\R^4=\R^3_y\times \R_v$, similarly to what happens for D6 branes. We then obtain
\es{}{
N_{\text{D3}}\equiv \frac{1}{(2\pi\sqrt{\alpha'})^4g_s}\int_{\Sigma_{\text{D3}}}\tilde{F}_5=\frac{k}{2\pi^2g_s}\mathcal{I}_{\text{D3}}[B]\,,\quad
\mathcal{I}_{\text{D3}}[B]\equiv \int_{S^3_{\text{D3}}}\left.\diff B\right|_{u=0}\,.
}
The same observations made for the solution in type IIA also apply here.

\subsection{Comments on the solutions}

We conclude this section with some comments on the solutions that we have found, comparing them with our initial goal and discussing the problems that we have encountered.

We begin by emphasizing a point that we have already made at the start of this section. Namely, in order to realize geometrically the $SU(2)_R$ symmetry of the 3d/4d $\mathcal{N}=2$ QFTs dual to our supergravity backgrounds, one is forced to study solutions that implicitly correspond to adding D-branes to the distribution of NS5 branes on a segment discussed around \eqref{NS5bar}. One issue with this is that the worldsheet CFT associated with that distribution is not known and currently there is no understanding of the associated boundary states, which makes it hard to give an interpretation of our solutions in terms of branes of the noncritical string theory. On the other hand, D-branes in the cigar background \eqref{cigartargetspace} have been investigated thoroughly in the literature (see Section \ref{sec:boundarystates} for references), but the dual distribution of NS5 branes on a circle \eqref{10dbranescircle} only preserves a $U(1)\subset SU(2)_R$, thus breaking the R-symmetry of the dual field theory explicitly.

In our two-derivative supergravity approximation,
we are inevitably describing smeared distributions with a large number $k$ NS5 branes. When adding D-branes to the picture, as we did in this section, these are also going to be smeared, complicating the interpretation of the dual field theory. This smearing is also likely to be responsible for the fact that, as we discussed, the ratio between the number of D-branes turns out to be irrational for both the IIA and the IIB solution. This situation should be contrasted with more standard case of holography, such as AdS$_5\times S^5$: in that case, one achieves small curvature by taking a large number of D3 branes localized at the same point in the transverse space. On the other hand, here we need a large number of smeared NS5 branes to begin with, which makes it hard to obtain solutions of 7d supergravity with localized D-branes. It might be possible that such solutions can be found directly in ten dimensions, in which case they would belong to the classification of \cite{Macpherson:2016xwk}, but such question is beyond the scope of this paper. 

From the perspective of noncritical superstring theory, it is natural to imagine that one should be able to tune the two paramters $N_c$ and $N_f$ in Tables \ref{tab:ciagarbranes4d} and \ref{tab:ciagarbranes3d} (representing the number of colors and flavors of the dual gauge theories, respectively) independently. Keeping in mind the 4d case for definiteness, a concrete goal would be that of obtaining string backgrounds describing $\mathcal{N}=2$ pure SYM ($N_f=0$) or SCQCD ($N_f=2N_c$), which could allow in the future to explore the large $N_c$ regime of those theories using a 2d sigma model. This does not seem to be possible in our solutions, and moreover while our $\R^{1,3}$ ansatz\footnote{Let us focus on the 4d case for definiteness, since completely analogous comments apply to the 3d case as well.} allows, at least in principle, for AdS$_5$ backgrounds, the conditions imposed by supersymmetry seem to rule out such conformal solutions. This seems to be at odds with the expectation that $\mathcal{N}=2$ SCQCD could be studied in this context, but it should not be surprising since the branes are partially smeared and quantizing the color and flavor fluxes independently seems problematic, as already mentioned.

At a technical level, the reason for the lack of independence between $N_c$ and $N_f$ can be traced back to the couplings between RR fields in the two-derivative supergravity. Among the RR fields at the lowest level that we described in Section \ref{sec:spectrum}, the only ones compatible with the isometries of $\R^{1,3}$ are the four 7d massive vectors, so that the same field is necessarily associated with color and flavor branes. This is reflected in the fact that the scalar $b$ and the gauge field $A$ in \eqref{actionSU2RRISO4} are coupled {\it \`a la} St\"uckelberg, so that effectively one has a massive vector. Moreover, from the uplift formulas of \cite{Cvetic:2000ah} it is easy to see that both the RR scalars and the RR vectors appear in the expression of both two- and four-form field strength in ten dimensions, thus making it impossible to have solutions where only one is vanishing, but not the other, as it can also be seen comparing the expressions \eqref{ND6} and \eqref{ND4} for the number of branes. As discussed in Section \ref{sec:boundarystates} from the worldsheet perspective and at the end of Section \ref{sec:RR_IIBNC} from the point of view of supergravity, what seems to distinguish between color and flavor branes is the strength of the backreaction on the supergravity fields for large values of the coordinate $\rho$. This makes it particularly important to find exact solutions to the PDE \eqref{PDE4d} (or its 3d analogue \eqref{PDE3d}), as they would allow to gain further insight on the configuration of branes as well as the dual field theory by considering the asymptotic expansion of $K$ both far from and near the branes.

\section{Outlook}\label{sec:outlook}

Noncritical superstrings offer a promising  route to holographic dualities for the more ``minimal'' gauge theories outside the universality class of ${\cal N}=4$ SYM, such as ${\cal N}=1$ and ${\cal N}=2$ SQCD. With this motivation, we have thoroughly revisited the spacetime description of noncritical superstring theory, focusing in this paper on the cases
with 6d super Poincar\'e invariance. Building on  previous literature, we have identified the effective supergravities that govern the lowest modes of the noncritical IIA and IIB strings.
A salient point  is that a manifestly supersymmetric description is only possible in seven dimensions. We have recognized the effective theories as the maximally supersymmetric $SO(4)$ and $ISO(4)$ gauged supergravities, respectively for the noncritical IIA and IIB strings. The simplest vacuum solution of these theories is the linear dilaton background, which preserves 6d super Poincar\'e invariance with half-maximal supersymmetry, in agreement with the expected counting of supercharges.
As we have emphasized throughout, the two-derivative approximation is not really justified for the noncritical models, but  can be viewed as the first term of a systematic higher-derivative expansion controlled by $1/k$, where $k \geq 2$ is an integer (the number of NS5 branes in a certain duality frame) parametrizing a family of string backgrounds, with the noncritical case corresponding to $k=2$.  This perspective is supported by the qualitative agreement between the supergravity solutions we discuss (strictly speaking only valid for large $k$) and the linearized backreaction we computed from the worldsheet in terms of the boundary states of the $k=2$ noncritical theory.

We have discussed BPS solutions of the seven-dimensional gauged supergravities and their uplift to ten dimensions by the consistent truncation ansatz. Solutions with only NSNS flux are under complete analytic control, but we have only a partial understanding of the backgrounds with RR flux, which are our  target for holographic applications. We have found an intriguing class of solutions,
parametrized by the solution of a simple-looking PDE,
which merits further study. 
We have argued that this class describes a Hanany-Witten setup with continuous distributions of D-branes which are moreover partially smeared, but it would be interesting to find explicit solutions of the PDE to solidify this interpretation.
In hindsight, 
it is not too surprising that we were not able to find solutions with localized branes. The same issue has been encountered in several other attempts to find holographic duals of four-dimensional gauge theories with eight supercharges~\cite{Polchinski:2000mx,Bertolini:2000dk,Gauntlett:2001ps,Bigazzi:2001aj,Bertolini:2001qa,Billo:2001vg,DiVecchia:2002ks}.

In the search for a supergravity description of $\mathcal{N}=2$ gauge theories it was natural to demand that the $SU(2)_R$ symmetry be realized geometrically, which lead us to focus on a distribution of NS5 branes on a segment. It would be interesting to develop an understanding of the dual CFT and the associated boundary states, as it would allow us to obtain a more precise intuitive picture of what the solutions of Section \ref{sec:RR} exactly describe.
More generally,  while the cigar CFT (T-dual to a distribution of NS5 branes on a circle) has received much attention in the literature, other distributions of NS5 branes appear to be on the same footing, at least from a supergravity perspective. However, not much is known in the literature about the associated worldsheet CFTs -- see \cite{MariosPetropoulos:2005rtq, Fotopoulos:2007rm, Prezas:2008ua, Fotopoulos:2010wc} for comments in this directions.

We are currently pursuing~\cite{wip} an analogous approach to $d=4$ noncritical superstring theories, targeting gauge theories with four supercharges.\footnote{The distributions of NS5 branes studied in \cite{Brennan:2020bju} should be relevant for this purpose.} This case is more promising for a number of reasons. The intuition developed in this paper suggests that the correct description for the $d=4$ noncritical superstring should be in terms of a {\it five}-dimensional gauged supergravity. This is a great technical simplification because
an ansatz with super Poincar\'e invariance in $\R^{1,3}$
leads to ordinary differential equations in the only transverse direction. What's more, there is now a clean separation between color branes filling $\R^{1,3}$
and flavor branes filling the entire 5d spacetime  (recall Table \ref{tab:cigar_branes_6d} in the Introduction)
and as such modify the supergravity action. This should allow to control separately the number of colors and of flavors, and in particular to target  {\it pure} ${\cal N}=1$ SYM theory.
Finally we can hope that the enhançon singularity observed in various attempts to study $\mathcal{N}=2$ gauge theories holographically \cite{Johnson:1999qt, Peet:1999nh, Polchinski:2000mx} should not be problematic in $\mathcal{N}=1$ theories, as already observed in alternative holographic realizations of the two theories \cite{DiVecchia:2002ks}.

From a narrower technical perspective, part of our work can be viewed as an investigation of the worldsheet mechanism behind consistent truncations in supergravity. While consistent truncations associated with AdS backgrounds and the corresponding gauged supergravities have received more attention in the literature due to the great accomplishments of the AdS/CFT correspondence, the presence of RR fluxes notoriously complicates the problem of performing worldsheet computations in such models. Here, on the other hand, we only have NSNS flux while still having a consistent truncation to a gauged supergravity. This allows a thorough investigation of the spectrum from a worldsheet perspective, which makes the selection of the states participating in the truncation significantly more transparent: in our case, all vertex operators are singlets of the bosonic $SU(2)_{k-2}$ CFT associated with the three-sphere in the geometry. It will be interesting to consider other cases in which a similar analysis is also possible, identifying the worldsheet mechanism underlying consistent truncations.

We regard this work as the first installment of a long-term research program. 
The next goal (perhaps more likely to be achieved in models with four supercharges)
is to find  supergravity solutions for the noncritical brane setup
that capture at least the qualitative physics of the dual gauge theories. One would then incorporate higher-derivative corrections as a systematic $1/k$ expansions, where $k$ labels a family of gauge theories that reduces to the target  theory for the noncritical value of $k$. 
The ultimate aspiration is a
full-fledged worldsheet description of the strongly coupled sigma model with RR flux.
The supergravity approach that we have begun to develop here cannot be the final tool for noncritical holography, 
but it should serve as an essential stepping stone.

\vspace{1mm}

\section*{Acknowledgements}

We thank Ofer Aharony, Nikolay Bobev, Davide Cassani, Diego Delmastro, Lorenz Eberhardt, Pietro Benetti Genolini, Emanuel Malek, Dario Martelli, Emil Martinec, Yaron Oz, Adar Sharon, James Sparks, Oscar Varela, Daniel Waldram, Zohar Komargodski for useful discussions. AD acknowledges support from the Mafalda \& Reinhard Oehme Fellowship. The work of LR is supported in part by NSF grant PHY-2210533 and
by the Simons Foundation grant 681267 (Simons Investigator Award).

\newpage

\appendix 

\section{Picture changing and RR sector low-lying states}
\label{app:picture changing}

In Section~\ref{sec:spectrum} we constructed low-lying RR sector states in the $(-\tfrac{1}{2}, -\tfrac{1}{2})$ picture. In Section~\ref{sec:boundarystates} we also need their expression in the $(-\frac{3}{2}, -\frac{1}{2})$, which we construct in this appendix following~\cite{DiVecchia:1997vef, Billo:1998vr}. 

\paragraph{BRST charge.} In the following, it will be easier to work in the BRST formalism. Let us then introduce the BRST charge 
\begin{equation}
\mathcal Q= \mathcal Q_0 + \mathcal Q_1 + \mathcal Q_2 \,, 
\end{equation}
where
\begin{equation}
\begin{aligned}
\mathcal Q_0 & = \oint \frac{\text dz}{2 \pi \ii} c(z) [T^{\text m}(z) + T^{\beta \gamma} + \partial c(z)b(z)] \,, \\
\mathcal Q_1 & = -\frac{1 }{2} \oint \frac{\text dz}{2 \pi \ii} \ex^{\varphi(z)} \eta(z) G^{\text m}(z) \,, \\
\mathcal Q_2 & = \frac{1}{4} \oint \frac{\text dz}{2 \pi \ii} b(z) \eta(z) \partial \eta(z) \ex^{2 \varphi(z)} \,. 
\end{aligned}
\end{equation}
It is convenient to explicitly reintroduce the dependence on all the four polarizations entering \eqref{6d-all-polar} and to rewrite low-lying RR sector states in the $(-\tfrac{1}{2}, -\tfrac{1}{2})$ picture as
\begin{equation}
\mathcal O_{\mathtt s} = \sum_{a,b, \ell, m= \pm} (\mathcal F^{a, b, \ell,m}C)_{\hat \epsilon \hat \tau} \, c \,  \ex^{-\frac{\varphi}{2}} \ex^{\frac{\ii \ell}{2}(\vartheta + a H)} \, V_{\hat \epsilon} \, \tilde c \,  \ex^{-\frac{\tilde \varphi}{2}} \ex^{\frac{\ii m}{2}(\tilde \vartheta - \mathtt s \, b \tilde H)} \, \tilde V_{\hat \tau} \, \ex^{\mathtt j\rho} \, \ex^{\ii q \cdot X} \,, 
\label{RR-12-12}
\end{equation}
where the various components of $\mathcal F^{a,b, \ell, m}C$ can be identified with tensor products of the various polarization spinors entering eq.~\eqref{6d-all-polar} while $\mathtt s = +1$ and $\mathtt s = -1$ denote the type IIA and type IIB noncritical superstring theory, respectively. It is easy to check that requiring the state \eqref{RR-12-12} to be BRST closed implies as expected the mass-shell condition \eqref{massshell_k7d} and the equations of motion for the ``field strengths'' $\mathcal F^{a,b, \ell, m}$\footnote{Here and in the following, to lighten the notation we frequently omit the labels $\ell, m$.}
\begin{equation}
\begin{aligned}
    \ii \,  \slashed q \mathcal F^{a,b} - a(\mathtt j + Q \tfrac{1+a}{2}) \mathcal F^{-a,b} & = 0 \,, \\
    \ii \,   \mathcal F^{a,b} \slashed q + b(\mathtt j + Q \tfrac{1- \mathtt s \, b}{2}) \mathcal F^{a,-b} & = 0 \,. \\
\end{aligned}
\label{6d-eom-F}
\end{equation}

Proceeding along the lines of \cite{Billo:1998vr}, we consider the following ansatz for the low-lying RR sector states \eqref{RR-12-12} in the $(-\frac{3}{2}, -\frac{1}{2})$ picture, 
\begin{equation}
\mathcal O_{\mathtt s}^{\left(-\frac{3}{2},-\frac{1}{2}\right)} = 2 \sqrt 2 \, a_M \sum_{M=0}^\infty W^{(M)} \,, 
\label{other-ansatz}
\end{equation}
where
\begin{equation}
    W^{(0)} = \sum_{a,b,\ell, m = \pm} \left(\mathcal A^{-a, b, \ell, m} C \right)_{\hat \epsilon \hat \tau} \,  c \, \ex^{-\frac{3}{2}\varphi} \, \ex^{\frac{\ii \ell}{2}(\vartheta + a H)} \, V_{\hat \epsilon} \, \tilde c \, \ex^{-\frac{\tilde \varphi}{2} } \, \ex^{\frac{\ii m}{2}(\tilde \vartheta - \mathtt s \, b \tilde H)} \, \tilde V_{\hat \tau} \, \ex^{\mathtt j \rho} \, \ex^{\ii q \cdot X}  \,, 
    \label{W0}
\end{equation}
and for $M \geq 1$
\begin{align}
    W^{(2M)} & = \sum_{a,b,\ell, m = \pm} \left(\mathcal A^{-a,b, \ell, m} C \right)_{\hat \epsilon \hat \tau}  \, \partial^{2M} \xi \dots \partial \xi \, c \, \ex^{(-\frac{3}{2}-2M) \varphi} \, \ex^{\frac{\ii \ell}{2}(\vartheta + a H)} \, V_{\hat \epsilon} \nonumber \\
    & \qquad \times \, \bar \partial^{2M-1} \tilde \eta \dots \bar \partial \tilde \eta \, \tilde \eta \,  \tilde c \, \ex^{(-\frac{1}{2}+2M)\tilde \varphi} \ex^{\frac{\ii m}{2}(\tilde \vartheta - \mathtt s \, b \tilde H)} \, \tilde V_{\hat \tau} \, \ex^{\mathtt j \rho} \, \ex^{\ii q \cdot X} \,, \\[0.2cm]
    W^{(2M-1)} & = \sum_{a,b,\ell, m = \pm} \left(\mathcal B^{a,-b, \ell, m} C \right)_{\hat \epsilon{\hat \tau}}  \, \partial^{2M-1} \xi \dots \partial \xi \, c \, \ex^{(-\frac{1}{2}-2M) \varphi} \, \ex^{\frac{\ii \ell}{2}(\vartheta + a H)} \, V_{\hat \epsilon} \nonumber \\
    & \qquad \times \, \bar \partial^{2M-2} \tilde \eta \dots \bar \partial \tilde \eta \, \tilde \eta \,  \tilde c \, \ex^{(-\frac{3}{2}+2M)\tilde \varphi} \ex^{\frac{\ii m}{2}(\tilde \vartheta - \mathtt s \, b \tilde H)} \, \tilde V_{\hat \tau} \, \ex^{\mathtt j \rho} \, \ex^{\ii q \cdot X}  \,, 
\end{align}
while the numerical coefficients $a_M$ will be specified later. 

\paragraph{BRST invariance.} Let us first verify that the ansatz \eqref{other-ansatz} is BRST invariant. Noticing that 
\begin{equation}
    \partial^{M-1} \eta \dots \partial \eta \, \eta \,  c \, \ex^{(-\frac{1}{2}+M)\varphi} \qquad \text{and} \qquad  \bar \partial^M \tilde \xi \dots \bar \partial \tilde \xi \, \tilde c \, \ex^{(-\frac{3}{2}-M)\tilde \varphi}
    \label{ghost-part-ansatz}
\end{equation}
have conformal dimension $-\frac{5}{8}$ for any $M$, it is easy to check that the commutators with $\mathcal Q_0$ and $\tilde{\mathcal Q}_0$ imply the mass-shell condition \eqref{mass-shell}. The same holds for $W^{(0)}$. Moreover, since the OPEs of $b \eta \partial \eta e^{2 \varphi}$ and $\tilde b \tilde \eta \bar \partial \tilde \eta e^{2 \tilde \varphi}$ respectively with the first and second state in \eqref{ghost-part-ansatz} are regular, the commutators of \eqref{other-ansatz} with $\mathcal Q_2$ and $\tilde{\mathcal Q}_2$ vanishes identically. The commutator 
\begin{align}
[\mathcal Q_1, W^{(0)}] = 0 
\end{align}
also vanishes identically and the only non-trivial commutators are
\begingroup
\allowdisplaybreaks
\begin{align}
[\mathcal Q_1, W^{(2M)}] = & -\frac{(2M)!}{2 \sqrt 2} \sum_{a,b,\ell,m = \pm} \partial^{2M-1} \xi \dots \partial \xi \, c \, \ex^{(-\frac{1}{2}-2M)\varphi} \, \ex^{\frac{\ii \ell}{2}(\vartheta + a H)} V_{\hat \epsilon} \nonumber \\
   & \qquad \left((\slashed q \mathcal A^{-a,b} - \ii \, a( \mathtt j + Q \tfrac{1+a}{2})\mathcal A^{a,b})C \right)_{\hat \epsilon{\hat \tau}} \, \bar \partial^{2M-1} \tilde \eta \dots \bar \partial \tilde \eta \, \tilde \eta \,  \tilde c \nonumber \\
   & \qquad \ex^{(-\frac{1}{2}+2M)\tilde \varphi} \ex^{\frac{\ii m}{2}(\tilde \vartheta -\mathtt s \, b \tilde H)} \, \tilde V_{\hat \tau} \, \ex^{\mathtt j \rho} \, \ex^{\ii q \cdot X} \,, \\
\Bigl[\tilde{\mathcal  Q}_1, W^{(2M-1)} \Bigr] = & -\frac{1}{2 \sqrt 2 (2M-1)!} \sum_{a,b, \ell,m = \pm}  \partial^{2M-1} \xi \dots \partial \xi \, c \,  \ex^{(-\frac{1}{2}-2M)\varphi} \ex^{\frac{\ii \ell}{2}(\vartheta + a H)} \, V_{\hat \epsilon} \nonumber \\
    & \qquad \left( (\mathcal B^{a,-b} \slashed q  + \ii \, b (\mathtt j+Q\tfrac{1-\mathtt s \, b}{2}) \, \mathcal B^{a,b})C \right)_{\hat \epsilon{\hat \tau}}\, \bar \partial^{2M-1} \tilde \eta \dots \bar \partial \tilde \eta \, \tilde \eta \,  \tilde c \nonumber \\
    & \qquad \ex^{(-\frac{1}{2}+2M)\tilde \varphi} \, \ex^{\frac{\ii m}{2}(\tilde \vartheta - \mathtt s \, b \tilde H)} \, \tilde V_{\hat \tau} \, \ex^{\mathtt j\rho} \, \ex^{\ii q \cdot X} \,, 
\end{align}
\endgroup
for $M\geq 1$ and 
\begingroup
\allowdisplaybreaks
\begin{align}
\Bigl[\tilde{\mathcal  Q}_1, W^{(2M)} \Bigr] = & \, \frac{1}{2 \sqrt 2 (2M)!} \sum_{a,b, \ell,m = \pm} \partial^{2M} \xi \dots \partial \xi \, c \,  \ex^{(-\frac{3}{2}-2M)\varphi} \, \ex^{\frac{\ii \ell}{2}(\vartheta + a H)} \, V_{\hat \epsilon} \nonumber \\
    & \qquad \left( (\mathcal A^{-a,b} \slashed q  - \ii \, b(\mathtt j+Q \tfrac{1+ \mathtt s \, b}{2}) \mathcal A^{-a,-b})C \right)_{\hat \epsilon{\hat \tau}} \, \bar \partial^{2M} \tilde \eta \dots  \bar \partial \tilde \eta \, \tilde \eta \,  \tilde c \nonumber \\
    & \qquad \ex^{(\frac{1}{2}+2M)\tilde \varphi} \, \ex^{\frac{\ii m}{2}(\tilde \vartheta - \mathtt s \, b \tilde H)} \, \tilde V_{\hat \tau} \, \ex^{\mathtt j\rho} \,  \ex^{\ii q \cdot X} \,, \\
\Bigl[\mathcal  Q_1, W^{(2M+1)} \Bigr] = & \, \frac{(2M+1)!}{2 \sqrt 2} \sum_{a,b, \ell,m = \pm} \partial^{2M} \xi \dots \partial \xi \, c \,  \ex^{(-\frac{3}{2}-2M)\varphi} \ex^{\frac{\ii \ell}{2}(\vartheta + a H)} \, V_{\hat \epsilon} \nonumber \\ 
    & \qquad \left( (\slashed q \mathcal B^{a,-b} + \ii \, a (\mathtt j + Q \tfrac{1-a}{2}) \mathcal B^{-a, -b})C \right)_{\hat \epsilon{\hat \tau}} \, \bar \partial^{2M} \tilde \eta \dots  \bar \partial \tilde \eta \, \tilde \eta \,  \tilde c \nonumber \\
    & \qquad \ex^{(\frac{1}{2}+2M)\tilde \varphi} \, \ex^{\frac{\ii m}{2}(\tilde \vartheta - \mathtt s \, b \tilde H)}\, \tilde V_{\hat \tau} \, \ex^{\mathtt j\rho} \,  \ex^{\ii q \cdot X} \ ,
\end{align}
\endgroup
for $M\geq 0$. Choosing the coefficients $a_M$ as 
\begin{equation}
    a_{M+1} = - \frac{a_M}{M!(M+1)!} \,, 
\end{equation}
we find 
\begin{align}
    a_{2M-1} \Bigl[\tilde{\mathcal  Q}_1, W^{(2M-1)} \Bigr]  + a_{2M}  [\mathcal Q_1, W^{(2M)}] &= 0 \,,  \\
      a_{2M} \Bigl[\tilde{\mathcal  Q}_1, W^{(2M)} \Bigr]  + a_{2M+1}  [\mathcal Q_1, W^{(2M+1)}] &= 0 \,, 
\end{align}
provided that 
\begin{equation}
\begin{aligned}
    \slashed q\mathcal A^{-a,b} - \ii \, a( \mathtt j + Q \tfrac{1+a}{2})\mathcal A^{a,b} &= \mathcal B^{a,-b} \slashed q  + \ii \, b (\mathtt j+Q \tfrac{1- \mathtt s \, b}{2}) \, \mathcal B^{a,b} \,, \\
    \mathcal A^{-a,b} \slashed q  - \ii \, b(\mathtt j + Q \tfrac{1+ \mathtt s \, b}{2}) \mathcal A^{-a,-b} &= \slashed q \mathcal B^{a,-b} + \ii \, a (\mathtt j + Q \tfrac{1-a}{2}) \mathcal B^{-a, -b} \,. 
\end{aligned}
\label{BRST-closed-32-12}
\end{equation}
Hence, eqs.~\eqref{mass-shell} and \eqref{BRST-closed-32-12} are sufficient for the state \eqref{other-ansatz} to be BRST closed. 

\paragraph{Picture change.} We should also require that the vertex operator \eqref{other-ansatz} is related to \eqref{RR-12-12} by picture changing, i.e.~that 
\begin{equation}
    \mathcal O_{\mathtt s} = [\mathcal Q + \tilde{\mathcal Q}, \xi(z) \mathcal O_{\mathtt s}^{\left(-\frac{3}{2},-\frac{1}{2}\right)}] \,. 
\end{equation}
Only $W^{(0)}$ contributes and one obtains the condition 
\begin{equation}
     \mathcal F^{a,b} = \Bigl( \ii \,  \slashed q \mathcal A^{-a,b} + \, a (\mathtt j + Q \tfrac{1+a}{2})\mathcal A^{a,b} \Bigr) \,.
     \label{F=kA-6d}
\end{equation}

\subsection{Interpreting the conditions}

\paragraph{7d language.} Let us repackage eqs.~\eqref{6d-eom-F}, \eqref{BRST-closed-32-12} and \eqref{F=kA-6d} in the more compact 7d notation introduced in Section~\ref{sec:spectrum}. We decompose $\mathcal F$, $\mathcal A$ and $\mathcal B$ into a sum over polarization and write {\it e.g.}
\begin{equation}
 \mathcal F = \sum_{a,b = \pm} \mathcal F^{a,b}  \,, \qquad  \mathcal F^{a,b} = \frac{1 + a \gamma^7_{6d}}{2} \mathcal F \frac{1 - b \gamma^7_{6d}}{2} \,,
\end{equation}
It is easy to check that
\begin{equation}
    a(\mathtt j+Q\tfrac{1-a}{2}) \mathcal F^{a,b} = \left( (\mathtt j + \tfrac{Q}{2}) \gamma^7_{6d} - \tfrac{Q}{2} \right) \frac{1 + a \gamma^7_{6d}}{2} \mathcal F \frac{1 - b \gamma^7_{6d}}{2}
\end{equation}
and using that $\{\slashed q, \gamma^7_{6d} \} = 0$, eqs.~\eqref{6d-eom-F}, \eqref{BRST-closed-32-12} and \eqref{F=kA-6d} can be rewritten as 
\begin{align}
     \left( \ii \, {\slashed p} - \tfrac{Q}{2} \right) \mathcal F & = \mathcal F \left( \ii \, {\slashed p} - \mathtt s\tfrac{Q}{2} \right) = 0  \,, \label{Feom} \\
    (\ii \, {\slashed p} +\tfrac{Q}{2}) \mathcal A &=   \mathcal B (\ii \, {\slashed p} +\mathtt s\tfrac{Q }{2}) \,, \label{AD1} \\ 
    \mathcal A (\ii \, {\slashed p} +\mathtt s\tfrac{Q}{2})  &=    (\ii \, {\slashed p} +\tfrac{Q}{2}) \mathcal B \,,  \label{AD2} \\
    \mathcal F &= \left( \ii \, {\slashed p} + \tfrac{Q}{2} \right) \mathcal A \,. \label{FA}
\end{align}

\paragraph{Identification of the potentials $\mathcal A$ and $\mathcal B$.} From equations \eqref{Feom}--\eqref{FA}, we see a good ansatz is
\es{DAansatz}{
\mathcal{B}=\mathtt{s}\mathcal{A}\,.
}
Replacing this into \eqref{AD2} leads to the condition
\es{Agauge}{
\slashed{{p}}\mathcal{A}=\mathtt{s}\mathcal{A}\slashed{{p}}\,,
}
which is now a constraint on the differential forms arising in the decomposition of $\mathcal{A}$. Moreover, note that the ansatz \eqref{DAansatz} makes the two equations \eqref{AD1} and \eqref{AD2} equivalent and is hence consistent. We also have 
\es{F=dA=Ad}{
\mathcal{F}=\left(\ii\,\slashed{{p}}+\tfrac{Q}{2}\right)\mathcal{A}=\mathcal{A}\left(\mathtt{s}\,\ii\,\slashed{{p}}+\tfrac{Q}{2}\right)\,,
}
which determine the forms in $\mathcal{F}$ in terms of those in $\mathcal{A}$ and are compatible by virtue of \eqref{Agauge}. Note also that \eqref{Feom} requires that ${p}^2=\tfrac{Q^2}{4}$ for $\mathcal{F}$ as well as for $\mathcal{A}$, so we shall always use this. To have a more concrete understanding of these constraints we shall now analyze them component by component, which requires to consider the two values of $\mathtt{s}$ independently. Note that we can decompose $\mathcal{A}$ using the Fierz identity \eqref{7dFierz} in a completely analogous way to $\mathcal{F}$, just replacing the differential forms $F_{(n)}$ appearing in \eqref{7dFierz} with a new set of forms $A_{(n)}$, $n=0,\ldots,3$.

\paragraph{Type IIA: $\boldsymbol{\mathtt{s}=+1}$.} In this case the condition \eqref{Agauge} gives
\es{Agauge_IIA}{
\diff \mathcal{A}_{(1)}=0\,,\quad
\star\diff \star \mathcal{A}_{(2)}=0\,,\quad
\diff \mathcal{A}_{(3)}=0\,,
}
which combined with \eqref{F=dA=Ad} lead to
\es{Feom_IIA}{
\mathcal{F}_{(0)}&=\frac{Q}{2}\mathcal{A}_{(0)}-\star\diff\star \mathcal{A}_{(1)}\,,\quad 
\mathcal{F}_{(1)}=\frac{Q}{2}\mathcal{A}_{(1)}+\diff \mathcal{A}_{(0)}\,,\\
\mathcal{F}_{(2)}&=\frac{Q}{2}\mathcal{A}_{(2)}-\star \diff \star \mathcal{A}_{(3)}\,,\quad
\mathcal{F}_{(3)}=\frac{Q}{2} \mathcal{A}_{(3)}+\diff \mathcal{A}_{(2)}\,.
}
One can check that the equations above, combined with the mass-shell condition, are enough to guarantee that \eqref{Feom} are satisfied. We can also go one step further and notice that we can further reduce the number of degrees of freedom carried by $\mathcal{A}$ in this description. To see this, let
\es{}{
\tilde{\mathcal{A}}_1=\mathcal{A}_{(1)}+\frac{2}{Q}\diff \mathcal{A}_{(0)}\,,\quad
\tilde{\mathcal{A}}_3=\mathcal{A}_{(3)}+\frac{2}{Q}\diff \mathcal{A}_{(2)}\,,
}
which using again the mass-shell condition as well as \eqref{Agauge_IIA} allow to rewrite the conditions \eqref{Feom_IIA} as
\es{}{
\mathcal{F}_{(0)}&=-\star \diff\star \tilde{\mathcal{A}}_{(1)}\,,\quad
\mathcal{F}_{(1)}=\frac{Q}{2}\tilde{\mathcal{A}}_{(1)}\,,\\
\mathcal{F}_{(2)}&=-\star\diff\star\tilde{\mathcal{A}}_{(3)}\,,\quad
\mathcal{F}_{(3)}=\frac{Q}{2}\tilde{\mathcal{A}}_{(3)}\,,
}
which are equivalent to just setting $\mathcal{A}_{(0)}=\mathcal{A}_{(2)}=0$ in \eqref{Feom_IIA}. This is clearly compatible with \eqref{Agauge_IIA}, since we are changing closed forms by exact terms. Finally, note that using the mass-shell condition combined with the conditions \eqref{Agauge_IIA}, we can observe that $\diff \tilde{\mathcal{A}}_1=\diff\tilde{\mathcal{A}}_3=0$ can be solved simply by
\es{}{
\tilde{\mathcal{A}}_{(1)}=\frac{4}{Q^2}\diff \mathcal{F}_{(0)}\,,\quad
\tilde{\mathcal{A}}_{(3)}=\frac{4}{Q^2}\diff \mathcal{F}_{(2)}\,.
}

\paragraph{Type IIB: $\boldsymbol{\mathtt{s}=-1}$.} In this case the condition \eqref{Agauge} gives
\es{Agauge_IIB}{
\diff \mathcal{A}_{(0)}=0\,,\quad
\star\diff \star \mathcal{A}_{(1)}=0\,,\quad
\diff \mathcal{A}_{(2)}=0\,,\quad
\star\diff \star \mathcal{A}_{(3)}=0\,,
}
where the first can be replaced with $\mathcal{A}_{(0)}=0$, and combined with \eqref{F=dA=Ad} this leads to
\es{Feom_IIB}{
\mathcal{F}_{(0)}&=0\,,\quad 
\mathcal{F}_{(1)}=\frac{Q}{2}\mathcal{A}_{(1)}-\star \diff \star \mathcal{A}_{(2)}\,,\\
\mathcal{F}_{(2)}&=\frac{Q}{2}\mathcal{A}_{(2)}-\diff \mathcal{A}_{(1)}\,,\quad
\mathcal{F}_{(3)}=\frac{Q}{2} \mathcal{A}_{(3)}-\star \diff \mathcal{A}_{(3)}\,.
}
One can check that the equations above, combined with the mass-shell condition, are enough to guarantee that \eqref{Feom} are satisfied. Also in this case, one can further reduce the number of degrees of freedom carried by $\mathcal{A}$ in this description. To see this, let
\es{}{
\tilde{\mathcal{A}}_{(2)}=\mathcal{A}_{(2)}-\frac{2}{Q}\diff \mathcal{A}_{(1)}\,,\quad
\tilde{\mathcal{A}}_{(3)}=\mathcal{A}_{(3)}\,,
}
which using again the mass-shell condition as well as \eqref{Agauge_IIB} allow to rewrite the conditions \eqref{Feom_IIB} as
\es{}{
\mathcal{F}_{(0)}&=0\,,\quad
\mathcal{F}_{(1)}=-\star \diff \star\tilde{\mathcal{A}}_{(2)}\,,\\
\mathcal{F}_{(2)}&=\frac{Q}{2}\tilde{\mathcal{A}}_{(2)}\,,\quad
\mathcal{F}_{(3)}=\frac{Q}{2} \tilde{\mathcal{A}}_{(3)}-\star \diff \tilde{\mathcal{A}}_{(3)}\,,
}
which are equivalent to just setting $\mathcal{A}_{(1)}=0$ in \eqref{Feom_IIB}, which clearly is compatible with \eqref{Agauge_IIB} since we are changing closed forms by exact terms. Finally, note that using the mass-shell condition combined with the conditions \eqref{Agauge_IIB}, we can observe that $\diff \tilde{\mathcal{A}}_2=0$ can be solved simply by
\es{}{
\tilde{\mathcal{A}}_{(2)}=-\frac{4}{Q^2}\diff \mathcal{F}_{(1)}\,.
}

\section{Group theory}\label{app:grouptheory}

\subsection{SO(5) vs USp(4)}\label{app:so5usp4}

In the formulation of maximal 7d supergravity, it is often useful to be able to convert from $SO(5)$ to $USp(4)$ indices when working with representations of the R-symmetry group. This is particularly true when dealing with fermions, which transform in spinorial representations of $SO(5)$, or ordinary representations of its double cover $USp(4)$. We use indices $M,N =1,\ldots, 5$ for the fundamental of $SO(5)$ and $a,b=1,\ldots,4$ for the fundamental of $USp(4)$, and adopt the same notation and conventions as \cite{Samtleben:2005bp}, which we summarize and expand here for convenience. $USp(4)$ indices are raised and lowered using complex conjugation, which for pseudoreal representations is equivalent to acting with the invariant two-form $\Omega$ satisfying 
\begin{align}
    \Omega_{ab}=\Omega_{[ab]}\,,\quad (\Omega_{ab})^*=\Omega^{ab}\,,\quad \Omega_{ab}\Omega^{cb}=\delta^{c}_a\,,
\end{align}
and in particular we pick
\begin{align}
    \Omega_{ab}=\Omega^{ab}=\ii \sigma^2\otimes \sigma^1=
    \begin{pmatrix}
        0 & 0 & 0 & +1\\
        0 & 0 & +1 & 0\\
        0 & -1 & 0 & 0\\
        -1 & 0 & 0 & 0\\
    \end{pmatrix}_{ab}\,.
\end{align}
To convert $SO(5)$ into $USp(4)$ indices we introduce a set of gamma matrices $(\gammaflavor_M)_a^{\,\,\,b}$, which we define as (with this specific position for the $USp(4)$ indices)
\begin{align}
    \gammaflavor_A=\sigma^A\otimes \sigma^3\,,\quad (A=1,2,3)\,,\quad
    \gammaflavor_4=1\otimes \sigma^1\,,\quad \gammaflavor_5=1\otimes \gammaflavor^2\,.
\end{align}
We note that their version with both indices raised, $(\gammaflavor_M)^{ab}=\Omega^{ac}(\gammaflavor_M)_c^{\,\,\,b}$, is antisymmetric and $\Omega$-traceless in the $USp(4)$ indices: 
\begin{align}\label{sigmaantitraceless}
    (\gammaflavor_M)^{ab}+(\gammaflavor_M)^{ba}=0=\Omega_{ab}(\gammaflavor_M)^{ab}\,.
\end{align}
Besides the obvious Clifford algebra
\begin{align}
    (\gammaflavor_M)_a^{\,\,\,c} (\gammaflavor_M)_a^{\,\,\,c}(\gammaflavor_N)_c^{\,\,\,b}=2\delta_{MN}\delta_a^{\,\,\,b}\,,
\end{align}
they satisfy
\begin{align}\label{completenessSigma}
    (\gammaflavor_M)_{ab} (\gammaflavor^M)_{cd}= 2\delta_a^{\,\,\,c} \delta_b^{\,\,\,d}-2\delta_a^{\,\,\,d} \delta_b^{\,\,\,c}- \Omega_{ab}\Omega^{cd}\,,
\end{align}
where the $SO(5)$ index $M$ is raised with the flat Euclidean metric. To give a simple example, a vector
\begin{align}
    v^M=(v^1,\ldots,v^5)\,,
\end{align}
can be expressed as a $4\times 4$ matrix with $USp(4)$ indices as
\begin{align}
    v^{ab}=v^M(\gammaflavor_M)^{ab}=
   \frac{1}{2} \begin{pmatrix}
        0 & -(v^1+\ii v^2) & v^4+\ii v^5 & v^3\\
        v^1+\ii v^2 & 0 & -v^3 & v^4-\ii v^5\\
        -(v^4+\ii v^5) & v^3 & 0 & v^1-\ii v^2\\
        -v^3 & -(v^4-\ii v^5) & -(v^1-\ii v^2) & 0
    \end{pmatrix}\,,
\end{align}
which is antisymmetric and $\Omega$-traceless according to \eqref{sigmaantitraceless}, thus giving the correct number (5) of degrees of freedom. Using \eqref{completenessSigma}, one also has the inverse relation
\begin{align}
    v^M=\frac{1}{2}(\gammaflavor^M)_{ab}v^{ab}\,.
\end{align}

\subsection{SL(5) $\to$ SL(4) $\otimes$ SO(1,1)}\label{app:SL5toSL4}

The choice of gauge group in both 7d theories breaks $SL(5)$ invariance as $SL(5)\to SL(4)\otimes SO(1,1)$: the $SO(4)$ part that is the common maximal compact subgroup to both gauge groups is a subgroup of $SL(4)$ that is gauged. In terms of generators
\begin{align}
    (t_{MN})_P^{\,\,\,Q}=\delta_{MP}\delta_N^{\,\,\,Q}-\tfrac{1}{5}\delta_{MN}\delta_P^{\,\,\,Q}\,,
\end{align}
we identify the $SL(4)$ generators as those obtained restricting the values of the indices $M,N$ to $M,N=1,\ldots,4$. The remaining generators are $t^{i}\equiv t_{\alpha 0}$ with $i=1,\ldots,4$ transforming in the fundamental representation of $SL(4)$ and $t^0\equiv 5\,t_{55}$ generating $SO(1,1)$. The non-compact $\mathbb{R}^4$ generated by the $t^{i}$ is only gauged in the $ISO(4)$ theory. Finally, the generator $t^0$ is associated with the 7d dilaton field $\phi_0$.

\subsection{The SO(4) gauged symmetry}\label{app:so4}

The 7d theories that we consider share a common compact subgroup of the gauge group: an $SO(4)$ that is embedded in $SL(5)$ in a way that naturally follows from \eqref{sl5toso4}, for which we use indices $i,j=1,\ldots,4$. Its generators are
\begin{align}
    (T_{ij})_{kl}=\delta_{ik}\delta_{jl}-\d_{il}\d_{jk}\,,
\end{align}
satisfying
\begin{align}
[T_{ij},T_{kl}]=-\d_{ik}T_{jl}-\d_{jl}T_{ik}+\d_{il}T_{jk}+\d_{jk}T_{il}\,,
\end{align}
and since $SO(4)=SU(2)_+\otimes SU(2)_-$ we can identify two sets of commuting $SU(2)$ generators
\begin{align}
    J^{\pm}_1=T_{13}\pm T_{42}\,,\quad
    J^{\pm}_2=\pm T_{14}+ T_{23}\,,\quad
    J^{\pm}_3=T_{14}\pm T_{34}\,,
\end{align}
which satisfy
\begin{align}
    [J^{\pm}_i,J^{\pm}_j]=-2\epsilon_{ijk}J^{\pm}_k\,,
    \qquad 
    [J^{\pm}_i,J^{\mp}_j]=0\,.
\end{align}
The diagonal $SU(2)_D$ subgroup of $SO(4)$ is generated by
\begin{align}
    J^D_i=J^+_i+J^-_i\,,
\end{align}
and corresponds to $SO(3)\simeq SU(2)/\mathbb{Z}_2$ rotations in the directions 1,2,3 (with the Cartan $J^D_3$ generating rotations in the 1-2 plane).

\section{Type II A/B conventions}\label{app:typeII}

When considering solutions of type IIA/B supergravity, we always work in the string frame. Our conventions for the equations of motion and supersymmetry variations are those of \cite{Hamilton:2016ito}, which we present here in a more physicist-friendly notation for the reader's convenience, alongside with some general conventions used here and throughout the paper. Lorentz indices for spacetimes of various dimensions have already appeared in various places and we remind the reader that we have used:
\begin{itemize}
    \item $\alpha,\beta,\ldots \in \{ 0,\ldots,5\}$ for 6d indices,
    \item $\mu,\nu,\ldots \in \{ 0,\ldots, 6\}$ for 7d indices,
    \item $\hat \mu, \hat \nu,\ldots \in \{ 0,\ldots, 7\}$ for 8d indices,
\end{itemize}
while here we are going to use
\es{}{
A,B,\ldots \in \{ 0,\ldots, 9 \}\,,
}
for ten-dimensional indices.

\subsection{General conventions}

We write the actions in terms of the ten-dimensional Newton's constant
\es{}{
16\pi\,G^{(10)}_N=2\kappa^2_{10}=(2\pi)^7\,(\alpha')^4\,,
}
in terms of the Regge slope $\alpha'=\ell_s^2$. When quantizing the flux associated with a D$p$-branes, we consider the magnetic RR flux $F_{8-p}$, which we integrated over a compact cycle $\Sigma_{8-p}$ to define the number $N_{\Dp}$ of $\Dp$-branes via
\es{}{
N_{\Dp}=\frac{1}{2\kappa^2_{10}T_{Dp}}\int_{\Sigma_{8-p}}F_{8-p}=\frac{1}{(2\pi\ell_s)^{p-7}\,g_s}\int_{\Sigma_{8-p}}F_{8-p}\,,
}
in terms of the $\Dp$-brane tension $T_{\Dp}=[(2\pi\ell_s)^p\,g_s\,\ell_s]^{-1}$ and the string coupling $g_s$. On the other hand, the tension for NS5 branes has an additional factor of $g_s$, $T_{NS5}=[(2\pi\ell_s)^p\,g_s^2\,\ell_s]^{-1}$, and the flux quantization condition reads
\es{}{
N_{\NS5}=\frac{1}{2\kappa^2_{10}T_{\NS5}}\int_{\Sigma_{3}}H_3=\frac{1}{(2\pi\ell_s)^2}\int_{\Sigma_3}H_3\,.
}

Given a $p$-form $\omega_p$, we introduce the contractions
\es{}{
|\omega_p|^2\equiv \frac{1}{p!}(\omega_p)_{A_1\ldots A_p}(\omega_p)^{A_1\ldots A_p}\,,\quad
\langle \omega_p,\omega_p\rangle_{AB}\equiv \frac{1}{(p-1)!}(\omega_p)_{A C_2\ldots C_p}(\omega_p)_{B}^{\,\,\,C_2\ldots C_p}\,,
}
while for their contractions with elements of the Clifford algebra we introduce ($k<p$)
\es{}{
(\slashed{\omega}_p)_{A_1\ldots A_k}\equiv \frac{1}{(p-k)!}(\omega_p)_{A_1\ldots A_k B_{k+1}\ldots B_p}\Gamma^{B_{k+1}\ldots B_p}\,,
}
where $\Gamma$ are Clifford algebra elements and we use $\Gamma_{A}$ for 10d gamma matrices, with $\Gamma_{\star}$ the chirality matrix.

\subsection{Type IIA}\label{app:typeIIAconventions}

The bosonic action for type IIA supergravity in our conventions reads
\es{SIIA}{
S^{(IIA)}&=S^{(IIA)}_{NSNS}+S^{(IIA)}_{RR}+S^{(IIA)}_{CS}\,,\\
S^{(IIA)}_{NSNS}&=\frac{1}{2\kappa^2_{10}}\int \diff^{10}x\,\sqrt{-g}\,\ex^{-2\Phi}\,\left[R+4|\diff\Phi|^2-\frac{1}{2}|H_3|^2\right]\,,\\
S^{(IIA)}_{RR}&=-\frac{1}{4\kappa^2_{10}}\int \diff^{10}x\,\sqrt{-g}\,\left[|F_2|^2+|\tilde{F}_4|^2\right]\,,\\
S^{(IIA)}_{CS}&=-\frac{1}{4\kappa^2_{10}}\int B_2\wedge F_4\wedge F_4\,,
}
in terms of the dilaton $\Phi$ and the field strengths
\es{}{
H_3=\diff B_2\,,\quad
F_2=\diff C_1\,,\quad
F_4=\diff C_3\,,
}
while the gauge-invariant combination entering the equations of motion is
\es{}{
\tilde{F}_4=F_4-C_1\wedge H_3\,.
}
The Einstein and dilaton equations of motion can be written as
\es{}{
R_{AB}=&-2\de_A\Phi\de_B\Phi+\frac{1}{2}\langle H_3,H_3\rangle_{AB}\\
&+\frac{1}{2}\ex^{2\Phi}\left[\langle F_2,F_2\rangle_{AB}-\frac{1}{2}g_{AB}|F_2|^2+\langle \tilde{F}_4,\tilde{F}_4\rangle_{AB}-\frac{1}{2}g_{AB}|\tilde{F}_4|^2)\right]\,,\\
\Box\Phi=&2|\diff\Phi|^2-\frac{1}{2}|H_3|^2+\frac{1}{4}\ex^{2\Phi}\,\left[3\,|F_2|^2+|\tilde{F}_4|^2\right]\,,
}
while the Maxwell equations for the fluxes read
\es{}{
\diff\star F_2&=H_3\wedge\star \tilde{F}_4\,,\\
\diff \star \tilde{F}_4&=-H_3\wedge\tilde{F}_4\,,\\
\diff\left[\ex^{-2\Phi}\star H_3\right]&=\frac{1}{2}\tilde{F}_4\wedge\tilde{F}_4-F_2\wedge\star \tilde{F}_4\,.
}
The conditions for supersymmetry, which come from the vanishing of the gravitino and dilatino supersymmetry variations, are
\es{}{
0&=\left[\nabla_A-\frac{1}{4}(\slashed{H}_3)_A\,\Gamma_\star+\frac{1}{8}\ex^{\Phi}\left(-\slashed{F}_2\,\Gamma_\star+\slashed{\tilde{F}}_4\right)\,\Gamma_A\right]\epsilon\,,\\
0&=\left[\slashed{\partial}\Phi\,\Gamma_\star-\frac{1}{2}\slashed{H}_3-\frac{1}{4}\ex^{\Phi}\left(3\slashed{F}_2+\slashed{\tilde{F}}_4\,\Gamma_\star\right)\right]\,\epsilon\,,
}
where $\epsilon$ is a 10d Dirac spinor.

All the solutions we are interested in have 
\es{0wedgeIIA}{
\tilde{F}_4\wedge\tilde{F}_4=H_3\wedge \tilde{F}_4=0\,,
}
which facilitates the introduction of dual potentials. In particular, in the main text we consider solutions describing a system of NS5-D4-D6 branes, for which the magnetic potentials are $B_2$, $C_3$ and $C_1$, respectively, but it is useful to consider the associated electric potentials too. We start with the electric potential $C_5$ for D4 branes, which we introduce observing that under the assumptions \eqref{0wedgeIIA} we have $\diff\star \tilde{F}_4=0$, which locally allows us to write
\es{}{
\star \tilde{F}_4\equiv F_6=\diff C_5\,.
}
We choose to define the other electric pontentials in terms of $C_5$. The electric potential $C_7$ associated with D6 branes can be introduced via
\es{}{
\star F_2-H_3\wedge C_5\equiv F_8=\diff C_7\,,
}
while the electric potential $B_6$ for NS5 branes is defined by
\es{}{
\ex^{-2\Phi}\star H_3-F_2\wedge C_5\equiv H_7=\diff B_6\,.
}

\subsection{Type IIB}\label{app:typeIIBconventions}

The bosonic action for type IIB supergravity in our conventions reads
\es{SIIB}{
S^{(IIB)}&=S^{(IIB)}_{NSNS}+S^{(IIB)}_{RR}+S^{(IIB)}_{CS}\,,\\
S^{(IIB)}_{NSNS}&=\frac{1}{2\kappa^2_{10}}\int \diff^{10}x\,\sqrt{-g}\,\ex^{-2\Phi}\,\left[R+4|\diff\Phi|^2-\frac{1}{2}|H_3|^2\right]\,,\\
S^{(IIB)}_{RR}&=-\frac{1}{4\kappa^2_{10}}\int \diff^{10}x\,\sqrt{-g}\,\left[|F_1|^2+|\tilde{F}_3|^2+\frac{1}{2}|\tilde{F}_5|^2\right]\,,\\
S^{(IIB)}_{CS}&=-\frac{1}{4\kappa^2_{10}}\int C_4\wedge H_3\wedge F_3\,,
}
in terms of the dilaton $\Phi$ and the fields strengths
\es{}{
H_3=\diff B_2\,,\quad
F_1=\diff C_0\,,\quad
F_3=\diff C_2\,,\quad
F_5=\diff C_4\,,
}
while the gauge invariant combinations that enter the equations of motion are
\es{}{
\tilde{F}_3=F_3-C_0\,H_3\,,\quad
\tilde{F}_5=F_5-\frac{1}{2}H_3\wedge C_2+\frac{1}{2}F_3\wedge B_2\,,
}
and moreover we demand that $\tilde{F}_5$ is self-dual:
\es{}{
\star\tilde{F}_5=\tilde{F}_5\,.
}
The Einstein and dilaton equations of motion can be written as
\es{}{
R_{AB}=&-2\de_A\Phi\de_B\Phi+\frac{1}{2}\langle H_3,H_3\rangle_{AB}\\
&+\frac{1}{2}\ex^{2\Phi}\left[\langle F_1,F_1\rangle_{AB}-\frac{1}{2}g_{AB}|F_1|^2+\langle \tilde{F}_3,\tilde{F}_3\rangle_{AB}-\frac{1}{2}g_{AB}|\tilde{F}_3|^2+\frac{1}{2}\langle \tilde{F}_5,\tilde{F}_5\rangle_{AB}\right]\,,\\
\Box\Phi=&2|\diff\Phi|^2-\frac{1}{2}|H_3|^2+\frac{1}{2}\ex^{2\Phi}\,\left[2|F_1|^2+|\tilde{F}_3|^2\right]\,,
}
while the Maxwell equations for the fluxes read
\es{}{
\diff \star F_1&=-H_3\wedge\star\tilde{F}_3\,,\\
\diff \star \tilde{F}_3&=-H_3\wedge\tilde{F}_5\,,\\
\diff \star \tilde{F}_5&=H_3 \wedge\tilde{F}_3\,,\\
\diff \left[\ex^{-2\Phi}\star H_3\right]&=\tilde{F}_3\wedge\tilde{F}_5+F_1\wedge\star \tilde{F}_3\,.
}
To write the conditions for supersymmetry, it is convenient to write
\es{eps12}{
\epsilon=
\begin{pmatrix}
    \epsilon_1\\
    \epsilon_2
\end{pmatrix}\,,
}
where $\epsilon_{1,2}$ are 10d spinors. In terms of this, we can express the conditions for supersymmetry as
\es{}{
0&=\left[\nabla_{A}+\frac{1}{4}(\slashed{H}_3)_{A}\,\sigma^3+\frac{1}{8}\ex^{\Phi}\left(\ii\,\slashed{F}_1\,\sigma^2-\slashed{\tilde{F}}_3\,\sigma^1+\frac{\ii}{2}\slashed{\tilde{F}}_5\,\sigma^2\right)\Gamma_{A}\right]\,\epsilon\,,\\
0&=\left[\slashed{\partial}\Phi+\frac{1}{2}\slashed{H}_3\,\sigma^3-\frac{1}{2}\ex^{\Phi}\left(2\ii\,\slashed{F}_1\,\sigma^2-\slashed{\tilde{F}}_3\,\sigma^1\right)\right]\epsilon\,,
}
where the Pauli matrices $\sigma^i$, $i=1,2,3$, act on the two components of $\epsilon$ in \eqref{eps12}.

In the study of explicit solutions, we find it useful to introduce two four-form potentials $C_4^{\el}$ and $C_4^{\ma}$, respectively electric and magnetic. In terms of these we have $F_5^{\el}=\diff C_4^{\el}$ and $F_5^{\ma}=\diff C_4^{\ma}$, and owing to the fact that our solutions satisfy
\es{}{
F_3\wedge F_5^{\ma}=H_3\wedge F_5^{\ma}=F_5^{\ma}\wedge F_5^{\ma}=F_5^{\el}\wedge F_5^{\el}=0\,,
}
as well as $\tilde{F}_3=F_3$ since $C_0=0$, we can introduce
\es{}{
\tilde{F}_5^{\el}=F_5^{\el}\,,\quad \tilde{F}_5^{\ma}=F_5^{\ma}-H_3\wedge C_2\,,
}
in terms of which we finally have the self-dual combination $\tilde{F}_5$ entering the equations of motion and Killing spinor equations as
\es{}{
\tilde{F}_5=\tilde{F}_5^{\el}+\tilde{F}_5^{\ma}\,,
}
since $\tilde{F}_5^{\el}=\star \tilde{F}_5^{\ma}$. In terms of these objects, Maxwell's equations for the four-form potential can be written as
\es{}{
\diff \tilde{F}_5^{\el}=0\,,\quad
\diff \tilde{F}_5^{\ma}=H_3\wedge F_3\,.
}

\section{Solutions with smeared branes}\label{app:smearedsol}

In this Appendix we present explicit solutions of ten-dimensional type IIA and type IIB supergravity which describe an intersecting system of NS5-D4-D6 branes (in IIA) or NS5-D3-D5 branes (in IIB). In these solutions the D-branes are completely smeared in the $\R^4$ transverse to the NS5 branes, in a way that respects the isometries of the specific arrangement of NS5 branes considered. These are not directly interesting for holography, since for that purpose one is interested in solutions where branes are (sufficiently) localized. Solutions of this type arise among the simplest within the ansatz considered in Section \ref{sec:RR}: consider the ansatz
\es{}{
K(r,\vec{y})=f(r)g(\vec{y})\,,
}
in \eqref{PDE4d} or \eqref{PDE3d} (with $\vec{y}$ a two- or three-dimensional vector, respectively). Then, for $f(r)=\sqrt{a+b\,r}$ we find that $g$ is constrained by
\es{}{
\Delta_y g(\vec{y})=0\,,
}
where $\Delta_y$ is the Laplacian on $\R^2_y$ or $\R^3_y$, according to whether we are considering type IIA or type IIB. Replacing this in \eqref{IIAsol_rpsi} or \eqref{IIBsol_rpsi} gives a smeared solution as described above, where $f(r)$ fixes the distribution of NS5 branes and the harmonic function $g(\vec{y})$ determines the distribution of D-branes. The distribution $f(r)=\sqrt{a+b\,r}$ is the one describing NS5 branes on a segment (see \eqref{NS5bar}), but it is straightforward to generalize this setup to a more general solution of type IIA/IIB supergravity which does not in general arise from the uplift of a solution in lower dimensions, and generically has no other symmetries than super-Poincaré invariance on $\R^{1,3}_x$/$\R^{1,2}_x$ (in type IIA/IIB) with eight supercharges. 

In the string frame and expressing the RR fluxes in terms of electric potentials, the IIA solution reads
\es{IIAsmeared}{
\ds_{10}&=F^{-1}\,\ds(\R^{1,3}_x)+F\,\ds(\R^2_y)+H\,\ds(\R^4_z)\,,\\
\ex^{2\Phi}&=\frac{H}{F^2}\,,\\
B_6&=\frac{F}{H}\vol(\R^{1,3}_x)\wedge\vol(\R^2_y)\,,\\
C_5&=F^{-1}\vol(\R^{1,3}_x)\wedge\diff \Lambda\,,\\
C_7&=\frac{H}{F}\vol(\R^{1,3}_x)\wedge\star_z\diff \Lambda\,,
}
where $F\equiv F(\vec{y})$ and $H\equiv H(\vec{z})$,  $\tilde{F}$ is defined via
\es{}{
\star_y\diff \tilde{F}=\diff F\,,
}
and
\es{Lambdasmeared}{
\Lambda\equiv \vec{\lambda}\cdot \vec{z}\,,\quad ||\vec{\lambda}||^2=1\,,
}
so that the constant vector $\vec{\lambda}$ selects a direction in $\R^4_z$ parametrized by four Cartesian coordinates $\vec{z}$. The configuration given in \eqref{IIAsmeared} solves the equations of motion of type IIA and preserves eight supercharges provided that the functions $F$ and $H$ satisfy
\es{}{
\Delta_yF=0\,,\quad
\Delta_zH=0\,.
}
The interpretation of the solution is as follows: $H$ is an harmonic function in the $\R^4_z$ transverse to a set of NS5 branes, which are fully localized and whose location is identified by the singularities of $H$\footnote{As usual, one should really think of the equation $\Delta_z H=0$ as valid away from sources.}. The choice of $H$ selects a certain distribution of NS5 branes in $\R^4_z$ and the D4 and D6 branes are completely smeared in those directions, following the distribution of the NS5 branes. For instance, if the choice of $H$ preserves certain isometries in $\R^4_z$, so will the distribution of D4 and D6 branes. For this reason, the function $F$ governing the distribution of D-branes is only harmonic on $\R^2_y$, which is where the branes are localized. Moreover, given that $\diff \Lambda$ and $\star_z \diff\Lambda$ are orthogonal, from the electric potentials we can see that D4 and D6 branes are orthogonal. Hence, with a rotation in $\R^4_z$ we can set $z_1\equiv u$ and call $\R^3_v$ the space parametrized by $z_2,z_3,z_4$, which reproduces the branes setup of Table \ref{tab:IIAbranes4d}.

An interesting question is which solutions of this type can arise from the uplift of a 7d gauged supergravity solution. The answer is that the function $H$ is selected precisely in the same way as described in Section \ref{sec:NS5distributions}, where purely NSNS solutions with only scalars are discussed. On the other hand, the additional presence of the function $F$ and of the RR potentials arises from a choice of massive RR vectors. There are four of them in the 7d $ISO(4)$ gauged supergravity (after a gauge choice is made that eliminate the scalars), which we can think of as aligned along unit-norm four-vector. The latter is nothing but $\vec{\lambda}$ introduced in \eqref{Lambdasmeared}.

Applying a T-duality along one of the $\R^{1,3}_x$ directions and then guessing a form of the solution which is localized along that direction, it is straightforward to obtain an analogue of \eqref{IIAsmeared} describing a system of NS5-D3-D5 branes in type IIB supergravity, with the same features as the NS5-D4-D6 system described above. With the same caveats as above, the correspodning solution reads
\es{IIBsmeared}{
\ds_{10}&=F^{-1}\,\ds(\R^{1,2}_x)+F\,\ds(\R^3_y)+H\,\ds(\R^4_z)\,,\\
\ex^{2\Phi}&=\frac{H}{F}\,,\\
B_6&=\frac{F}{H}\vol(\R^{1,2}_x)\wedge\vol(\R^3_y)\,,\\
C_4^{\el}&=F^{-1}\vol(\R^{1,2}_x)\wedge\diff \Lambda\,,\\
C_6&=\frac{H}{F}\vol(\R^{1,2}_x)\wedge\star_z\diff \Lambda\,,
}
with the same conditions as above and the only difference that now $F\equiv F(\vec{y})$ depends on the three Cartesian coordinates $\vec{y}$ of $\R^3_y$ and is subject to $\Delta_yF=0$ where $\Delta_y$ is now the 3d Laplacian.

Finally, we note that the solutions \eqref{IIAsmeared} and \eqref{IIBsmeared} can be obtained performing a sequence of T-dualities on the solutions of \cite{Massar:1999sb,Kallosh:2001zc,Bergshoeff:2003sy}, where an analogous system with F1 strings as well as D0-D8 branes is analyzed. T-duality of course give a smearing at the level of supergravity, but it is straightforward to guess the localized form of the solution (at least along the T-duality directions). The same F1-D0-D8 system is also considered \cite{Roest:2004aqa}, where the author mentions the expectation that an analogous solution should be possible as a result of an uplift from 7d gauged supergravity. This is precisely what we realized here.

\section{Asymptotic expansion and Laplace approximation}
\label{app:Laplace-method}

In this appendix we review the Laplace method \cite{Miller:2006, Wojdylo:2004, Gergo} to compute the asymptotic expansion of integrals of the form 
\begin{equation}
    I(\rho) = \int_a^b e^{-\rho f(x)} g(x) \text dx \,, 
    \label{I(rho)}
\end{equation}
for $\rho \to \infty$. We assume $f(x)$ and $g(x)$ to be continuous functions and that as $x \to a^+$, 
\begin{align}
    f(x) & \sim f(a) + \sum_{k=0}^\infty a_k (x-a)^{k+\alpha} \,,  \label{f-Taylor} \\
    g(x) & \sim \sum_{k=0}^\infty b_k (x-a)^{k+\beta -1} \,, \label{g-Taylor}
\end{align}
with $\alpha >0$ and $\text{Re}(\beta) > 0$. Moreover, we assume that the only minimum of $f(x)$ in $[a,b]$ occurs at $x=a$. Under suitable assumptions (see {\it e.g.}~\cite{Miller:2006} for the precise statement), then
\begin{equation}
    I(\rho) \sim e^{-\rho f(a)} \sum_{n=0}^\infty \Gamma\Bigl(\frac{n + \beta}{\alpha}\Bigr) \frac{c_n}{\rho^{\frac{n + \beta}{\alpha}}} \,. 
    \label{Laplace-approx}
\end{equation}
The coefficients $c_n$ depend on the Taylor coefficients $a_k$ and $b_k$ entering eqs.~\eqref{f-Taylor} and \eqref{g-Taylor}. For example, 
\begin{equation}
   c_0 = \frac{b_0}{\alpha \, a_0^{\beta/\alpha}} \,, \qquad c_1 = \frac{1}{a_0^{(\beta+1)/ \alpha}} \left( \frac{b_1}{\alpha} - \frac{(\beta+1)a_1 b_0}{\alpha^2 a_0} \right) \,. 
\end{equation}
and 
\begin{equation}
  c_2 = \frac{1}{a_0^{(\beta+2)/\alpha}} \left( \frac{b_2}{\alpha} - \frac{(\beta+2) a_1 b_1}{\alpha^2 a_0} +((\alpha + \beta +2)a_1^2 - 2 \alpha a_0 a_2) \frac{(\beta+1)b_0}{2 \alpha^2 a_0^2}\right)  \ .
\end{equation}
and they can be computed as \cite{Wojdylo:2004, Gergo}
\begin{equation}
c_n = \frac{1}{\alpha} \sum_{k=0}^n \frac{b_{n-k}}{k!}\left[\frac{d^k}{dx^k} \left( \frac{(x-a)^\alpha}{f(x)-f(a)} \right)^{(n+\beta)/\alpha} \right]_{x=a} \,. 
\label{cn-Lapalace}
\end{equation}

\newpage

\providecommand{\href}[2]{#2}\begingroup\raggedright\endgroup

\end{fmffile}

\end{document}